\documentclass[10pt,twoside,a4paper]{article}

\usepackage{multicol}
\usepackage{amsmath}
\usepackage{amsfonts}
\usepackage{mathtools}
\usepackage{bm}
\usepackage{fancyhdr}
\usepackage{hyperref}

\renewcommand{\sectionmark}[1]%
 {\markboth{\thesection:\ #1}{}}

\renewcommand{\Re}{\mathrm{Re}}%

\numberwithin{equation}{section}

\begin{document}

\thispagestyle{empty}

\title{\bfseries Extended Comment on \\
``One-Range Addition Theorems for \\ Coulomb Interaction
Potential and Its Derivatives'' \\
by I.\ I.\ Guseinov \\
(Chem.\ Phys.\ Vol.\ 309 (2005), pp.\ 209 - 213)}

\author{Ernst Joachim Weniger \\
Institut f\"ur Physikalische und Theoretische Chemie \\
Universit\"at Regensburg, D-93040 Regensburg, Germany}

\date{Submitted to the Los Alamos Preprint Server: \\ 
31 December 2007 (Version 3)}

\maketitle

\begin{abstract}
  Addition theorems are principal tools that express a function $f
  (\bm{r} \pm \bm{r}')$ in terms of products of other functions that only
  depend on either $\bm{r}$ or $\bm{r}'$. The best known example of such
  an addition theorem is the Laplace expansion of the Coulomb potential
  which possesses a characteristic two-range form. Guseinov [Chem. Phys.
  {\bf 309}, 209 - 213 (2005)] derived one-range addition theorems for
  the Coulomb potential via the limit $\beta \to 0$ in previously derived
  one-range addition theorems for the Yukawa potential $\exp \bigl(
  -\beta \vert \bm{r}-\bm{r}'\vert \bigr) /\vert \bm{r}-\bm{r}'\vert$. At
  first sight, this looks like a remarkable achievement, but from a
  mathematical point of view, Guseinov's work is at best questionable and
  in some cases fundamentally flawed. One-range addition theorems are
  expansions in terms of functions that are complete and orthonormal in a
  given Hilbert space, but Guseinov replaced the complete and orthonormal
  functions by nonorthogonal Slater-type functions and rearranged the
  resulting expansions. This is a dangerous operation whose validity must
  be checked. It is shown that the one-center limit $\bm{r}' = \bm{0}$ of
  Guseinov's rearranged Yukawa addition theorems as well as of several
  other addition theorems does not exist. Moreover, the Coulomb potential
  does not belong to any of the Hilbert spaces implicitly used by
  Guseinov. Accordingly, one-range addition theorems for the Coulomb
  potential diverge in the mean. Instead, these one-range addition
  theorems have to interpreted as expansions of generalized functions in
  the sense of Schwartz that converge weakly in suitable functionals.
\end{abstract}


\newpage

\tableofcontents

\newpage


\begin{multicols}{2}

\typeout{==> Section: Introduction}
\section{Introduction}
\label{Sec:Intro}

In many subfields of physics and physical chemistry, an essential step
towards a solution of a problem consists in a separation of variables.
Principal mathematical tools, that can accomplish such a separation of
variables, are so-called addition theorems. These are expansions of a
given function $f (\bm{r} \pm \bm{r}')$ with $\bm{r}, \bm{r}' \in
\mathbb{R}^3$ in products of other functions that only depend on either
$\bm{r}$ or $\bm{r}'$. A review of the relevant literature with an
emphasis on electronic structure calculations can be found in
\cite{Weniger/2000a,Weniger/2002,Weniger/2005}. Applications of addition
theorems in classical physics are described in books by Martin
\cite{Martin/2006} and Jones \cite{Jones/1985}.

In atomic or molecular calculations, we are predominantly interested in
irreducible spherical tensors 
\begin{equation}
  \label{Def_IrrSphericalTensor}
F_{\ell}^{m} (\bm{r}) \; = \;
f_{\ell} (r) \, Y_{\ell}^{m} (\bm{r}/r) \, ,
\end{equation}
that can be represented as products of a radial function $f_{\ell} (r)$
multiplied by a surface spherical harmonic $Y_{\ell}^{m} (\theta, \phi)$
defined by (\ref{Def_Ylm}). The convenient orthonormality and
completeness properties of spherical harmonics make it highly desirable
that the functions of either $\bm{r}$ or $\bm{r}'$, which occur in the
expansion of $f (\bm{r} \pm \bm{r}')$, are also irreducible spherical
tensors of the type of (\ref{Def_IrrSphericalTensor}). Thus, addition
theorems are expansions in terms of products of spherical harmonics with
arguments $\theta, \phi = \bm{r}/r$ and $\theta', \phi' = \bm{r}'/r'$,
respectively.

The best known example of such an addition theorem is the Laplace
expansion of the Coulomb potential:
\begin{align}
  \label{LapExp}
  \frac{1}{\vert {\bm{r}} \pm {\bm{r}}' \vert} & \; = \; 4\pi \,
  \sum_{\lambda=0}^{\infty} \, \frac{(\mp 1)^{\lambda}}{2\lambda+1} \,
  \sum_{\mu=-\lambda}^{\lambda} \, \bigl[ \mathcal{Y}_{\lambda}^{\mu}
  ({\bm{r}_{<}}) \bigr]^{*} \, \mathcal{Z}_{\lambda}^{\mu} ({\bm{r}}_{>})
  \, ,
  \notag \\
  & \qquad \vert \bm{r}_{<} \vert = \min (r,r')\, , \vert \bm{r}_{>}
  \vert = \max (r,r') \, .
\end{align}
Here, ${\mathcal{Y}}_{\ell}^{m}$ and ${\mathcal{Z}}_{\ell}^{m}$ denote
regular and irregular solid harmonics defined by (\ref{Def_RegSolHar})
and (\ref{Def_IrregSolHar}), respectively.

The Laplace expansion (\ref{LapExp}) leads to a separation of the
variables ${\bm{r}}$ and ${\bm{r}}'$ and decouples integration variables
in expectation values of the Coulomb potential. However, the right-hand
side of (\ref{LapExp}) depends on ${\bm{r}}$ and ${\bm{r}}'$ only
indirectly via the two vectors $\bm{r}_{<}$ and $\bm{r}_{>}$ that satisfy
$\vert \bm{r}_{<} \vert < \vert \bm{r}_{>} \vert$. Hence, the Laplace
expansion has a two-range form, depending on the relative length of
$\bm{r}$ and $\bm{r}'$. This is a complication, which occurs frequently
among addition theorems and which can lead to nontrivial technical
problems.

As discussed in more details in
\cite{Weniger/2000a,Weniger/2002,Weniger/2005}, an addition theorem for
$f (\bm{r} \pm \bm{r}')$ can be derived by performing a three-dimensional
Taylor expansion (see for example \cite[p.\
181]{Biedenharn/Louck/1981a}):
\begin{equation}
  \label{ExpDifOp}
f (\bm{r} \pm \bm{r}') \; = \; \sum_{n=0}^{\infty} \,
\frac{(\pm \bm{r}' \cdot \nabla)^n}{n!} \, f (\bm{r})
\; = \;
\mathrm{e}^{\pm \bm{r}' \cdot \nabla} \, f (\bm{r}) \, .
\end{equation}
Thus, the translation operator
\begin{equation}
  \label{CartTransOp}
\mathrm{e}^{\bm{r}' \cdot \nabla} \; = \;
\mathrm{e}^{x' \partial /\partial x} \,
\mathrm{e}^{y' \partial /\partial y} \,
\mathrm{e}^{z' \partial /\partial z}
\end{equation}
generates $f (\bm{r} + \bm{r}')$ by performing a three-dimensional Taylor
expansion of $f$ around $\bm{r}$ with shift vector $\bm{r}'$. Since the
variables $\bm{r}$ and $\bm{r}'$ are separated, the series expansion
(\ref{ExpDifOp}) is indeed an addition theorem.

We could also expand $f$ around $\bm{r}'$ and use $\bm{r}$ as the shift
vector. This would produce an addition theorem for $f (\bm{r} \pm
\bm{r}')$ in which the roles of $\bm{r}$ and $\bm{r}'$ are interchanged.
Both approaches are mathematically legitimate and equivalent if $f$ is
analytic in the sense of complex analysis at $\bm{r}$, $\bm{r}'$, and
$\bm{r} \pm \bm{r}'$ for essentially arbitrary vectors $\bm{r}, \bm{r}'
\in \mathbb{R}^3$. Unfortunately, this is normally not true. Most
functions of interest in atomic and molecular electronic structure theory
are either singular or not analytic at the origin $\bm{r} = (0, 0, 0)$.
Obvious examples are the Coulomb potential, which is singular at the
origin, or the $1s$ hydrogen eigenfunction, which possesses a cusp at the
origin. In fact, all exponentially decaying function sets discussed in
this article are not analytic at the origin.

Nonanalyticity is not a mathematical sophistication that can safely be
ignored in practical applications. The radius of convergence of the
Taylor expansion of a function is determined by the location of its
nearest singularity or pole. If a function $f$ is not analytic at the
origin, an addition theorem derived via (\ref{ExpDifOp}) can only
converge if $\vert \bm{r} \vert > \vert \bm{r}' \vert$, which gives rise
to a two-range form. 

A two-range form can be quite inconvenient. The use of such an addition
theorem in a multicenter integral ultimately leads to indefinite
integrals of special functions, whose efficient and reliable evaluations
can be very difficult (I struggled with these problems long ago in my
diploma thesis \cite{Weniger/1977}, which was published in condensed form
in \cite{Steinborn/Weniger/1977}). In contrast, a $1s$ Gaussian $\exp (-
\beta r^2)$ with $\beta > 0$ is analytic for all $\bm{r} \in
\mathbb{R}^3$. This implies that its addition theorem, which can be
obtained via (\ref{ExpDifOp}), has a one-range form (see for example
\cite[Eq.\ (9)]{Kaufmann/Baumeister/1989} or also \cite[Eq.\
(17)]{ChowChiu/Moharerrzadeh/2001}).

The fact that addition theorems derived via (\ref{ExpDifOp}) are just
rearranged three-dimensional Taylor series automatically implies that
they converge \emph{pointwise}.

It is one of the central results of modern mathematics that convergence
is not an abstract or global property. It depends very much on how we
measure it. Pointwise convergence is a very demanding requirement.
Moreover, it is not really necessary if we only want to use addition
theorems in multicenter integrals. Therefore, it makes sense to wonder,
whether computational benefits can be achieved by relaxing the
requirement of pointwise convergence and by demanding instead a weaker
form of convergence.

This is indeed possible. Often, one-range addition theorems can be
constructed that are essentially expansions of $f (\bm{r} \pm \bm{r}')$
in terms of complete orthonormal function sets. Such an addition theorem
converges in the mean with respect to the norm of the corresponding
Hilbert space. This had already been done by several authors (see for
example \cite{Filter/1978,Filter/Steinborn/1980,%
  Steinborn/Filter/1980,Steinborn/1982,Novosadov/1983,Yamaguchi/1983,%
  Weniger/Steinborn/1984,Weniger/1985,Homeier/Weniger/Steinborn/1992a}
and references therein). Guseinov also derived and applied one-range
addition theorems in several articles
\cite{Guseinov/1977b,Guseinov/1980a,Guseinov/1985a}.  

In recent years, Guseinov and his coworkers made an industry out of
one-range addition theorems of exponentially decaying functions.  They
published -- in addition to several other articles on different topics --
a very long list of articles on the derivation and application of
one-range addition theorems
\cite{Guseinov/2001a,Guseinov/2002b,Guseinov/2002c,Guseinov/2002d,%
  Guseinov/2003b,Guseinov/2003c,Guseinov/2003d,Guseinov/2003e,%
  Guseinov/2004a,Guseinov/2004b,Guseinov/2004c,Guseinov/2004d,%
  Guseinov/2004e,Guseinov/2004f,Guseinov/2004i,Guseinov/2004k,%
  Guseinov/2005a,Guseinov/2005b,Guseinov/2005c,Guseinov/2005d,%
  Guseinov/2005e,Guseinov/2005f,Guseinov/2005g,Guseinov/2006a,%
  Guseinov/2007e,Guseinov/2007f,Guseinov/2007g,%
  Guseinov/2006b,Guseinov/Aydin/Mamedov/2003,Guseinov/Mamedov/2001c,%
  Guseinov/Mamedov/2002d,Guseinov/Mamedov/2003,Guseinov/Mamedov/2004b,%
  Guseinov/Mamedov/2004d,Guseinov/Mamedov/2004e,Guseinov/Mamedov/2004g,%
  Guseinov/Mamedov/2004h,Guseinov/Mamedov/2005c,Guseinov/Mamedov/2005d,%
  Guseinov/Mamedov/2005g,Guseinov/Mamedov/Orbay/2000,%
  Guseinov/Mamedov/Rzaeva/2001,Guseinov/Mamedov/Rzaeva/2002,%
  Guseinov/Mamedov/Suenel/2002,Guseinov/Mamedov/Oener/Hueseyin/2001,%
  Guseinov/Mamedov/Oezdogan/Orbay/1999,Guseinov/Mamedov/Orbay/2000,%
  Guseinov/Rzaeva/Mamedov/Orbay/Oezdogan/Oener/1999}, that are often
remarkably similar and that do not give due credit to the previous work
of others.

At first sight, this long list of articles looks quite impressive.
Unfortunately, from a mathematical point of view Guseinov's treatment of
one-range addition theorems is at best questionable and in some cases
fundamentally flawed. Guseinov and his coworkers ignored basic facts of
Hilbert space and approximation theory as well as all questions of
convergence and existence. Such a cavalier attitude is not uncommon among
scientists, but it is completely unacceptable in the context of addition
theorems and multicenter integrals which are essentially a mathematical
topic.

As discussed in this article, Guseinov's approach leads to serious
problems, which are particularly evident in the case of Guseinov's
derivation of a one-range addition theorem for the Coulomb potential
\cite{Guseinov/2005a}. However, as discussed in more details later,
serious mathematical problems persist also in all other articles dealing
with one-range addition theorem. Therefore, the title of this article is
somewhat misleading. In this article, I will try to provide a reasonably
balanced and detailed discussion of one-range addition theorems.
Accordingly, my criticism is not limited to \cite{Guseinov/2005a} and
extends to all articles by Guseinov and coworkers dealing with one-range
addition theorems
\cite{Guseinov/2001a,Guseinov/2002b,Guseinov/2002c,Guseinov/2002d,%
  Guseinov/2003b,Guseinov/2003c,Guseinov/2003d,Guseinov/2003e,%
  Guseinov/2004a,Guseinov/2004b,Guseinov/2004c,Guseinov/2004d,%
  Guseinov/2004e,Guseinov/2004f,Guseinov/2004i,Guseinov/2004k,%
  Guseinov/2005a,Guseinov/2005b,Guseinov/2005c,Guseinov/2005d,%
  Guseinov/2005e,Guseinov/2005f,Guseinov/2005g,Guseinov/2006a,%
  Guseinov/2007e,Guseinov/2007f,Guseinov/2007g,%
  Guseinov/2006b,Guseinov/Aydin/Mamedov/2003,Guseinov/Mamedov/2001c,%
  Guseinov/Mamedov/2002d,Guseinov/Mamedov/2003,Guseinov/Mamedov/2004b,%
  Guseinov/Mamedov/2004d,Guseinov/Mamedov/2004e,Guseinov/Mamedov/2004g,%
  Guseinov/Mamedov/2004h,Guseinov/Mamedov/2005c,Guseinov/Mamedov/2005d,%
  Guseinov/Mamedov/2005g,Guseinov/Mamedov/Orbay/2000,%
  Guseinov/Mamedov/Rzaeva/2001,Guseinov/Mamedov/Rzaeva/2002,%
  Guseinov/Mamedov/Suenel/2002,Guseinov/Mamedov/Oener/Hueseyin/2001,%
  Guseinov/Mamedov/Oezdogan/Orbay/1999,Guseinov/Mamedov/Orbay/2000,%
  Guseinov/Rzaeva/Mamedov/Orbay/Oezdogan/Oener/1999}.

Guseinov's starting point for his one-range addition theorems of the
Coulomb potential was a class of one-range addition theorems
\cite{Guseinov/2001a,Guseinov/2002b,Guseinov/2002c,Guseinov/2002d} for
Slater-type functions which in unnormalized form can be expressed as
follows:
\begin{equation}
  \label{Def_STF}
\chi_{N, L}^{M} (\beta, \mathbf{r}) \; = \;
(\beta r)^{N-L-1} \, \mathrm{e}^{- \beta r} \,
\mathcal{Y}_{L}^{M} (\beta \bm{r}) \, .
\end{equation}
Here, $N$ is some kind of generalized principal quantum number, $L$ and
$M$ are the usual angular momentum quantum numbers, and $\beta$ is a
positive scaling parameter. In the vast majority of articles dealing with
Slater-type functions, $N$ is assumed to be a positive integer $\ge 1$,
i.e., $N \in \mathbb{N}$. However, several authors -- among them Guseinov
and coworkers
\cite{Guseinov/2002b,Guseinov/2002d,Guseinov/2003a,Guseinov/2003b,%
  Guseinov/2003d,Guseinov/2003e,Guseinov/2004b,Guseinov/2004c,%
  Guseinov/2005f,Guseinov/Mamedov/2002d,Guseinov/Mamedov/2003,%
  Guseinov/Mamedov/2004b,Guseinov/Mamedov/2004d,Guseinov/Mamedov/2004e,%
  Guseinov/Mamedov/2004g,Guseinov/Mamedov/2004h,Guseinov/Mamedov/2005c,%
  Guseinov/Mamedov/2005d,Guseinov/2005e,Guseinov/Mamedov/Suenel/2002,%
  Guseinov/2003a,Guseinov/2004g,Guseinov/2004j,Guseinov/Mamedov/2002a,%
  Guseinov/Mamedov/2002e,Guseinov/Mamedov/2005h,%
  Guseinov/Mamedov/Suenel/2002} -- considered also nonintegral principal
quantum numbers $N \in \mathbb{R} \setminus \mathbb{N}$.

Guseinov's one-range addition theorems for Slater-type functions with
nonintegral principal quantum numbers
\cite{Guseinov/2002b,Guseinov/2002c,Guseinov/2002d} contain as the
special case $N=L=M=0$ addition theorems for the Yukawa potential $\exp
(- \beta r)/r$ \cite{Yukawa/1935}, which may be viewed to be an
exponentially screened Coulomb potential. Guseinov derived his one-range
addition theorems for the Coulomb potential by exploiting the obvious
relationship $1/r = \lim_{\beta \to 0} \exp (- \beta r)/r$ in the
one-range addition theorems for the Yukawa potential.

At first sight, Guseinov's derivation of a class of one-range addition
theorems for the Coulomb potential looks like a remarkable achievement.
So far, the Laplace expansion (\ref{LapExp}) has routinely been used in
atomic and molecular electronic structure calculations, but its
characteristic two-range form can easily lead to nontrivial technical
problems.  Obviously, a one-range addition theorem for the Coulomb
potential should simplify subsequent integrations in multicenter
integrals substantially.

As already remarked above, one-range addition theorems are essentially
expansions of a given function $f (\bm{r} \pm \bm{r}')$ belonging to a
suitable Hilbert space in terms of functions that are complete and
orthonormal in this Hilbert space. It is an essential feature of
orthogonal expansions in general and of one-range addition theorems in
special that they only converge in the mean with respect to the norm of
the underlying Hilbert space, but not necessarily pointwise (see for
example \cite{Askey/Wainger/1965}).

The fact, that orthogonal expansions only converge in the mean if the
function, which is to be expanded, is an element of the corresponding
Hilbert space, has an obvious consequence: Since the Coulomb potential
does not belong to any of the Hilbert spaces, which Guseinov implicitly
used and which all involve an integration over the whole $\mathbb{R}^3$,
Guseinov's one-range addition theorems for the Coulomb potential all
diverge in the mean.

But this is not the only problem. For the derivation of his addition
theorems, Guseinov had first expanded Slater-type functions $\chi_{N,
  L}^{M} (\beta, \mathbf{r} \pm \bm{r}')$ with arbitrary principal
quantum numbers $N \in \mathbb{R}$ in terms of the functions
$\prescript{}{k}{\Psi}_{n, \ell}^{m} (\beta, \bm{r})$ defined by
(\ref{Def_Psi_Guseinov}), yielding the addition theorems
(\ref{Gus_OneRangeAddTheorSTF_k}). The functions
$\prescript{}{k}{\Psi}_{n, \ell}^{m} (\beta, \bm{r})$ with $k = -1, 0, 1,
2, \dots$, which Guseinov had introduced in \cite[Eq.\
(1)]{Guseinov/2002b}, are -- as discussed in Section \ref{Sec:LagTypeFun}
-- complete and orthonormal in the weighted Hilbert space $L_{r^k}^{2}
(\mathbb{R}^3)$ defined by (\ref{HilbertL_r^k^2}). As discussed in more
details in Section (\ref{Sec:LagTypeFun}), Guseinov's functions
generalize some other, well established function sets also based on the
generalized Laguerre polynomials.

For some reasons, which had never had been explained comprehensively and
which I thus do not really understand, Guseinov considered it to be
advantageous to replace in his addition theorems
(\ref{Gus_OneRangeAddTheorSTF_k}) his complete and orthonormal functions
$\prescript{}{k}{\Psi}_{n, \ell}^{m} (\beta, \bm{r})$ by nonorthogonal
Slater-type functions with integral principal quantum numbers and to
rearrange the order of summations of his expansions. In this way,
Guseinov obtained the expansions
(\ref{Gus_RearrOneRangeAddTheorSTF_STF_k}) of Slater-type functions with
in general nonintegral principal quantum numbers in terms of Slater-type
functions with integral principal quantum numbers located at a different
center.

Due to the intrinsic complexity of addition theorems, it is by no means
easy to decide whether such a rearrangement is legitimate and whether the
resulting expansions (\ref{Gus_RearrOneRangeAddTheorSTF_STF_k}) are
mathematically meaningful. In general, this is an open question, but if
the principal quantum number of $\chi_{N, L}^{M} (\beta, \mathbf{r} \pm
\bm{r}')$ is nonintegral, $N \in \mathbb{R} \setminus \mathbb{N}$, an
affirmative answer is possible: The one-center limit $\bm{r}' = \bm{0}$
in (\ref{Gus_RearrOneRangeAddTheorSTF_STF_k}) does not exist, which means
that Guseinov's rearranged addition theorems
(\ref{Gus_RearrOneRangeAddTheorSTF_STF_k}) do not exist for the whole
argument set $\mathbb{R}^3 \times \mathbb{R}^3$.

This conclusion is highly consequential: The rearranged addition theorems
(\ref{Gus_RearrOneRangeAddTheorSTF_STF_k}) with $N=L=M=0$ were the
starting point for Guseinov's derivation of one-range addition theorems
for the Coulomb potential. Thus, Guseinov obtained his one-range addition
theorems for the Coulomb potential via addition theorems for the Yukawa
potential that do not exist for the whole argument set.

Guseinov was not the first one who had derived a divergent expansion of
the Coulomb potential in terms of a complete and orthonormal function
set. Salmon, Birss, and Ruedenberg \cite{Salmon/Birss/Ruedenberg/1968}
derived a bipolar expansion of the Coulomb potential in terms of the
Gaussian-type eigenfunctions of a three-dimensional isotropic harmonic
oscillator defined by (\ref{Def_OscillFun}) which are complete and
orthonormal in the Hilbert space $ L^{2} (\mathbb{R}^3)$ of square
integrable functions defined by (\ref{HilbertL^2}) (see for instance
\cite[Section V]{Weniger/1985}). However, Silverstone and Kay
\cite{Silverstone/Kay/1969} demonstrated that this bipolar expansion
diverges. This observation was later confirmed by Ruedenberg and Salmon
\cite{Ruedenberg/Salmon/1969}.

Because of their divergence, one might dismiss Guseinov's one-range
addition theorems of the Coulomb potential to be practically useless,
just as the expansion of Salmon, Birss, and Ruedenberg
\cite{Salmon/Birss/Ruedenberg/1968} had been dismissed. However, this
would be premature and the situation is actually more complicated but
also less hopeless than it may appear at first sight. As discussed in
Section \ref{Sec:WeakConvAddTheor_CoulombPot}, these addition theorems
can be interpreted as weakly convergent expansions of generalized
functions, that are mathematically meaningful and yield convergent
results when used in suitable functionals.

It is quite obvious that Guseinov had failed to understand the
mathematical theory behind one-range addition theorems. Thus is bad
enough. However, the referees of Guseinov's numerous recent articles on
one-range addition theorems and related topics
\cite{Guseinov/2001a,Guseinov/2002b,Guseinov/2002c,Guseinov/2002d,%
  Guseinov/2003b,Guseinov/2003c,Guseinov/2003d,Guseinov/2003e,%
  Guseinov/2004a,Guseinov/2004b,Guseinov/2004c,Guseinov/2004d,%
  Guseinov/2004e,Guseinov/2004f,Guseinov/2004i,Guseinov/2004k,%
  Guseinov/2005a,Guseinov/2005b,Guseinov/2005c,Guseinov/2005d,%
  Guseinov/2005e,Guseinov/2005f,Guseinov/2005g,Guseinov/2006a,%
  Guseinov/2007e,Guseinov/2007f,Guseinov/2007g,%
  Guseinov/2006b,Guseinov/Aydin/Mamedov/2003,Guseinov/Mamedov/2001c,%
  Guseinov/Mamedov/2002d,Guseinov/Mamedov/2003,Guseinov/Mamedov/2004b,%
  Guseinov/Mamedov/2004d,Guseinov/Mamedov/2004e,Guseinov/Mamedov/2004g,%
  Guseinov/Mamedov/2004h,Guseinov/Mamedov/2005c,Guseinov/Mamedov/2005d,%
  Guseinov/Mamedov/2005g,Guseinov/Mamedov/Orbay/2000,%
  Guseinov/Mamedov/Rzaeva/2001,Guseinov/Mamedov/Rzaeva/2002,%
  Guseinov/Mamedov/Suenel/2002,Guseinov/Mamedov/Oener/Hueseyin/2001,%
  Guseinov/Mamedov/Oezdogan/Orbay/1999,Guseinov/Mamedov/Orbay/2000,%
  Guseinov/Rzaeva/Mamedov/Orbay/Oezdogan/Oener/1999} apparently also
failed to understand this theory. Therefore, I will try to give a
compact, but hopefully comprehensive treatment of one-range addition
theorems and of the mathematical tools, which are needed for their
derivation, interpretation, and application.

Section \ref{Sec:HilbertSpace} gives a compact review of the for our
purposes most fundamental properties of Hilbert spaces in the context of
approximation theory and of orthogonal expansions. Section
\ref{Sec:OneRanAddTheor} discusses the derivation and the basic
properties of one-range addition theorems. In Section
\ref{Sec:LagTypeFun}, complete and orthonormal Laguerre-type function
sets and their corresponding Hilbert spaces are discussed. Section
\ref{AddTheor_ETF} discusses the derivation of one-range addition
theorems for Slater-type functions in terms of Guseinov's functions
$\prescript{}{k}{\Psi}_{n, \ell}^{m} (\beta, \bm{r})$ using techniques
that were applied by Filter and Steinborn \cite{Filter/Steinborn/1980}
for the derivation of related addition theorems. Section
\ref{Sec:AddTheor_STF} discusses Guseinov's highly questionable
derivation of one-range addition theorems for Slater-type functions.
Section \ref{Sec:WeakConvAddTheor_CoulombPot} discusses how divergent
one-range addition theorems of the Coulomb potential can be interpreted
as weakly convergent expansions that are mathematically meaningful in
suitable functionals. Section \ref{Sec:DiffTechYlmNabla} discusses
alternatives to the differentiation techniques, which Guseinov had used
for the generation of new one-range addition theorems and which are
neither convenient from a technical point of view nor mathematically
justified. This article is concluded by a short summary in Section
\ref{Sec:SummConclu}. Finally, there is Appendix \ref{App:YlmGaunt}
listing some basic facts about spherical harmonics and Gaunt
coefficients.

\typeout{==> Section: Basic Hilbert Space Theory}
\section{Basic Hilbert Space Theory}
\label{Sec:HilbertSpace}

Let us assume that $\mathcal{V}$ is a \emph{vector space} over the
complex numbers $\mathbb{C}$ that possesses an \emph{inner product} $(
\cdot \vert \cdot ) \colon \mathcal{V} \times \mathcal{V} \to \mathbb{C}$
(see for instance \cite[p.\ 36]{Reed/Simon/1980}). If $\mathcal{V}$ is
complete with respect to the norm $\Vert \cdot \Vert \colon \mathcal{V}
\to \mathbb{R}_{+}$ defined by
\begin{equation}
  \label{DefNorm_V}
\Vert u \Vert \; = \; \sqrt{(u \vert u)} \, ,
\end{equation}
i.e., if every Cauchy sequence in $\mathcal{V}$ converges with respect to
(\ref{DefNorm_V}) to an element of $\mathcal{V}$, then $\mathcal{V}$ is
called a Hilbert space. In the mathematical literature, convergence in
the means with respect to the norm $\Vert \cdot \Vert$ is frequently
called \emph{strong} convergence, in contrast to \emph{weak} convergence
which will be discussed later in Section
\ref{Sec:WeakConvAddTheor_CoulombPot}.

Hilbert spaces play a major role in various branches of mathematics and
mathematical physics and in particular also in approximation theory. If
$f$ is an element of some Hilbert space $\mathcal{H}$ and if $\{
\varphi_n \}_{n=0}^{\infty}$ is linearly independent and complete in
$\mathcal{H}$, then we can construct approximations
\begin{equation}
  \label{f_FinAppr}
f_N \; = \; \sum_{n=0}^{N} C_{n}^{(N)} \varphi_n
\end{equation}
to $f$, where $N$ is a finite integer. The coefficients $C_{n}^{(N)}$ are
chosen in such a way that the mean square deviation $\Vert f - f_N
\Vert^2 = (f - f_N \vert f - f_N)$ becomes minimal.

It is one of the central results of approximation theory that $\Vert f -
f_N \Vert^2$ becomes minimal if $\{ \varphi_n \}_{n=0}^{\infty}$ is an
\emph{orthonormal} function set satisfying $(\varphi_n \vert
\varphi_{n'}) = \delta_{n {n'}}$ for all indices $n$ and $n'$ and if the
coefficients are chosen according to $C_{n}^{(N)} = (\varphi_n \vert f)$
(see for example \cite[Theorem 9 on p.\ 51]{Davis/1989}). Since the
coefficients $(\varphi_n \vert f)$ do not depend on the truncation order
$N$ in $f_N$, $f$ possesses an expansion
\begin{equation}
  \label{Expand_f_CONS}
f \; = \; \sum_{n=0}^{\infty} \, (\varphi_n \vert f) \, \varphi_n
\end{equation}
in terms of the orthogonal function set $\{ \varphi_n \}_{n=0}^{\infty}$
that converges in the mean with respect to the norm $\Vert \cdot \Vert$
of the Hilbert space $\mathcal{H}$.

This result emphasizes the central role of orthogonal expansions both in
Hilbert spaces as well as in approximation theory. If the functions $\{
\varphi_n \}_{n=0}^{\infty}$ are only normalized satisfying $(\varphi_n
\vert \varphi_n) = 1$ but not orthogonal, then it is in general only
possible to construct finite approximations $f_N$ of the type of
(\ref{f_FinAppr}) by minimizing the mean square deviation $\Vert f - f_N
\Vert^2 = (f - f_N \vert f - f_N)$, but we cannot tacitly assume that the
coefficients $C_n^{(N)}$ in $f_N$ converge to a well defined limit $C_n =
C_n^{(\infty)}$ as $N \to \infty$. Accordingly, the existence of an
expansion
\begin{equation}
  \label{Expand_f_formal}
f \; = \; \sum_{n=0}^{\infty} \, C_n \, \varphi_n
\end{equation}
is not guaranteed if the functions $\{ \varphi_n \}_{n=0}^{\infty}$ are
not orthogonal. This fact is well documented both in the mathematical
literature (see for example \cite[Theorem 10 on p.\ 54]{Davis/1989} or
\cite[Section 1.4]{Higgins/1977}) as well as in the literature on
electronic structure calculations
\cite{Klahn/1975,Klahn/Bingel/1977a,Klahn/Bingel/1977b,%
  Klahn/Bingel/1977c,Klahn/1981,Klahn/Morgan/1984}). 

Thus, completeness of a function set $\{ \varphi_n \}_{n=0}^{\infty}$ in
a Hilbert space $\mathcal{H}$ does not suffice to guarantee the existence
of formal expansions in terms of these functions: These expansions may or
they may not exist. Horrifying examples of pathologies of nonorthogonal
expansions can be found in \cite[Section III.I]{Klahn/1981}.

There is another, practically very consequential aspect of orthogonal
expansions of the type of (\ref{Expand_f_CONS}); The coefficients
$(\varphi_n \vert f)$ of orthogonal expansions satisfy Parseval's
equality (see for example \cite[Eq.\ (II.2) on p.\ 45]{Reed/Simon/1980})
\begin{equation}
  \label{ParsevalEquality}
\Vert f \Vert^2 \; = \; 
\sum_{n=0}^{\infty} \, \vert (\varphi_n \vert f) \vert^2 \, .
\end{equation}
Thus, the coefficients $(\varphi_n \vert f)$ are bounded in magnitude and
they have to vanish as $n \to \infty$. This may well be the main reason
why orthogonal expansions tend to be computationally well behaved. In the
case of nonorthogonal expansions of the type of (\ref{Expand_f_formal})
it can instead happen that the expansion coefficients $C_n$ are
unbounded, have alternating signs, and increase in magnitude with
increasing index (see for example \cite[Appendix E on pp.\ 162 -
164]{Filter/1978} or \cite[Table I on p.\ 166]{Klahn/1981}). Such a
behavior can easily lead to a cancellation of significant digits and to a
catastrophic accumulation of rounding errors.

If we remove a single function from a complete and orthonormal function
set, it becomes incomplete. In the case of nonorthogonal expansions of
the type of (\ref{Expand_f_formal}), the situation is much more
complicated since nonorthogonal function sets are in general
\emph{overcomplete} as well as \emph{almost linearly dependent} (see for
example \cite{Klahn/Bingel/1977c,Klahn/1981} and references therein).
Again, this can be the source of serious numerical problems.

Of course, there are situations in which nonorthogonal expansions offer
computational advantages (see fore example the discussion in
\cite{Daubechies/Grossmann/Meyer/1986}). However, in the vast majority of
all cases, orthogonal expansions are clearly superior. Consequently, one
should should not voluntarily abandon the highly useful feature of
orthogonality unless there are truly compelling reasons.

The natural inner product for effective one-particle wave functions $f, g
\colon \mathbb{R}^3 \to \mathbb{C}$ in atomic and molecular electronic
structure calculations on the basis of the Hartree-Fock-Roothaan
equations \cite{Hartree/1928,Fock/1930,Roothaan/1951} is given by the
integral
\begin{equation}
  \label{InnerProd}
( f \vert g )_2 \; = \;
\int \bigl[ f (\bm{r}) \bigr]^{*} \, g (\bm{r}) \,
\mathrm{d}^3 \bm{r} \, .
\end{equation}
As always in this article, integration extends over the whole
$\mathbb{R}^3$. The corresponding norm satisfies
\begin{equation}
  \label{Norm_2}
\Vert f \Vert_2 \; = \; \sqrt{( f \vert f )_2} \, .
\end{equation}
Accordingly, the Hilbert space 
\begin{align}
  \label{HilbertL^2}
  L^{2} (\mathbb{R}^3) & \; = \; \Bigl\{ f \colon \mathbb{R}^3 \to
  \mathbb{C} \Bigm\vert \, \int \, \vert f (\bm{r}) \vert^2 \,
  \mathrm{d}^3 \bm{r} < \infty \Bigr\} 
  \notag \\[1\jot]
  & \; = \; \bigl\{ f \colon \mathbb{R}^3 \to \mathbb{C} \big\vert \,
  \Vert f \Vert_2 < \infty \bigr\}
\end{align}
of square-integrable one-particle wave functions provides the natural
setting for atomic and molecular electronic structure calculations in
general as well as for the derivation of one-range addition theorems in
special.

The whole formalism of inner products, norms, and function spaces can be
generalized to include weight functions $w \colon \mathbb{R}^3 \to
\mathbb{R}_{+}$. If $w (\bm{r}) \ge 0$ is such a positive weight
function, we can define an inner product with respect to the weight
function $w$ for functions $f, g \colon \mathbb{R}^3 \to \mathbb{C}$
according to
\begin{equation}
  \label{InnerProd_w}
( f \vert g )_{w, 2} \; = \;
\int \bigl[ f (\bm{r}) \bigr]^{*} \, w (\bm{r}) \,
g (\bm{r}) \, \mathrm{d}^3 \bm{r} \, .
\end{equation}
On the basis of the inner product (\ref{InnerProd_w}), the norm of a
function $f \colon \mathbb{R}^3 \to \mathbb{C}$ with respect to the
weight function $w$ is defined according to
\begin{equation}
  \label{Norm_w_2}
\Vert f \Vert_{w, 2} \; = \; \sqrt{( f \vert f )_{w, 2}} \, ,
\end{equation}
and the Hilbert space $L_{w}^{2} (\mathbb{R}^3)$ of square integrable
functions with respect to the weight function $w$ is defined via the norm
(\ref{Norm_w_2}) according to
\begin{align}
  \label{HilbertL_w^2}
  L_{w}^{2} (\mathbb{R}^3) & \; = \; \Bigl\{ f \colon \mathbb{R}^3 \to
  \mathbb{C} \Bigm\vert \, \int \, w (\bm{r}) \, \vert f (\bm{r})
  \vert^2 \,
  \mathrm{d}^3 \bm{r} < \infty \Bigr\} \notag \\[1\jot]
  & \; = \; \bigl\{ f \colon \mathbb{R}^3 \to \mathbb{C} \big\vert \,
  \Vert f \Vert_{w, 2} < \infty \bigr\} \, .
\end{align}

By augmenting the inner product (\ref{InnerProd}) by a nontrivial weight
function $w (\bm{r}) \ne 1$ according to (\ref{InnerProd_w}), it should
at least in principle be possible to accomplish some fine-tuning in
approximation processes. So, it looks like an obvious idea to use instead
of $L^{2} (\mathbb{R}^3)$ a weighted Hilbert $L_{w}^{2} (\mathbb{R}^3)$
that is better adapted to the problem under consideration. 

Unfortunately, there are some principal problems which must not be
ignored. In general, we neither have $L_{w}^{2} (\mathbb{R}^3) \subset
L^{2} (\mathbb{R}^3)$ nor $L^{2} (\mathbb{R}^3) \subset L_{w}^{2}
(\mathbb{R}^3)$. Thus, the two Hilbert spaces $L^{2} (\mathbb{R}^3)$ and
$L_{w}^{2} (\mathbb{R}^3)$ are mathematically inequivalent. Accordingly,
the expansion of a function $f \in L^{2} (\mathbb{R}^3)$ in terms of a
function set, that is complete and orthonormal in $L_{w}^{2}
(\mathbb{R}^3)$, does not necessarily lead to a convergent result.
Moreover, expansions of a given function in terms of function sets, that
are complete and orthonormal in either $L^{2} (\mathbb{R}^3)$ or in
$L_{w}^{2} (\mathbb{R}^3)$, respectively, can also be computationally
different, since they can have substantially different rates of
convergence (see the simple example discussed in Section
\ref{Sec:LagTypeFun}).

Even the Hilbert space $L^{2} (\mathbb{R}^3)$ of square integrable
functions is -- loosely speaking -- too large for electronic structure
calculations on the basis of the Hartree-Fock-Roothaan equations.  All
effective one-particle wave functions must belong to $L^{2}
(\mathbb{R}^3)$, but the converse is not necessarily true, i.e., there
are elements of $L^{2} (\mathbb{R}^3)$ that cannot be used in electronic
structure calculations. An example is Yukawa potential $\exp (- \beta
r)/r$. It belongs to $L^{2} (\mathbb{R}^3)$, but it cannot be used as a
trial function in electronic structure calculations since the expectation
values of the kinetic energy and of the Coulomb potential do not exist.

The necessary exclusion of unsuitable elements from $L^{2}
(\mathbb{R}^3)$ can be accomplished via some other generalizations of
$L^{2}$ spaces, the so-called \emph{Sobolev spaces} (see for example
\cite{Sobolev/1963,Adams/1975,Mazja/1985}), which are of considerable
importance not only in electronic structure theory. The in the context of
electronic structure calculations most important Sobolev space
$W_{2}^{(1)} (\mathbb{R}^3)$ can be defined via the following inner
product for differentiable functions $f, g \colon \mathbb{R}^3 \to
\mathbb{C}$:
\begin{equation}
  \label{Sobolev_InnerProd}
\langle f \vert g \rangle_{2, 1} \; = \;
\int \bigl[ f (\bm{r}) \bigr]^{*} \, \frac{\eta^2 - \nabla^2}{2\eta^2}
\, g (\bm{r}) \, \mathrm{d}^3 \bm{r} \, .
\end{equation}
Here, $\nabla = \bigl( \partial/\partial x, \partial/\partial y,
\partial/\partial z \bigr)$ is the three-dimensional gradient operator,
and $\eta$ is a real constant. As usual, integration extends over the
whole $\mathbb{R}^3$.

On the basis of the inner product (\ref{Sobolev_InnerProd}), the
following Sobolev-type norm of a differentiable function $f \colon
\mathbb{R}^3 \to \mathbb{C}$ can be defined:
\begin{equation}
  \label{Sobolev_Norm_2}
\langle \! \langle f \rangle \! \rangle_{2, 1} \; = \;
\sqrt{\langle f \vert f \rangle_{2, 1}} \, .
\end{equation}
Both the inner product $\langle f \vert g \rangle_{2, 1}$ as well as the
norm $\langle \! \langle f \rangle \! \rangle_{2, 1}$ depend on the real
parameter $\eta$. Since it can be shown that the norms defined via
(\ref{Sobolev_Norm_2}) are equivalent for all $\eta \in \mathbb{R}
\setminus {0}$ and thus give rise to the same topology, their dependence
on $\eta$ is not explicitly indicated.

The Sobolev space $W_{2}^{(1)} (\mathbb{R}^3)$ is defined via the norm
(\ref{Sobolev_Norm_2}) according to
\begin{align}
  \label{Sobolev_W_2^1}
  & W_{2}^{(1)} (\mathbb{R}^3)
  \notag \\
  & \; = \; \Bigl\{ f \colon \mathbb{R}^3 \to
  \mathbb{C} \Bigm\vert \, \int \bigl[ f (\bm{r}) \bigr]^{*} \,
  \frac{\eta^2 - \nabla^2}{2\eta^2} \, f (\bm{r}) \,
  \mathrm{d}^3 \bm{r} < \infty \Bigr\} \notag \\[1\jot]
  & \; = \; \bigl\{ f \colon
 \mathbb{R}^3 \to \mathbb{C} \big\vert \, \langle
  \! \langle f \rangle \! \rangle_{2, 1} < \infty \bigr\} \, .
\end{align}
Obviously, the Sobolev space $W_{2}^{(1)} (\mathbb{R}^3)$ is also a
Hilbert space. In addition, $W_{2}^{(1)} (\mathbb{R}^3)$ is also a
\emph{proper} subspace of the Hilbert space $L^{2} (\mathbb{R}^3)$, i.e.,
$W_{2}^{(1)} (\mathbb{R}^3) \subset L^{2} (\mathbb{R}^3)$. Nevertheless,
approximation processes in $L^{2} (\mathbb{R}^3)$ and $W_{2}^{(1)}
(\mathbb{R}^3)$, respectively, can differ substantially. Loosely
speaking, we can say that an approximation scheme in $L^{2}
(\mathbb{R}^3)$ approximates $f$ in the mean, but in $W_{2}^{(1)}
(\mathbb{R}^3)$ both $f$ as well as $\nabla f$ are approximated in the
mean. Thus, in Sobolev spaces certain possible pathologies can be
avoided. This can have far-reaching consequences in quantum mechanical
calculations which are essentially special approximation processes. In
\cite[Section 9]{Klahn/Bingel/1977a}, it was shown that completeness of a
one-particle basis in $L^{2} (\mathbb{R}^3)$ does not suffice to
guarantee the convergence of computations based on the Rayleigh-Ritz
variational principle. This is only guaranteed if the one-particle basis
is complete in $W_{2}^{(1)} (\mathbb{R}^3)$ (see also
\cite{Fonte/1981,Hill/1998}).

\typeout{==> Section: One-Range Addition Theorems}
\section{One-Range Addition Theorems}
\label{Sec:OneRanAddTheor}
  
Let us assume that $f \in L^2 (\mathbb{R}^3)$ and that the functions $\{
\varphi_{n, \ell}^{m} (\bm{r}) \}_{n, \ell, m}$ are complete and
orthonormal in $L^2 (\mathbb{R}^3)$. It makes sense to assume that the
functions $\{ \varphi_{n, \ell}^{m} (\bm{r}) \}_{n, \ell, m}$ are
irreducible spherical tensors of the type of
(\ref{Def_IrrSphericalTensor}). Thus, the index $n$ can be viewed to be
some kind of generalized principal quantum number, and $\ell$ and $m$ are
the usual angular momentum quantum numbers.

An addition theorem for $f (\bm{r} \pm \bm{r}')$, which converges in the
mean with respect to the norm of $L^2 (\mathbb{R}^3)$, can be constructed
by expanding $f$ in terms of the orthonormal functions $\{ \varphi_{n,
  \ell}^{m} (\bm{r}) \}_{n, \ell, m}$: 
\begin{subequations}
  \label{OneRangeAddTheor}
  \begin{align}
     \label{OneRangeAddTheor_a}
    f (\bm{r} \pm \bm{r}') & \; = \; \sum_{n \ell m} \, C_{n, \ell}^{m}
    (f; \pm \bm{r}') \, \varphi_{n, \ell}^{m} (\bm{r}) \, ,
    \\
     \label{OneRangeAddTheor_b}
    C_{n, \ell}^{m} (f; \pm \bm{r}') & \; = \; \int \, \bigl[
    \varphi_{n, \ell}^{m} (\bm{r}) \bigr]^{*} \, f (\bm{r} \pm \bm{r}')
    \, \mathrm{d}^3 \bm{r} \, .
  \end{align}
\end{subequations}
The summation limit in (\ref{OneRangeAddTheor}) depend on the exact
definition of the function set $\{ \varphi_{n, \ell}^{m} (\bm{r}) \}_{n,
  \ell, m}$. This article always uses the convention
\begin{equation}
\sum_{n \ell m} \; = \;
\sum_{n=1}^{\infty} \, \sum_{\ell=0}^{n-1} \, \sum_{m=-\ell}^{\ell} \, ,
\end{equation}
which is in agreement with the usual convention for the bound state
hydrogen eigenfunctions.

The expansion (\ref{OneRangeAddTheor}) is a one-range addition theorem
since the variables ${\bm{r}}$ and ${\bm{r}}'$ are completely separated:
The dependence on $\bm{r}$ is entirely contained in the functions
$\varphi_{n, \ell}^{m} (\bm{r})$, whereas $\bm{r}'$ occurs only in the
expansion coefficients $C_{n, \ell}^{m} (f; \pm \bm{r}')$ which are
overlap integrals.

If the overlap integrals $C_{n, \ell}^{m} (f; \pm \bm{r}')$ can be
expanded in terms of the functions $\varphi_{n, \ell}^{m} (\bm{r}')$
according to
\begin{subequations}
  \label{OverlapExpand}
  \begin{align}
    \label{OverlapExpand_a}
    C_{n, \ell}^{m} (f; \pm \bm{r}') & \; = \; \sum_{n' \ell' m'} \,
    T_{n' \ell' m'}^{n \ell m} (f; \pm) \, \varphi_{n', \ell'}^{m'}
    (\bm{r}') \, ,
    \\
    \label{OverlapExpand_b}
    T_{n' \ell' m'}^{n \ell m} (f; \pm) & \; = \; \int \, \bigl[
    \varphi_{n', \ell'}^{m'} (\bm{r}') \bigr]^{*} \, C_{n, \ell}^{m} (f;
    \pm \bm{r}') \, \mathrm{d}^3 \bm{r}' \, ,
  \end{align}
\end{subequations}
then the addition theorem (\ref{OneRangeAddTheor}) assumes a completely
symmetrical form:
\begin{equation}
  \label{SymOneRangeAddTheor}
f (\bm{r} \pm \bm{r}') \; = \;
\sum_{\substack{n \ell m \\ n' \ell' m'}} \,
T_{n' \ell' m'}^{n \ell m} (f; \pm) \, \varphi_{n, \ell}^{m} (\bm{r}) \,
\varphi_{n', \ell'}^{m'} (\bm{r}') \, .
\end{equation}

From a purely formal point of view, the derivation of one-range addition
theorems is a triviality. The challenging part is the construction of
computationally convenient mathematical expressions for the overlap
integrals $C_{n, \ell}^{m} (f; \pm \bm{r}')$ in (\ref{OneRangeAddTheor})
or the coefficients $T_{n' \ell' m'}^{n \ell m} (f; \pm)$ in
(\ref{SymOneRangeAddTheor}). In realistic applications, we cannot not
tacitly assume that the use of one-range addition theorems in a
multicenter integral necessarily leads to rapidly convergent expansions.
Therefore, we must be able to compute $C_{n, \ell}^{m} (f; \pm \bm{r}')$
and/or $T_{n' \ell' m'}^{n \ell m} (f; \pm)$ efficiently and reliably
even for possibly very large indices (see for instance the convergence
rates reported in \cite{Trivedi/Steinborn/1982}).

Symmetrical one-range addition theorems of the kind of
(\ref{SymOneRangeAddTheor}), which converge in the mean with respect to
the norm of the Hilbert space $L^2 (\mathbb{R}^3)$, were constructed by
Filter and Steinborn \cite[Eqs.\ (5.11) and
(5.12)]{Filter/Steinborn/1980} and later applied by Kranz and Steinborn
\cite{Kranz/Steinborn/1982} and by Trivedi and Steinborn
\cite{Trivedi/Steinborn/1982}.

As already remarked above, a nontrivial weight function $w (\bm{r}) \ne
1$ in an inner product can give more weight to those regions of space in
which $f$ is large, while deemphasizing the contribution from those
regions in which $f$ is small.  Accordingly, the inclusion of a suitable
weight function $w$ can improve convergence. It is thus an in principle
obvious idea to construct one-range addition theorems that converge with
respect to the norm of a suitable weighted Hilbert space $L_{w}^{2}
(\mathbb{R}^3)$ defined in (\ref{HilbertL_w^2}). Let us therefore assume
that $f$ belongs to $L_{w}^{2} (\mathbb{R}^3)$. Then we can construct a
one-range addition theorem by expanding $f (\bm{r} \pm \bm{r}')$ with
respect to a function set $\{ \psi_{n, \ell}^{m} (\bm{r}) \}_{n, \ell,
  m}$ that is complete in $L_{w}^{2} (\mathbb{R}^3)$ and orthonormal with
respect to the modified inner product (\ref{InnerProd_w}):
\begin{align}
  \label{SymOneRangeAddTheor_w}
& f (\bm{r} \pm \bm{r}')
\notag \\ 
& \qquad \; = \;
\sum_{\substack{n \ell m \\ n' \ell' m'}} \,
\mathbf{T}_{n' \ell' m'}^{n \ell m} (f, w; \pm) \,
\psi_{n, \ell}^{m} (\bm{r}) \, \psi_{n', \ell'}^{m'} (\bm{r}') \, .
\end{align}
For the derivation of (\ref{SymOneRangeAddTheor_w}), we only have to
replace (\ref{OneRangeAddTheor}) by
\begin{subequations}
  \label{OneRangeAddTheor_w}
  \begin{align}
    \label{OneRangeAddTheor_w_a}
    & f (\bm{r} \pm \bm{r}') \; = \; \sum_{n \ell m} \, \mathbf{C}_{n,
      \ell}^{m} (f, w; \pm \bm{r}') \, \psi_{n, \ell}^{m} (\bm{r}) \, ,
    \\
    \label{OneRangeAddTheor_w_b}
    & \mathbf{C}_{n, \ell}^{m} (f, w; \pm \bm{r}')
     \notag \\
    & \qquad \; = \; \int \, \bigl[ \psi_{n,
      \ell}^{m} (\bm{r}) \bigr]^{*} \, w (\bm{r}) \, f (\bm{r} \pm
    \bm{r}') \, \mathrm{d}^3 \bm{r} \, ,
  \end{align}
\end{subequations}
and (\ref{OverlapExpand}) by
\begin{subequations}
  \label{OverlapExpand_w}
  \begin{align}
    \label{OverlapExpand_w_a}
    & \mathbf{C}_{n, \ell}^{m} (f, w; \pm \bm{r}')
    \notag \\
    & \qquad \; = \; \sum_{n' \ell' m'} \, \mathbf{T}_{n' \ell' m'}^{n
      \ell m} (f, w; \pm) \, \psi_{n', \ell'}^{m'} (\bm{r}') \, ,
    \\
    \label{OverlapExpand_w_b}
    & \mathbf{T}_{n' \ell' m'}^{n \ell m} (f, w; \pm)
    \notag \\
    & \quad \; = \; \int \, \bigl[ \psi_{n', \ell'}^{m'} (\bm{r}')
    \bigr]^{*} \, w (\bm{r}') \, \mathbf{C}_{n, \ell}^{m} (f, w; \pm
    \bm{r}') \, \mathrm{d}^3 \bm{r}' \, .
  \end{align}
\end{subequations}

As discussed in Section \ref{Sec:LagTypeFun}, Guseinov and coworkers
\cite{Guseinov/2002c,Guseinov/2002d,Guseinov/2003b,Guseinov/2003c,%
  Guseinov/2003d,Guseinov/2003e,Guseinov/2004a,Guseinov/2004b,%
  Guseinov/2004c,Guseinov/2004d,Guseinov/2004e,Guseinov/2004f,%
  Guseinov/2004i,Guseinov/2004k,Guseinov/2005a,Guseinov/2005b,%
  Guseinov/2005c,Guseinov/2005d,Guseinov/2005e,Guseinov/2005f,%
  Guseinov/2005g,Guseinov/2006a,Guseinov/2006b,Guseinov/2007e,%
  Guseinov/2007f,Guseinov/2007g,Guseinov/Aydin/Mamedov/2003,%
  Guseinov/Mamedov/2002d,Guseinov/Mamedov/2003,Guseinov/Mamedov/2004b,%
  Guseinov/Mamedov/2004d,Guseinov/Mamedov/2004e,Guseinov/Mamedov/2004h,%
  Guseinov/Mamedov/2005c,Guseinov/Mamedov/2005d,Guseinov/Mamedov/2005g}
used for the derivation of one-range addition theorems and of related
mathematical objects predominantly the functions
$\prescript{}{k}{\Psi}_{n, \ell}^{m} (\beta, \bm{r})$ defined in
(\ref{Def_Psi_Guseinov}) that are complete and orthonormal in the
weighted Hilbert space $L_{r^k}^{2} (\mathbb{R}^3)$ defined in
(\ref{HilbertL_r^k^2}).  As long as the functions, which are to be
expanded, belong to the weighted Hilbert space $L_{r^k}^{2}
(\mathbb{R}^3)$, this is a completely legitimate approach that leads to
addition theorems of the type of (\ref{SymOneRangeAddTheor_w}) which
converge with respect to the norm (\ref{Norm_r^k_2}) of $L_{r^k}^{2}
(\mathbb{R}^3)$.

Unfortunately, the discussion in Section \ref{Sec:LagTypeFun} shows that
the use of a weight function $r^k$ with $k \ge 1$ does not necessarily
lead to better results. Moreover, it is very difficult or even
practically impossible to decide on the basis of simple \emph{a priori}
considerations for which $k$ the weight function $r^k$ produces best
results. Thus, it is by no means clear how we can actually profit from
this additional degree of freedom.

There is also another annoying practical problem which can easily occur
in multicenter integrals: Let us assume that $f (\bm{r} \pm \bm{r}')$ is
expanded in terms of a function set $\{ \psi_{n, \ell}^{m} (\bm{r})
\}_{n, \ell, m}$ that is complete in a weighted Hilbert space $L_{w}^{2}
(\mathbb{R}^3)$ and orthonormal with respect to the corresponding inner
product (\ref{InnerProd_w}). If we use this one-range addition theorem in
a multicenter integral containing $f (\bm{r} \pm \bm{r}')$, the weight
function $w (r)$, which makes $\{ \psi_{n, \ell}^{m} (\bm{r}) \}_{n,
  \ell, m}$ orthonormal, normally does not occur there.  Consequently,
subsequent integrations involving $f (\bm{r} \pm \bm{r}')$ are not
necessarily simplified by orthogonality.  Thus, it may be more difficult
to apply a one-range addition theorems of the type of
(\ref{SymOneRangeAddTheor_w}) based on a nontrivial weight function $w
(\bm{r}) \ne 1$ than an addition theorem of the type of
(\ref{SymOneRangeAddTheor}). It is not the complexity of an addition
theorem, that really matters, but the complexity of the series expansion,
which we obtain by inserting an addition theorem into a multicenter
integral.

If we compare one-range and two-range addition theorems, it is obvious
that one-range addition theorems greatly facilitate subsequent
integrations. Thus, they seem to be clearly superior to two-range
addition theorems of the type of the Laplace expansion (\ref{LapExp}).

However, one-range addition theorems also have disadvantages, and a
balanced assessment of the relative merits of one-range and two-range
addition theorems is by no means easy. Firstly, one-range addition
theorems usually have a more complicated structure than two-range
addition theorems (typically, they contain one additional infinite
summation). Secondly, one-range addition theorems normally do not
converge pointwise, but only in the mean with respect to the norm of the
underlying Hilbert space. This can be quite advantageous, since it makes
it possible to expand functions with singularities and/or
discontinuities. If, however, singularities and/or discontinuities are
present, then a nonsmooth function has to be approximated by smooth
functions. Very often, this leads to slow convergence.

The probably most severe disadvantage of one-range addition theorems
compared to two-range addition theorems is that the approach sketched
above cannot be applied to all functions of interest. As remarked in
Section \ref{Sec:HilbertSpace}, inner products for effective one-particle
wave functions in atomic and molecular electronic structure calculations
all involve integrations over the whole $\mathbb{R}^{3}$.  Thus, an
effective one-particle wave function $f$ has to belong to the Hilbert
space $L^2 (\mathbb{R}^3)$ or also to a suitable weighted Hilbert space
$L_w^2 (\mathbb{R}^3)$. Many functions of considerable relevance such as
the Coulomb potential or the irregular solid harmonic
$\mathcal{Z}_{\ell}^{m} (\bm{r})$ are only locally integrable, not square
integrable with respect to an integration over the whole $\mathbb{R}^{3}$
and thus do not belong to $L^2 (\mathbb{R}^3)$ or to any other weighted
Hilbert spaces discussed in this article. If we nevertheless want to
formulate one-range addition theorems for these functions by expanding
them in terms of a complete and orthonormal function set, we are faced
with the problem that these one-range addition theorems diverge in the
mean.

\typeout{==> Section: Laguerre-Type Function Sets and Their Hilbert
  Spaces}
\section{Laguerre-Type Function Sets and Their Hilbert Spaces}
\label{Sec:LagTypeFun}
 
The surface spherical harmonics $Y_{\ell}^{m} (\theta, \phi)$ are
complete and orthonormal with respect to an integration over the surface
of the unit sphere in $\mathbb{R}^3$ (an explicit proof can for instance
be found in \cite[Section III.7.6]{Prugovecki/1981}). Since more complex
Hilbert spaces can be constructed by forming tensor products of simpler
Hilbert spaces (see for example \cite[Section II.6.5]{Prugovecki/1981}),
we only have to find suitable radial functions that are complete and
orthogonal with respect to an integration from $0$ to $\infty$ (see also
\cite[Lemma 6 on p.\ 31]{Klahn/Bingel/1977b}).  Thus, we more or less
automatically arrive at function sets based on the \emph{generalized
  Laguerre polynomials}.

The generalized Laguerre polynomials $L_{n}^{(\alpha)} (x)$ with $\Re
(\alpha) > - 1$ are orthogonal polynomials associated with the
integration interval $[0, \infty)$ and the weight function $w (x) =
x^{\alpha} \mathrm{exp}^{-x}$. They possess the following explicit
expressions \cite[p.\ 240]{Magnus/Oberhettinger/Soni/1966}:
\begin{subequations}
  \label{Def_GLagPol}
  \begin{align}
    \label{GLag_FinSum}
    L_{n}^{(\alpha)} (x) & \; = \; \sum_{\nu=0}^{n} \, (-1)^{\nu} \,
    \binom{n+\alpha}{n-\nu} \, \frac{x^{\nu}}{\nu!}
    \\
    \label{GLag_1F1}
    & \; = \; \frac{(\alpha+1)_n}{n!} \, {}_1 F_1 (-n; \alpha+1; x) \, .
  \end{align}
\end{subequations}
These expressions can be used to extend the definition of
$L_{n}^{(\alpha)} (x)$ to complex values of $\alpha$. The generalized
Laguerre polynomials can also be defined via the following Rodrigues
relationship \cite[p.\ 241]{Magnus/Oberhettinger/Soni/1966}:
\begin{equation}
  \label{GLag_Rodrigues}
L_{n}^{(\alpha)} (x) \; = \; x^{-\alpha} \, \frac{\mathrm{e}^{x}}{n!} \,
\frac{\mathrm{d}^n}{\mathrm{d} x^n} \,
\bigl[ \mathrm{e}^{-x} x^{n+\alpha} \bigr] \, .
\end{equation}

The generalized Laguerre polynomials satisfy for $\Re (\alpha) > - 1$
and $m, n \in \mathbb{N}_0$ the following orthogonality relationship
\cite[p.\ 241]{Magnus/Oberhettinger/Soni/1966}:
\begin{equation}
  \label{GLag_Orthogonality}
\int_{0}^{\infty} \, x^{\alpha} \, \mathrm{e}^{-x} \,
L_{m}^{(\alpha)} (x) \, L_{n}^{(\alpha)} (x) \, \mathrm{d} x \; = \;
\frac{\Gamma (\alpha+n+1)}{n!} \, \delta_{m n} \, .
\end{equation}
Henceforth, the condition $\Re (\alpha) > - 1$, which is necessary for
the existence of this and related integrals, will always be tacitly
assumed.

The completeness of the generalized Laguerre polynomials in the weighted
Hilbert space
\begin{align}
  \label{HilbertL^2_Lag}
  & L^{2}_{\mathrm{e}^{-x} x^{\alpha}} (\mathbb{R}_{+}) \; = \;
  \notag \\
  & \; = \; \Bigl\{ f \colon \mathbb{R}_{+} \to \mathbb{C} \Bigm\vert \,
  \int_{0}^{\infty} \,\mathrm{e}^{-x} \, x^{\alpha} \, \vert f (x)
  \vert^2 \, \mathrm{d} x < \infty \Bigr\}
\end{align}
is a classic result of mathematical analysis (see for example
\cite[p.\ 33]{Higgins/1977}, \cite[pp.\ 349 - 351]{Sansone/1977}, or
\cite[pp.\ 235 - 238]{Tricomi/1970}).

A different convention for Laguerre polynomials is frequently used in the
quantum mechanical literature. For example, Bethe and Salpeter \cite[Eq.\
(3.5)]{Bethe/Salpeter/1977} define \emph{associated Laguerre functions}
$\bigl[L_{n}^{m} (x)\bigr]_{\text{BS}}$ with $n, m \in \mathbb{N}_0$ via
the Rodrigues-type relationships
\begin{subequations}
  \label{AssLagFun_BS}
  \begin{align}
    \label{AssLagFun_BS_1}
    \bigl[L_{n}^{m} (x)\bigr]_{\text{BS}} & \; = \;
    \frac{\mathrm{d}^{m}}{\mathrm{d} x^{m}} \, \bigl[L_{n}
    (x)\bigr]_{\text{BS}} \, ,
    \\
    \label{AssLagFun_BS_2}
    \bigl[L_{n} (x)\bigr]_{\text{BS}} & \; = \; \mathrm{e}^x \,
    \frac{\mathrm{d}^{n}}{\mathrm{d} x^{n}} \, \bigl[ \mathrm{e}^{-x}
    x^{n} \bigr] \, .
  \end{align}
\end{subequations}
Comparison of (\ref{GLag_Rodrigues}) and (\ref{AssLagFun_BS_2}) implies:
\begin{equation}
  \label{GLagPol_2_GLagPol_BS}
L_{n}^{(m)} (x) \; = \; 
\frac{(-1)^m}{(n+m)!} \, \bigl[L_{n+m}^{m} (x)\bigr]_{\text{BS}} \, .
\end{equation}
The convention of Bethe and Salpeter \cite{Bethe/Salpeter/1977} is also
used in the books by Condon and Shortley \cite[Eqs.\ (6) and (9) on p.\
115]{Condon/Shortley/1970} and by Condon and Odaba\c{s}i \cite[Eq.\ (2)
on p.\ 189]{Condon/Odabasi/1980} as well as in numerous articles.

In my opinion, the use of associated Laguerre functions
$\bigl[L_{n+m}^{m} (x)\bigr]_{\text{BS}}$ defined by (\ref{AssLagFun_BS})
is not recommendable. It follows from (\ref{GLagPol_2_GLagPol_BS}) that
these functions cannot express generalized Laguerre polynomials
$L_{n}^{(\alpha)}$ with nonintegral superscripts $\alpha$. This is both
artificial and unnecessary. For example, the eigenfunctions $\Omega_{n,
  \ell}^{m} (\beta, \bm{r})$ of the Hamiltonian $\beta^{-2} \nabla^2 -
\beta^2 r^2$ of the three-dimensional isotropic harmonic oscillator
contain generalized Laguerre polynomials with half-integral superscripts
(see for example \cite[Eq.\ (5.4)]{Weniger/1985} and references therein):
\begin{align}
  \label{Def_OscillFun}
  & \Omega_{n, \ell}^{m} (\beta, \bm{r}) \; = \; \beta^{3/2} \left[
    \frac{2 (n-\ell-1)!} {\Gamma (n+1/2)} \right]^{1/2}
  \notag \\
  & \qquad \times \, \mathrm{e}^{-\beta^2 r^2/2} \,
  L_{n-\ell-1}^{(\ell+1/2)} (\beta^2 r^2) \, \mathcal{Y}_{\ell}^{m}
  (\beta \bm{r}) \, .
\end{align}
These functions satisfy the orthonormality condition (see for example
\cite[Eq.\ (5.5)]{Weniger/1985})
\begin{equation}
  \label{OscillFun_OrthoNor}
\int \, \bigl[ \Omega_{n, \ell}^{m} (\beta, \bm{r}) \bigr]^{*} \,
\Omega_{n', \ell'}^{m'} (\beta, \bm{r}) \, \mathrm{d}^3 \bm{r}
\; = \; \delta_{n n'} \delta_{\ell \ell'} \delta_{m m'}
\end{equation}
and are complete and orthonormal in $L^{2} (\mathbb{R}^3)$ (see also
\cite[pp.\ 36 - 37]{Klahn/Bingel/1977b}).

In this article, only the mathematical notation for generalized Laguerre
polynomials is used. Additional conventions used in physics were
discussed by Kaijser and Smith \cite[Footnote 1 on p.\
48]{Kaijser/Smith/1977}.

Guseinov and coworkers use the convention for Laguerre polynomials that
is common in quantum mechanics (compare \cite[Eq.\ (4) on p.\
189]{Condon/Odabasi/1980} with \cite[Eq.\ (3)]{Guseinov/2002b}), although
they give the book by Gradshteyn and Rhyzhik
\cite{Gradshteyn/Rhyzhik/1994} as their primary reference on Laguerre
polynomials.  This is misleading: Gradshteyn and Rhyzhik use the
mathematical definition of the generalized Laguerre polynomials (compare
\cite[Eq.\ (8.970.1) on p.\ 1061]{Gradshteyn/Rhyzhik/1994} with
(\ref{GLag_Rodrigues})).

In principle, every function set, that is complete and orthonormal in
either $L^{2} (\mathbb{R}^3)$ or also in a suitable weighted Hilbert
space $L_{w}^{2} (\mathbb{R}^3)$, can be used for the construction of
one-range addition theorems. If we insist on an exponentially decaying
function set that is complete and orthonormal in $L^{2} (\mathbb{R}^3)$,
then the following functions, which were introduced by Hylleraas
\cite[Footnote ${}^{*}$ on p.\ 349]{Hylleraas/1929}, Shull and L\"owdin
\cite{Shull/Loewdin/1955}, and L\"{o}wdin and Shull \cite[Eq.\
(46)]{Loewdin/Shull/1956}, are the most natural choice:
\begin{align}
  \label{Def_LambdaFun}
  & \Lambda_{n, \ell}^{m} (\beta, \bm{r}) \; = \; (2 \beta)^{3/2} \left[
    \frac{(n-\ell-1)!} {(n+\ell+1)!} \right]^{1/2}
  \notag \\
  & \qquad \times \, \mathrm{e}^{-\beta r} \, L_{n-\ell-1}^{(2\ell+2)}
  (2 \beta r) \, \mathcal{Y}_{\ell}^{m} (2 \beta \bm{r}) \, .
\end{align}
Here, $\beta > 0$ is a scaling parameter.

Lambda functions are orthonormal with respect to a integration over the
whole $\mathbb{R}^3$:
\begin{equation}
  \label{Lambda_OrthoNor}
\int \, \bigl[ \Lambda_{n, \ell}^{m} (\beta, \bm{r}) \bigr]^{*} \,
\Lambda_{n', \ell'}^{m'} (\beta, \bm{r}) \, \mathrm{d}^3 \bm{r}
\; = \; \delta_{n n'} \delta_{\ell \ell'} \delta_{m m'} \, .
\end{equation}
Accordingly, they are complete and orthonormal in $L^{2} (\mathbb{R}^3)$.

Closely related to both bound-state hydrogen eigenfunctions and Lambda
functions are the following functions which were already used in 1928 by
Hylleraas \cite[Eq.\ (25) on p.\ 478]{Hylleraas/1928} and which are
commonly called Coulomb Sturmians or simply Sturmians:
\begin{align}
  \label{Def_SturmFun}
  & \Psi_{n, \ell}^{m} (\beta, \bm{r}) \; = \; (2 \beta)^{3/2} \,
    \left[ \frac{(n-\ell-1)!}{2n(n+\ell)!} \right]^{3/2}
  \notag \\[1\jot]
  & \qquad \times \, \mathrm{e}^{-\beta r} \, L_{n-\ell-1}^{(2\ell+1)}
  (2 \beta r) \, \mathcal{Y}_{\ell}^{m} (2 \beta \bm{r}) \, .
\end{align}
If we replace in (\ref{Def_SturmFun}) $\beta$ by $Z/n$, we obtain the
bound state eigenfunctions of a hydrogenlike ion with nuclear charge $Z$.
These eigenfunctions are orthonormal with respect to an integration over
the whole $\mathbb{R}^3$, but they are incomplete without the inclusion
of the continuum eigenfunctions (see for example
\cite{Weniger/Steinborn/1984} and references therein).

Sturmians -- or rather their three-dimensional Fourier transforms --
occur also in Fock's treatment of the hydrogen atom \cite{Fock/1935},
albeit in a somewhat disguised form \cite[Section VI]{Weniger/1985}.

It follows at once from (\ref{GLag_Orthogonality}) that the Sturmians
satisfy \cite[Eq.\ (4.7)]{Weniger/1985}:
\begin{equation}
  \label{SturmFun_1/r_OrthoGon}
\int \, \bigl[ \Psi_{n, \ell}^{m} (\beta, \bm{r}) \bigr]^{*} \,
\frac{1}{r} \,
\Psi_{n', \ell'}^{m'} (\beta, \bm{r}) \, \mathrm{d}^3 \bm{r}
\; = \; \frac{\beta}{n} \,
\delta_{n n'} \, \delta_{\ell \ell'} \, \delta_{m m'} \, .
\end{equation}
Accordingly, the Sturmians are a complete orthogonal set in the weighted
Hilbert space
\begin{align}
  \label{HilbertL_1/r^2}
  & L_{1/r}^{2} (\mathbb{R}^3)
  \notag \\
  & \qquad \; = \; \Bigl\{ f \colon \mathbb{R}^3 \to \mathbb{C} \Bigm\vert \,
  \int \, \frac{1}{r} \, \vert f (\bm{r}) \vert^2 \,
  \mathrm{d}^3 \bm{r} < \infty \Bigr\} \, .
\end{align}
Unfortunately, the Hilbert space $L_{1/r}^{2} (\mathbb{R}^3)$ is not
necessarily suited for quantum mechanical calculations since we neither
have $L^{2} (\mathbb{R}^3) \subset L_{1/r}^{2} (\mathbb{R}^3)$ nor
$L_{1/r}^{2} (\mathbb{R}^3) \subset L^{2} (\mathbb{R}^3)$. Thus, one
might arrive at the premature conclusion that Sturmians are not
particularly suited for our purposes. 

If, however, we combine the differential equation satisfied by the
Sturmians \cite[Eq.\ (4.9)]{Weniger/1985},
\begin{equation}
  \label{DiffEq_SturmFun}
\left[ \nabla^2 \, + \, \frac{2 \beta n}{r} \, - \, \beta^2 \right] \,
\Psi_{n, \ell}^{m} (\beta, \bm{r}) \; = \; 0 \, ,
\end{equation}
with the orthogonality relationship (\ref{SturmFun_1/r_OrthoGon}), we see
that the Sturmians satisfy the Sobolev-type orthonormality \cite[Eq.\
(4.10)]{Weniger/1985}
\begin{align}
  \label{SturmFun_Sobolev_OrthoNor}
  & \int \, \bigl[ \Psi_{n, \ell}^{m} (\beta, \bm{r}) \bigr]^{*} \,
  \frac{\beta^2 - \nabla^2}{2\beta^2} \,
  \Psi_{n', \ell'}^{m'} (\beta, \bm{r}) \, \mathrm{d}^3 \bm{r}
  \notag \\
  & \qquad \; = \;
  \delta_{n n'} \, \delta_{\ell \ell'} \, \delta_{m m'} \, .
\end{align}
Accordingly, Sturmians are complete and orthonormal in the Sobolev
space $W_{2}^{(1)} (\mathbb{R}^3$.

In \cite[Eq.\ (1)]{Guseinov/2002b}, Guseinov introduced a fairly large
class of complete and orthonormal functions which, if the mathematical
notation for generalized Laguerre polynomials is used, can be expressed
as follows:
\begin{align}
  \label{Def_Psi_Guseinov}
  & \prescript{}{k}{\Psi}_{n, \ell}^{m} (\beta, \bm{r}) \; = \; \left[
    \frac{(2\beta)^{k+3} (n-\ell-1)!}{(n+\ell+k+1)!} \right]^{1/2}
  \notag \\
  & \qquad \times \mathrm{e}^{-
    \beta r} \, L_{n-\ell-1}^{(2\ell+k+2)} (2 \beta r) \,
  \mathcal{Y}_{\ell}^{m} (2 \beta \bm{r}) \, .
\end{align}
The indices satisfy $n \in \mathbb{N}$, $k = -1, 0, 1, 2, \dots$, $\ell
\in \mathbb{N}_0 \le n - 1$, $-\ell \le m \le \ell$, and the scaling
parameter $\beta$ is positive. Guseinov and coworkers used these
functions quite extensively in the context of one-range addition theorems
\cite{Guseinov/2002c,Guseinov/2002d,Guseinov/2003b,Guseinov/2003c,%
  Guseinov/2003d,Guseinov/2003e,Guseinov/2004a,Guseinov/2004b,%
  Guseinov/2004c,Guseinov/2004d,Guseinov/2004e,Guseinov/2004f,%
  Guseinov/2004i,Guseinov/2004k,Guseinov/2005a,Guseinov/2005b,%
  Guseinov/2005c,Guseinov/2005d,Guseinov/2005e,Guseinov/2005f,%
  Guseinov/2005g,Guseinov/2006a,Guseinov/2006b,Guseinov/2007e,%
  Guseinov/2007f,Guseinov/2007g,Guseinov/Aydin/Mamedov/2003,%
  Guseinov/Mamedov/2002d,Guseinov/Mamedov/2003,Guseinov/Mamedov/2004b,%
  Guseinov/Mamedov/2004d,Guseinov/Mamedov/2004e,Guseinov/Mamedov/2004h,%
  Guseinov/Mamedov/2005c,Guseinov/Mamedov/2005d,Guseinov/Mamedov/2005g}.

If Guseinov had used the mathematical definition (\ref{Def_GLagPol}) of
the generalized Laguerre polynomials and not the unnecessarily
restrictive convention (\ref{AssLagFun_BS}), he probably would have
noticed that the parameter $k = -1, 0, 1, 2, \dots$ in his functions
$\prescript{}{k}{\Psi}_{n, \ell}^{m} (\beta, \bm{r})$ need not be
integral and can be generalized to real values $k \in [-1, \infty)$.  One
only has to replace the factorial $(n+\ell+k+1)!$ in
(\ref{Def_Psi_Guseinov}) by the gamma function $\Gamma (n+\ell+k+2)$.

Guseinov's functions satisfy the orthonormality relationship (compare
also \cite[Eq.\ (4)]{Guseinov/2002c})
\begin{align}
  \label{Psi_Guseinov_OrthoNor}
  & \int \, \bigl[ \prescript{}{k}{\Psi}_{n, \ell}^{m} (\beta, \bm{r})
  \bigr]^{*} \, r^k \, \prescript{}{k}{\Psi}_{n', \ell'}^{m'} (\beta,
  \bm{r}) \, \mathrm{d}^3 \bm{r}
  \notag \\
  & \qquad \; = \; \delta_{n n'} \, \delta_{\ell \ell'} \, \delta_{m m'}
  \, .
\end{align}
Accordingly, Guseinov's functions are a complete and orthonormal set in
the weighted Hilbert space $L_{r^k}^{2} (\mathbb{R}^3)$ with $k = -1, 0,
1, 2, \dots$, which is defined via the inner product
\begin{equation}
  \label{InnerProd_r^k}
  ( f \vert g )_{r^k, 2} \; = \;
  \int \bigl[ f (\bm{r}) \bigr]^{*} \, r^k \,
  g (\bm{r}) \, \mathrm{d}^3 \bm{r}
\end{equation}
and the norm
\begin{equation}
  \label{Norm_r^k_2}
\Vert f \Vert_{r^{k}, 2} \; = \; \sqrt{( f \vert f )_{r^k, 2}}
\end{equation}
according to
\begin{align}
  \label{HilbertL_r^k^2}
  L_{r^k}^{2} (\mathbb{R}^3) & \; = \; \Bigl\{ f \colon \mathbb{R}^3 \to
  \mathbb{C} \Bigm\vert \, \int \, r^k \, \vert f (\bm{r}) \vert^2 \,
  \mathrm{d}^3 \bm{r} < \infty \Bigr\} \notag \\[1\jot]
  & \; = \; \bigl\{ f \colon \mathbb{R}^3 \to \mathbb{C} \big\vert \, \Vert f
  \Vert_{r^k, 2} < \infty \bigr\} \, .
\end{align}

For $k=-1$, the functions $\prescript{}{k}{\Psi}_{n, \ell}^{m} (\beta,
\bm{r})$ are apart from a slightly different normalization factor
identical to the Sturmian functions defined by (\ref{Def_SturmFun}).
Thus, they are complete and orthonormal in the Hilbert space $L_{1/r}^{2}
(\mathbb{R}^3)$ defined by (\ref{HilbertL_1/r^2}) or -- when used in
combination with the inner product (\ref{Sobolev_InnerProd}) with
$\zeta=\beta$ -- complete and orthogonal in the Sobolev space
$W_{2}^{(1)} (\mathbb{R}^3)$, which is a proper subspace of the Hilbert
space $L^{2} (\mathbb{R}^3)$. It is important to notice that $L_{1/r}^{2}
(\mathbb{R}^3)$ and $W_{2}^{(1)} (\mathbb{R}^3)$ are \emph{not}
identical, although the Sturmians and thus Guseinov's functions with
$k=-1$ are complete and orthogonal in both spaces.

For $k=0$, the functions $\prescript{}{k}{\Psi}_{n, \ell}^{m} (\beta,
\bm{r})$ are identical to the Lambda functions defined by
(\ref{Def_LambdaFun}). Thus, they are complete and orthonormal in $L^{2}
(\mathbb{R}^3)$.

For $k = 1, 2, 3, \dots$, the weighted Hilbert spaces $L_{r^k}^{2}
(\mathbb{R}^3)$ are genuinely different from the Hilbert space $L^{2}
(\mathbb{R}^3)$ and we neither have $L^{2} (\mathbb{R}^3) \subset
L_{r^k}^{2} (\mathbb{R}^3)$ nor $L_{r^k}^{2} (\mathbb{R}^3) \subset L^{2}
(\mathbb{R}^3)$. The weight function $r^k$ in the inner product
(\ref{InnerProd_r^k}) admits functions to $L_{r^k}^{2} (\mathbb{R}^3)$
that are too singular at the origin to belong to $L^{2} (\mathbb{R}^3)$,
and it excludes some functions belonging to $L^{2} (\mathbb{R}^3)$ that
decay slowly like a fixed power $r^{-\alpha}$ as $r \to \infty$.

If $f \in L_{r^k}^{2} (\mathbb{R}^3)$, then it is guaranteed that the
expansion
\begin{subequations}
  \label{exp_f2gPsi}
  \begin{align}
    f (\bm{r}) & \; = \; \sum_{n \ell m} \,
    \prescript{}{k}{\mathcal{F}}_{n, \ell}^{m} (\beta; f) \,
    \prescript{}{k}{\Psi}_{n, \ell}^{m} (\beta, \bm{r}) \, ,
    \\
    \prescript{}{k}{\mathcal{F}}_{n, \ell}^{m} (\beta; f) & \; = \; \int
    \, \left[ \prescript{}{k}{\Psi}_{n, \ell}^{m} (\beta, \bm{r})
    \right]^{*} \, r^k \, f (\bm{r}) \, \mathrm{d}^{3} \bm{r} \, ,
  \end{align}
\end{subequations}
in terms of Guseinov's functions converges in the mean with respect to
the norm (\ref{Norm_r^k_2}) of $L_{r^k}^{2} (\mathbb{R}^3)$. If, however,
$f \in L^{2} (\mathbb{R}^3)$, the convergence of this expansion with
respect to the norm (\ref{Norm_r^k_2}) of $L_{r^k}^{2} (\mathbb{R}^3)$ is
not guaranteed. 

For fixed $k = -1, 0, 1, \dots$, the functions $\prescript{}{k}{\Psi}_{n,
  \ell}^{m} (\beta, \bm{r})$ are by construction complete and orthogonal
in the corresponding Hilbert space $L_{r^k}^{2} (\mathbb{R}^3)$, but for
$k \neq 0$ the expansion (\ref{exp_f2gPsi}) does not necessarily converge
with respect to (\ref{Norm_r^k_2}) for arbitrary functions $f$ belonging
to the for our purposes most important Hilbert space $L (\mathbb{R}^3)$
of square integrable functions. Thus, the convergence of
(\ref{exp_f2gPsi}) is only guaranteed if we have $f \in L^{2}
(\mathbb{R}^3) \cap L_{r^k}^{2} (\mathbb{R}^3)$.  This is a complications
which we have to take into account whenever we do expansions in terms of
Guseinov's functions $\prescript{}{k}{\Psi}_{n, \ell}^{m} (\beta,
\bm{r})$ with $k \neq 0$.

It is by no means clear whether and for which $k \neq 0$ a weighted
Hilbert space $L_{r^k}^{2} (\mathbb{R}^3)$ provides a proper setting for
bound state electronic structure calculations. It certainly makes sense
to be cautious and to avoid potential complications whenever possible.
Therefore, I would refrain from using the functions
$\prescript{}{k}{\Psi}_{n, \ell}^{m} (\beta, \bm{r})$ with $k \neq 0$
unless I see obvious computational benefits. Generalization for the sake
of generalization is rarely a good idea.

In addition, the rate of convergence of the expansion of a given function
in terms of Guseinov's functions $\prescript{}{k}{\Psi}_{n, \ell}^{m}
(\beta, \bm{r})$ may depend quite strongly on the choice of $k$. For $k
\ge 1$, the inner product (\ref{InnerProd_r^k}) gives less weight to the
region close to the origin and greater weight to the region away from the
origin, which may or may not improve convergence.

The numerical consequences of a variation of $k$ can be studied by
expanding the exponential
\begin{equation}
\mathrm{e}^{-x \beta r} \, (4\pi)^{-1/2} \; = \;
\mathrm{e}^{-x \beta r} \, Y_{0}^{0} (\bm{r}/r) \, , \qquad x > 0 \, ,
\end{equation}
in terms of Guseinov's functions:
\begin{subequations}
  \label{exp2gPsi_1}
  \begin{align}
    & \mathrm{e}^{-x \beta r} \, Y_{0}^{0} (\bm{r}/r)
    \notag \\
    & \quad \; = \; \sum_{n=1}^{\infty} \, \sum_{\ell=0}^{n-1} \,
    \sum_{m=-\ell}^{\ell} \, \prescript{}{k}{\mathcal{E}}_{n, \ell}^{m}
    (x, \beta) \, \prescript{}{k}{\Psi}_{n, \ell}^{m} (\beta, \bm{r}) \, ,
    \\
    & \prescript{}{k}{\mathcal{E}}_{n, \ell}^{m} (x, \beta)
    \notag \\
    & \quad \; = \; \int \, \left[ \prescript{}{k}{\Psi}_{n, \ell}^{m}
      (\beta, \bm{r}) \right]^{*} \, \mathrm{e}^{-x \beta r} \, Y_{0}^{0}
    (\bm{r}/r) \, \mathrm{d}^{3} \bm{r} \, .
  \end{align}
\end{subequations}
The orthonormality of the spherical harmonics implies:
\begin{align}
  \label{exp2gPsi_2}
 & \prescript{}{k}{\mathcal{E}}_{n, \ell}^{m} (x, \beta) \; = \; 
   \left[ \frac{(2\beta)^{k+3} (n-1)!}{(n+k+1)!} \right]^{1/2} \,
     \delta_{\ell 0} \, \delta_{m 0}
   \notag \\[1.5\jot]
   & \qquad \times \, \int_{0}^{\infty} \, \mathrm{e}^{-(1+x)\beta r} \,
   r^{k+2} \, L_{n-1}^{(k+2)} (2 \beta r) \, \mathrm{d} r \, .
\end{align}
This expression can be rewritten as follows:
\begin{align}
  \label{exp2gPsi_3}
  & \prescript{}{k}{\mathcal{E}}_{n, 0}^{0} (x, \beta) \; = \; 
    \left[ \frac{(n-1)!}{(2\beta)^{k+3} (n+k+1)!} \right]^{1/2}
  \notag \\[1.5\jot]
  & \qquad \times \, \int_{0}^{\infty} \, \mathrm{e}^{-(1+x)t/2} \,
  t^{k+2} \, L_{n-1}^{(k+2)} (t) \, \mathrm{d} t \, .
\end{align}
With the help of the integral \cite[Eq.\ 7.414.8 on p.\
850]{Gradshteyn/Rhyzhik/1994}
\begin{align}
  & \int_{0}^{\infty} \, \mathrm{e}^{-s t} \, t^{\alpha} \,
  L_{n}^{(\alpha)} (t) \, \mathrm{d} t \; = \;
  \frac{\Gamma (\alpha+n+1) (s-1)^n}{n! s^{\alpha+n+1}} \, ,
  \notag \\
  & \qquad \Re (\alpha) > - 1 \, , \qquad \Re (s) > 0 \, ,
\end{align}
we obtain
\begin{align}
  & \int_{0}^{\infty} \, \mathrm{e}^{- (1+x) t/2} \, t^{k+2} \,
  L_{n-1}^{(k+2)} (t) \, \mathrm{d} t
  \notag \\
  & \qquad \; = \; \left[ \frac{2}{x+1} \right]^{k+3} \,
  \frac{(n+k+1)!}{(n-1)!} \, \left[ \frac{x-1}{x+1} \right]^{n-1}
\end{align}
and
\begin{align}
 & \prescript{}{k}{\mathcal{E}}_{n, 0}^{0} (x, \beta) \; = \; 
   \left[ \frac{2}{x+1} \right]^{k+3}
  \notag \\
  & \qquad \left[ \frac{(n+k+1)!}{2\beta)^{k+3} (n-1)!} \right]^{1/2} \,
  \left[ \frac{x-1}{x+1} \right]^{n-1} \, .
\end{align}
This result shows that the expansion (\ref{exp2gPsi_1}) becomes trivial
for $x=1$ because then only the term with $n=1$ is different from zero.
However, it also shows that for $x \neq 1$ the rate of convergence of the
expansion (\ref{exp2gPsi_1}) decreases with increasing $k \in
\mathbb{N}$.

This example does not imply that Guseinov's functions necessarily produce
inferior convergence rates for higher values of $k$.  Most likely, other
examples can be constructed for which higher values of $k$ improve
convergence. However, this example shows that the inclusion of the
nontrivial weight function $r^k$ in an inner product can have significant
numerical consequences. Since it is not at all clear whether the weighted
Hilbert spaces $L_{r^k}^{2} (\mathbb{R}^3)$ with $k \neq 0$ provide a
proper setting for bound state electronic structure calculations, I
cannot recommend to routinely use Guseinov's functions
$\prescript{}{k}{\Psi}_{n, \ell}^{m} (\beta, \bm{r})$ with $k \neq 0$ for
the construction of one-range addition theorems.

This assessment does not necessarily apply to Guseinov's functions
$\prescript{}{k}{\Psi}_{n, \ell}^{m} (\beta, \bm{r})$ with $k=-1$, which
are up to a slightly different normalization factor identical to the
Sturmians defined by (\ref{Def_SturmFun}). Sturmians are complete and
orthogonal in the weighted Hilbert space $L_{1/r}^2 (\mathbb{R}^3)$,
whose usefulness in electronic structure calculations is not clear since
we neither have $L^{2} (\mathbb{R}^3) \subset L_{1/r}^{2} (\mathbb{R}^3)$
nor $L_{1/r}^{2} (\mathbb{R}^3) \subset L^{2} (\mathbb{R}^3)$. But
Sturmians are also complete and orthonormal in the Sobolev pace
$W_{2}^{(1)} (\mathbb{R}^3$ defined by (\ref{Sobolev_W_2^1}), which is a
proper subspace of $L^{2} (\mathbb{R}^3)$. As discussed in Section
\ref{Sec:HilbertSpace}, a basis set has to be complete in $W_{2}^{(1)}
(\mathbb{R}^3$ to guarantee the convergence of variational calculations.
Completeness in $L^{2} (\mathbb{R}^3)$ does not suffice.

Let me emphasize once more that Laguerre expansions converge in general
only in the mean, but not necessarily pointwise (see for example
\cite{Askey/Wainger/1965}). Additional conditions, which a function has
to satisfy in order to guarantee that its Laguerre expansion converges
pointwise, were discussed by Szeg\"{o} \cite[Theorem 9.1.5 on p.\
246]{Szegoe/1967} (see also \cite[Appendix]{Filter/Steinborn/1980}).

\typeout{==> Section: Addition Theorems for Exponentially Decaying
  Functions}
\section{Addition Theorems for Exponentially Decaying Functions}
\label{AddTheor_ETF}

Pointwise convergent two-range addition theorems for all the commonly
occurring exponentially decaying functions can be constructed via the
two-range addition theorems \cite{Weniger/2002,Weniger/Steinborn/1989b}
of the so-called $B$ functions that were introduced by Filter and
Steinborn \cite[Eq.\ (2.14)]{Filter/Steinborn/1978b}:
\begin{align}
  \label{Def:B_Fun}
B_{n,\ell}^{m} (\beta, \hm{r}) & \; = \;
[2^{n+\ell} (n+\ell)!]^{-1} \, \hat{k}_{n-1/2} (\beta r) \,
\mathcal{Y}_{\ell}^{m} (\beta \bm{r}) \, ,
\notag \\
 & \qquad \beta > 0 \, , \qquad n \in \mathbb{Z} \, .
\end{align}
Here, $\hat{k}_{n-1/2}$ is a reduced Bessel function of half-integral
order $n-1/2$ defined by \cite[Eqs.\ (3.1) and
(3.2)]{Steinborn/Filter/1975c}
\begin{equation}
   \label{Def:RBF}
\hat{k}_{\nu} (z) \; = \; (2/\pi)^{1/2} \, z^{\nu} \, K_{\nu} (z) \, ,
\end{equation}
and $K_{\nu} (z)$ is a modified Bessel function of the second kind
\cite[p.\ 66]{Magnus/Oberhettinger/Soni/1966}. If the order $\nu$ of a
reduced Bessel function is half-integral, it can be expressed as an
exponential multiplied by a terminating confluent hypergeometric series
${}_1 F_1$ (see for example \cite[Eq.\ (3.7)]{Weniger/Steinborn/1983b}).

$B$ functions are comparatively complicated mathematical objects, and it
is not obvious why they should offer any advantages compared to other
exponentially decaying function sets. However, $B$ functions possess a
three-dimensional Fourier transform of remarkable simplicity:
\begin{align}
  \label{FT_B_Fun}
  \bar{B}_{n,\ell}^{m} (\alpha, \bm{p}) & \; = \; (2\pi)^{-3/2} \, \int
  \, \mathrm{e}^{- \mathrm{i} \bm{p} \cdot \bm{r}} \, B_{n,\ell}^{m}
  (\alpha, \bm{r}) \, \mathrm{d}^3 \bm{r}
  \notag \\
  & \; = \; (2/\pi)^{1/2} \, \frac{\alpha^{2n+\ell-1}}{[\alpha^2 +
    p^2]^{n+\ell+1}} \, \mathcal{Y}_{\ell}^{m} (- \mathrm{i} \bm{p}) \, .
\end{align}
This is the most consequential and also the most often cited result of my
PhD thesis \cite[Eq.\ (7.1-6) on p.\ 160]{Weniger/1982}. Later,
(\ref{FT_B_Fun}) was published in \cite[Eq.\
(3.7)]{Weniger/Steinborn/1983a}. Independently and almost simultaneously,
(\ref{FT_B_Fun}) was also derived by Niukkanen \cite[Eqs.\ (57) -
(58)]{Niukkanen/1984c}.

The exceptionally simple Fourier transform (\ref{FT_B_Fun}) gives $B$
functions a unique position among exponentially decaying functions.  It
also explains why other exponentially decaying functions like Slater-type
functions with integral principal quantum numbers, bound state hydrogen
eigenfunctions, and other functions based on generalized Laguerre
polynomials such as Lambda functions, Sturmians, or Guseinov's functions
can all be expressed in terms of finite linear combinations of $B$
functions (details and further references can be found in \cite[Section
IV]{Weniger/1985} or \cite[Section 4]{Weniger/2002}). Two-range addition
theorems for the exponentially decaying functions mentioned above can be
written down immediately by forming finite linear combinations of the
corresponding addition theorems of $B$ functions
\cite{Weniger/2002,Weniger/Steinborn/1989b}.

$B$ functions also turned out to be extremely useful for the derivation
of one-range addition theorems. Filter and Steinborn \cite[Eqs.\ (5.11)
and (5.12)]{Filter/Steinborn/1980} derived symmetrical one-range addition
theorems of the type of (\ref{SymOneRangeAddTheor}) for Lambda and $B$
functions by expanding them in terms of Lambda functions:
\begin{align}
  \label{OneRangeAddTheor_Lambda}
  & \Lambda_{N L}^{M} (\beta, \bm{r} - \bm{r}') \; = \; \sum_{\ell_1
    \ell_2} \, \sum_{m_1} \, \langle L M \vert \ell_1 m_2 \vert \ell_2
  m_2 \rangle
  \notag \\
  & \qquad \times \, \sum_{n_1 n_2} \, T_{n_1 \ell_1}^{n_2 \ell_2 N L} \,
  \Lambda_{n_1 \ell_1}^{m_1} (\beta, \bm{r}) \, \Lambda_{n_2
    \ell_2}^{m_2} (\beta, \bm{r}') \, ,
  \\
  \label{OneRangeAddTheor_B}
  & B_{N L}^{M} (\beta, \bm{r} - \bm{r}') \; = \; \sum_{\ell_1 \ell_2} \,
  \sum_{m_1} \, \langle L M \vert \ell_1 m_2 \vert \ell_2 m_2 \rangle
  \notag \\
  & \qquad \times \, \sum_{n_1 n_2} \, a_{n_1 \ell_1}^{n_2 \ell_2 N+L L}
  \, \Lambda_{n_1 \ell_1}^{m_1} (\beta, \bm{r}) \, \Lambda_{n_2
    \ell_2}^{m_2} (\beta, \bm{r}') \, .
\end{align}
These addition theorems contain Gaunt coefficients defined by
(\ref{Def_Gaunt}). Filter and Steinborn \cite[Eqs.\ (4.9), (4.10), and
(4.33)]{Filter/Steinborn/1980} were able to derive explicit expressions
for the coefficients $T_{n_1 \ell_1}^{n_2 \ell_2 N L}$ and $a_{n_1
  \ell_1}^{n_2 \ell_2 N+L L}$ occurring in
(\ref{OneRangeAddTheor_Lambda}) and (\ref{OneRangeAddTheor_B}),
respectively. In \cite[Eq.\ (7.8)]{Weniger/1985} it was shown that the
explicit expression for the coefficients $T_{n_1 \ell_1}^{n_2 \ell_2 N
  L}$ in the addition theorem (\ref{OneRangeAddTheor_Lambda}) can also be
derived via the weakly convergent expansion of the plane wave in terms of
Lambda functions \cite[Eq.\ (4.38)]{Weniger/1985}. The same approach
works also for the coefficients $a_{n_1 \ell_1}^{n_2 \ell_2 N+L L}$ in
the $B$ function addition theorem (\ref{OneRangeAddTheor_B}) (see
\cite[Eq.\ (7.7)]{Weniger/1985}), and it should also work in the case of
Slater-type functions with integral principal quantum numbers defined by
(\ref{Def_STF}), although this has apparently not been done yet.

The derivation of the addition theorems (\ref{OneRangeAddTheor_Lambda})
and (\ref{OneRangeAddTheor_B}) by Filter and Steinborn was based on their
remarkably compact convolution theorem of $B$ functions \cite[Eq.\
(4.1)]{Filter/Steinborn/1978a} which, however, can be derived more easily
with the help of the Fourier transform (\ref{FT_B_Fun}) \cite[Section
V]{Weniger/Steinborn/1983a}: {\allowdisplaybreaks\begin{align}
  \label{ConvInt_Bnlm_ESP}
  & \int \, B_{n_1,\ell_1}^{m_1} (\beta, [\bm{r}-\bm{r}']) \, B_{n_2,
    \ell_2}^{m_2} (\beta, \bm{r}') \, \mathrm{d}^{3} \bm{r}'
  \notag \\
  & \qquad \; = \; \frac{4\pi}{\beta^3} \,
  \sum_{\ell=\ell_{\mathrm{min}}}^{\ell_{\mathrm{max}}} \! {}^{(2)} \,
  \langle \ell m_1+m_2 \vert \ell_1 m_1 \vert \ell_2 m_2 \rangle
  \notag \\
  & \qquad \qquad \times \, \sum_{t=0}^{\Delta \ell} \, (-1)^t \,
  {\binom{\Delta \ell} {t}}
  \notag \\
  & \qquad \qquad \qquad \times \, B_{n_1+n_2+\ell_1+\ell_2-\ell-t+1,
    \ell}^{m_1+m_2} (\beta, \bm{r}) \, .
\end{align}}%
The abbreviation $\Delta \ell$ is defined by (\ref{Def_Del_l}).
For an application of this highly convenient expression for the
construction of the addition theorem (\ref{OneRangeAddTheor_Lambda}),
Filter and Steinborn only had to express Lambda function in terms of $B$
functions. This can be accomplished with the help of the following
expression \cite[Eq.\ (3.3-35)]{Weniger/1982} (see also \cite[Eq.\ (3.17)
and Ref.\ \lbrack 23\rbrack\ on p.\ 2736]{Filter/Steinborn/1980}):
\begin{align}
  \label{GLag_FinSum_RBF}
  & \mathrm{e}^{-z} \, L_{n}^{(\alpha)} (2z) \; = \; (2n+\alpha+1)
  \notag \\
  & \qquad \sum_{\nu=0}^{n} \, \frac{(-2)^{\nu} \Gamma (n+\alpha+\nu+1)}
  {\nu! (n-\nu)! \Gamma (\alpha+2\nu+2)} \, \hat{k}_{\nu+1/2} (z) \, .
\end{align}
This yields a finite sum representation of Lambda functions
in terms of $B$ functions \cite[Eq.\ (3.18)]{Filter/Steinborn/1980}:
\begin{align}
  \label{Lambda_Bfun}
  & \Lambda_{n, \ell}^{m} (\beta, \bm{r}) \; = \;
    (2 \beta)^{3/2} \, 2^{\ell} \, \frac{(2n+1)}{(2\ell+3)!!} \,
    \left[ \frac{(n+\ell+1)!}{(n-\ell-1)!} \right]^{1/2}
  \notag \\[1\jot]
  & \quad \times \sum_{\nu=0}^{n-\ell-1} \, \frac{(-n+\ell+1)_{\nu} \,
    (n+\ell+2)_{\nu}}{\nu! \, (\ell+5/2)_{\nu}} \, B_{\nu+1,\ell}^{m}
  (\beta, \hm{r}) \, .
\end{align}
In this way, all overlap integrals of Lambda functions or between $B$ and
Lambda functions, which occur according to (\ref{OneRangeAddTheor}) as
intermediate steps in the derivation of the one-range addition theorems
(\ref{OneRangeAddTheor_Lambda}) and (\ref{OneRangeAddTheor_B}), can
because of (\ref{ConvInt_Bnlm_ESP}) be expressed in terms of $B$
functions.

Now, one only has to express all $B$ functions in terms of Lambda
functions. This can be done with the help of the following relationship
\cite[Eq.\ (B.2) on p.\ 214]{Buchholz/1969}:
\begin{equation}
  \label{RBF_SpecGLag}
\hat{k}_{n+1/2} (z) \; = \;
(-2)^{-n} \, n! \, \mathrm{e}^{-z} \, L_{n}^{(-2n-1)} (2z) \, .
\end{equation}
As is well known, a generalized Laguerre polynomial with superscript
$\beta$ can be expressed as a finite sum of generalized Laguerre
polynomials with superscript $\alpha$ \cite[p.\
249]{Magnus/Oberhettinger/Soni/1966}:
\begin{equation}
  \label{GLag_FinSum_2}
L_{n}^{(\beta)} (x) \; = \;  
\sum_{m=0}^{n} \, \frac{(\beta-\alpha)_m}{m!} \, L_{n-m}^{(\alpha)} (x) 
\, .
\end{equation}
This yields the following finite sum representation of a reduced Bessel
function with half integral order in terms of generalized Laguerre
polynomials with an essentially arbitrary superscripts $\alpha$:
\begin{align}
  \label{RBF_ExpGLag}
  & \hat{k}_{n+1/2} (z) \; = \; \frac{n!}{2^n} \, \mathrm{e}^{-z}
  \notag \\
  & \qquad \times \, \sum_{m=0}^{n} \, (-1)^m \, 
  \binom {2n+\alpha+1}{n-m} \, L_{m}^{(\alpha)} (2z) \, .
\end{align}
Filter and Steinborn \cite[Eq.\ (3.13)]{Filter/Steinborn/1980} gave an
expression which wrongly contains the additional factor $(-1)^n$.

Thus, we can express a $B$ function as a finite sum of Lambda functions
(see also \cite[Eq.\ (3.14)]{Filter/Steinborn/1980}):
\begin{align}
  \label{Bfun2Lambda}
  & B_{n,\ell}^{m} (\beta, \hm{r}) \; = \; (2\beta)^{-3/2}
  \frac{(n+2\ell+3)_{n-1}}{2^{2n+2\ell-1} \, (n+\ell)!}
  \notag \\
  & \qquad \times \, \sum_{\nu=0}^{n-1} \,
  \frac{(1-n)_{\nu}}{(n+2\ell+3)_{\nu}} \, 
  \left[ \frac{(\nu+2\ell+2)!} {\nu!} \right]^{1/2} 
  \notag \\
  & \qquad \qquad \times \, 
  \Lambda_{\nu+\ell+1, \ell}^{m} (\beta, \bm{r}) \, ,
  \qquad n \in \mathbb{N} \, , \quad \beta > 0 \, .
\end{align}

Starting from the Lambda function addition theorem
(\ref{OneRangeAddTheor_Lambda}), a symmetrical one-range addition theorem
of the type of (\ref{SymOneRangeAddTheor}) for Slater-type functions with
integral principal quantum numbers $n \in \mathbb{N}$ can be derived
easily. As is well known, an integral power $x^m$ with $m \in
\mathbb{N}_0$ can be expressed as a finite sum of generalized Laguerre
polynomials \cite[Eq.\ (2) on p.\ 207]{Rainville/1971}:
\begin{equation}
  \label{x^m_GlagPol}
x^m \; = \; (\alpha+1)_m \, \sum_{n=0}^{m} \,
\frac{(-m)_n}{(\alpha+1)_n} \, L_{n}^{(\alpha)} (x) \, .
\end{equation}
Thus, a Slater-type function with integral principal quantum number $n
\in \mathbb{N}$ defined by (\ref{Def_STF}) can be expressed as a finite
sum of Lambda functions \cite[Eq.\ (3.10)]{Filter/Steinborn/1980}:
\begin{align}
  & \chi_{n, \ell}^{m} (\beta, \bm{r}) \; = \; (2\beta)^{-3/2} \,
  \frac{(2\ell+3)_{n-\ell-1}}{2^{n-1}}
  \notag \\
  & \qquad \times \, \sum_{\nu=0}^{n-\ell-1} \,
  \frac{(-n+\ell+1)_{\nu}}{(2\ell+3)_{\nu}} \, \left[
    \frac{(\nu+2\ell+2)!} {\nu!} \right]^{1/2}
  \notag \\
  & \qquad \qquad \times \, \Lambda_{\nu+\ell+1, \ell}^{m} (\beta,
  \bm{r}) \, .
\end{align}
Accordingly, a symmetrical one-range addition theorem for Slater-type
functions with integral principal quantum numbers $n \in \mathbb{N}$ in
terms of Lambda functions can be written down immediately. The expansion
coefficients of this addition theorem are just simple finite sums of the
coefficients $T_{n_1 \ell_1}^{n_2 \ell_2 N L}$ in
(\ref{OneRangeAddTheor_Lambda}). 

Alternatively, one could just as well start from the $B$ function
addition theorem (\ref{OneRangeAddTheor_B}) since a Slater-type function
with an integral principal quantum number can also be expressed as a
finite sum of $B$ functions \cite[Eqs.\ (3.3) and
(3.4)]{Filter/Steinborn/1978b}.
\begin{align}
  \label{STF->Bfun}
  & \chi_{n, \ell}^{m} (\beta, \bm{r}) 
  \notag \\ 
  & \quad \; = \;
  2^n \, \sum_{\sigma \ge 0} \, (-1)^{\sigma} \,
  \frac{(-[n-\ell-1]/2)_{\sigma} \, (-[n-\ell]/2)_{\sigma}}{{\sigma}!}
  \notag \\
  & \quad \quad \qquad \times \,
  (n-\sigma)! \, B_{n-\ell-\sigma, \ell}^{m} (\beta, \bm{r}) \, .
\end{align}
If the principal quantum number $n$ is a positive integer, this expansion
terminates because of the Pochhammer symbols $(-[n-\ell-1]/2)_{\sigma}$
and $(-[n-\ell]/2)_{\sigma}$ after a finite number of steps. 

Thus, the expansion coefficients of the Slater addition theorem can also
be expressed as simple finite sums of the coefficients $a_{n_1
  \ell_1}^{n_2 \ell_2 N+L L}$ in (\ref{OneRangeAddTheor_B}).

The finite sum formula (\ref{x^m_GlagPol}) can be generalized to
nonintegral powers $x^{\mu}$ with $\mu \in \mathbb{C} \setminus
\mathbb{N}_0$. With the help of \cite[Eq.\ (7.414.7) on p.\
850]{Gradshteyn/Rhyzhik/1994}
\begin{align}
  \label{GR_7.414.7}
  & \int_{0}^{\infty} \, \mathrm{e}^{-st} \, t^{\beta} \,
  L_{n}^{(\alpha)} (t) \, \mathrm{d} t
  \notag \\
  & \qquad \; = \; \frac{\Gamma (\beta+1) \, \Gamma (\alpha+n+1)}{n! \,
    \Gamma (\alpha+1)} \, s^{-\beta-1} 
\notag \\ 
  & \qquad \qquad \times \, {}_2 F_1 (-n, \beta+1;
  \alpha+1; 1/s) \, ,
  \notag \\[1\jot]
  & \qquad \qquad \qquad \Re (\beta) > - 1 \, , \qquad \Re (s) > 0 \, ,
\end{align}
we obtain:
\begin{align}
  \label{GenPow2GLag}
  x^{\mu} & \; = \; \frac{\Gamma (\mu+\alpha+1)}{\Gamma (\alpha+1)} \,
  \sum_{n=0}^{\infty} \, \frac{(-\mu)_n}{(\alpha+1)_n} \,
  L_{n}^{(\alpha)} (x) \, ,
  \notag \\
  & \qquad \mu \in \mathbb{C} \setminus \mathbb{N}_0 \, , \qquad \Re
  (\mu+\alpha), \Re (\alpha) > - 1 \, .
\end{align}
If we set $\mu=m$ with $m \in \mathbb{N}_0$, the infinite series on the
right-hand side terminates because of the Pochhammer symbol $(-m)_n$ and
we obtain the finite sum (\ref{x^m_GlagPol}). However, there is a
fundamental difference between (\ref{x^m_GlagPol}) and
(\ref{GenPow2GLag}). The finite sum formula (\ref{x^m_GlagPol}) is a
relationship among polynomials. Therefore, it is valid pointwise for
arbitrary $x \in \mathbb{C}$. In the case of the infinite series
(\ref{GenPow2GLag}), we know that it converges in the mean in the radial
Hilbert space (\ref{HilbertL^2_Lag}), but we cannot assume that it
converges pointwise. Moreover, the index dependence of the series
coefficients on the right-hand side of (\ref{GenPow2GLag}) indicates that
the convergence of this expansion is slow if it does not terminate. These
convergence problems can be demonstrated by considering the special case
$\mu=-1$ and $x \to 0$ in (\ref{GenPow2GLag}). Then, the left-hand side
approaches $+\infty$. For an analysis of the behavior of the right-hand
side of (\ref{GenPow2GLag}) as $x \to 0$, we use (\ref{GLag_1F1}) and
obtain
\begin{equation}
L_{n}^{(\alpha)} (0) \; = \; \frac{(\alpha+1)_n}{n!} \, .
\end{equation}
Inserting this into (\ref{GLag_1F1}) yields:
\begin{align}
  \lim_{x \to 0} \, \frac{1}{x} & \; = \; \frac{1}{\alpha} \,
  \sum_{n=0}^{\infty} \, \frac{(1)_n}{(\alpha+1)_n} \,
  \frac{(\alpha+1)_n}{n!}
  \notag \\
  & \; = \; \frac{1}{\alpha} \, \sum_{n=0}^{\infty} \, \frac{(1)_n}{n!}
  \, .
\end{align}
Since $(1)_n=n!$, the series diverges for $x \to 0$ to $+\infty$,
although each individual term of the series (\ref{GenPow2GLag}) with
$\mu=-1$ is well behaved as $x \to 0$. Of course, pointwise convergence
-- or also divergence -- of an expansion like (\ref{GenPow2GLag}) for
nonintegral $\mu \in \mathbb{R} \setminus \mathbb{N}_0$ is not really
important if we want to use it in integrals. However, the at best slow
decay of the terms on the right-hand side of (\ref{GenPow2GLag})
indicates that expansions of a Slater-type function with nonintegral
principal quantum number $n \notin \mathbb{N}$ in terms of Lambda
functions or other exponentially decaying Laguerre-type function sets
converge slowly.

By combining (\ref{GenPow2GLag}) with the Lambda function addition
theorem (\ref{OneRangeAddTheor_Lambda}), a symmetrical one-range addition
theorem for a Slater-type functions with a nonintegral principal quantum
numbers in terms of Lambda functions can be formulated easily. The only
principal constraint on the nonintegral principal quantum number is that
the corresponding Slater-type function must be square integrable, i.e.,
it must belong to the Hilbert space $L^{2} (\mathbb{R}^3)$. However, the
expansion coefficients of this one range addition theorem are now given
by an infinite series involving the coefficients $T_{n_1 \ell_1}^{n_2
  \ell_2 N L}$ in (\ref{OneRangeAddTheor_Lambda}). As remarked above, we
have no \emph{a priori} reason to assume that the convergence of this
series would be rapid. So, unless somebody accomplishes a substantial
simplification -- for example by expressing an inner sum in closed form
via a summation theorem of a generalized hypergeometric series with unit
argument -- we are confronted with a one-range addition theorem for
Slater-type functions with nonintegral principal quantum numbers that is
significantly more complicated and less suited for practical work than
the corresponding addition theorem for Slater-type functions with
integral principal quantum numbers.

Let us now assume that we have a symmetrical one-range addition theorem
for Slater-type functions in terms of Lambda functions of the type of
(\ref{OneRangeAddTheor_Lambda}) or (\ref{OneRangeAddTheor_B}). Then we
only have to replace the Lambda functions by Guseinov's functions to
obtain an expansion of a Slater-type function in terms of Guseinov's
functions $\prescript{}{k}{\Psi}_{n, \ell}^{m} (\beta, \bm{r})$ with $k =
-1, 0, 1, 2, \dots$ defined by (\ref{Def_Psi_Guseinov}). With the help of
(\ref{GLag_FinSum_2}) we obtain the following expression for a Lambda
function as a finite sum of Guseinov's functions:
\begin{align}
  \label{Lambda2GusFun}
  \Lambda_{n, \ell}^{m} (\beta, \bm{r}) & \; = \; 
  (2\beta)^{-k/2} \sum_{\nu=0}^{\min (n-\ell-1, k)} \, \left[
    \frac{(n-\ell-\nu)_{\nu}}{(n+\ell+2)_{k-\nu}} \right]^{1/2}
  \notag \\
  & \qquad \times \,
  \frac{(-k)_{\nu}}{\nu!} \, \prescript{}{k}{\Psi}_{n-\nu, \ell}^{m}
  (\beta, \bm{r}) \, .
\end{align}
In this way, expansions in terms of Lambda functions can be transformed
easily to expansions in terms of Guseinov's functions.

An inverse relationship -- the expansion of Guseinov's functions in terms
of Lambda functions -- can also be derived via (\ref{GLag_FinSum_2}):
\begin{align}
  \label{GusFun2Lambda}
  \prescript{}{k}{\Psi}_{n, \ell}^{m} (\beta, \bm{r}) & \; = \;
  (2\beta)^{k/2} \, \sum_{\nu=0}^{n-\ell-1} \, \left[
    \frac{(n-\ell-\nu)_{\nu}}{(n+\ell-\nu+2)_{k+\nu}} \right]^{1/2}
  \notag \\
  & \qquad \times \,
  \frac{(k)_{\nu}}{\nu!} \, \Lambda_{n-\nu, \ell}^{m} (\beta, \bm{r}) \, .
\end{align}
With the help of this finite sum, we could -- starting from the addition
theorem (\ref{OneRangeAddTheor_Lambda}) for Lambda functions --
immediately write down a symmetrical one-range addition theorem for
Guseinov's function. Moreover, all Lambda functions in this addition
theorem could be replaced by Guseinov's functions via
(\ref{Lambda2GusFun}). 

The approach of Filter and Steinborn \cite{Filter/Steinborn/1980}, which
is based on the convolution theorem (\ref{ConvInt_Bnlm_ESP}) of $B$
functions, can also be used to construct from the scratch symmetrical
one-range addition theorems that are expansions in terms of Guseinov's
functions. Setting $w (r) = r^k$ and $\psi_{n, \ell}^{m} (\bm{r}) =
\prescript{}{k}{\Psi}_{n, \ell}^{m} (\beta, \bm{r})$ in
(\ref{SymOneRangeAddTheor_w}) yields the following one-range addition
theorem: 
{\allowdisplaybreaks\begin{subequations}
  \label{SymOneRangeAddTheor_GusFun}
  \begin{align}
    \label{SymOneRangeAddTheor_GusFun_a}
    & f (\bm{r} \pm \bm{r}') \; = \; \sum_{\substack{n \ell m \\ n'
        \ell' m'}} \, \prescript{}{k}{\mathbf{T}}_{n' \ell' m'}^{n \ell
      m} (f; \beta, \pm)
    \notag \\
    & \qquad \qquad \qquad \, \times \, 
    \prescript{}{k}{\Psi}_{n, \ell}^{m} (\beta, \bm{r}) \, 
    \prescript{}{k}{\Psi}_{n', \ell'}^{m'} (\beta, \bm{r}') \, ,
    \\
    \label{SymOneRangeAddTheor_GusFun_b}
    & \prescript{}{k}{\mathbf{T}}_{n' \ell' m'}^{n \ell m} (f; \beta,
    \pm)
    \notag \\
    & \qquad \; = \; \int \, \bigl[ \prescript{}{k}{\Psi}_{n',
      \ell'}^{m'} (\beta, \bm{r}') \bigr]^{*} \, (r')^k \,
    \prescript{}{k}{\mathbf{C}}_{n, \ell}^{m} (f; \beta, \pm \bm{r}')
    \, \mathrm{d}^3 \bm{r}' \, ,
    \\
    \label{SymOneRangeAddTheor_GusFun_c}
    & \prescript{}{k}{\mathbf{C}}_{n, \ell}^{m} (f; \beta, \pm \bm{r}')
    \notag \\
    & \qquad \; = \; \int \, \bigl[ \prescript{}{k}{\Psi}_{n, \ell}^{m}
    (\beta, \bm{r}) \bigr]^{*} \, r^k \, f (\bm{r} \pm \bm{r}') \,
    \mathrm{d}^3 \bm{r} \, .
  \end{align}
\end{subequations}}
If $f \in L_{r^k}^{2} (\mathbb{R}^3)$, this addition theorem converges in
the mean according to the norm (\ref{InnerProd_r^k}) of the weighted
Hilbert space $L_{r^k}^{2} (\mathbb{R}^3)$.

For the derivation of an addition theorem of the type of
(\ref{SymOneRangeAddTheor_GusFun}) for Guseinov's functions or $B$
functions via the convolution theorem (\ref{ConvInt_Bnlm_ESP}) of $B$
functions, we have to express Guseinov's functions in terms of $B$
functions. This can be done with the help of (\ref{GLag_FinSum_RBF}),
yielding
\begin{align}
  \label{GusFun_Bfun}
  & \prescript{}{k}{\Psi}_{n, \ell}^{m} (\beta,
  \bm{r}) \; = \; \left\{ \frac{\beta^{k+3} \, (n+\ell+k+1)!}{2^{k+1} \,
      (n-\ell-1)!}  \right\}^{1/2}
  \notag \\
  & \quad \times \frac{(2n+k+1) \, \Gamma (1/2) \, (\ell+1)!}  {\Gamma
    \bigl(\ell+2+k/2\bigr) \, \Gamma \bigl(\ell+[k+5]/2\bigr)}
  \notag \\
  & \qquad \times \sum_{\nu=0}^{n-\ell-1} \, \frac{(-n+\ell+1)_{\nu} \,
    (n+\ell+k+2)_{\nu} \, (\ell+2)_{\nu}}{\nu!  \,
    \bigl(\ell+2+k/2\bigr)_{\nu} \, \bigl(\ell+[k+5]/2\bigr)_{\nu}}
  \notag \\
  & \qquad \quad \times \, B_{\nu+1,\ell}^{m} (\beta, \hm{r}) \, .
\end{align}
With the help of this relationship, two-range addition theorems of
Guseinov's function can be written down immediately by forming finite
linear combinations of the corresponding addition theorems of $B$
functions \cite{Weniger/2002,Weniger/Steinborn/1989b}.

The weight function $r^k$ in (\ref{SymOneRangeAddTheor_GusFun_b}) and
(\ref{SymOneRangeAddTheor_GusFun_c}), which is responsible for the
orthonormality of Guseinov's functions according to
(\ref{Psi_Guseinov_OrthoNor}), can be absorbed with the help of
\cite[Eq.\ (6.1))]{Filter/Steinborn/1978a}
\begin{align}
  \label{IntPowRBF_HI}
  & z^{s} \hat{k}_{n-1/2} (z)
  \notag \\
  & \qquad \; = \; \sum_{\sigma \ge 0} \, (-2)^{\sigma} \,
   \frac{(-s/2)_{\sigma} \, (-n-[s-1]/2)_{\sigma}}{{\sigma}!} 
\notag \\
& \qquad \qquad \times \, 
  \hat{k}_{n+s-\sigma-1/2} (z) \, ,
  \notag \\
  & \qquad \qquad \qquad s = -1, 0, 1, 2, \dots \, ,
   \qquad n \in \mathbb{N} \, ,
\end{align}
yielding
\begin{align}
  \label{IntPowBfun}
  & r^{s} B_{n, \ell}^{m} (\beta, \bm{r})
  \notag \\
  & \qquad \; = \; (2/\beta)^s \, \sum_{\sigma \ge 0} \, (-1)^{\sigma} \,
  \frac{(-s/2)_{\sigma} \, (-n-[s-1]/2)_{\sigma}}
  {{\sigma}! \, (n+\ell+1)_{s-\sigma}}
\notag \\
  & \qquad \qquad \times \,  B_{n+s-\sigma, \ell}^{m} (\beta, \bm{r}) \, ,
  \notag \\
  & \qquad \qquad \qquad s = -1, 0, 1, 2, \dots \, , \qquad n \in
  \mathbb{N} \, .
\end{align}
If $s$ is an even integer, the Pochhammer symbol $(-s/2)_{\sigma}$ causes a
truncation of the summation after a finite number of steps, and if $s$ is
odd, this truncation is accomplished by the Pochhammer symbol
$(-n-[s-1]/2)_{\sigma}$.

Finally, we have to express $B$ functions as finite linear combinations
of Guseinov's functions. With the help of (\ref{RBF_ExpGLag}), we obtain:
\begin{align}
  \label{Bfun_GusFun}
  & B_{n,\ell}^{m} (\beta, \hm{r}) \; = \;
  \frac{(n+2\ell+k+3)_{n-1}}{2^{2n+2\ell-1} \, (n+\ell)!}
  \notag \\
  & \qquad \times \, \sum_{\nu=0}^{n-1} \,
  \frac{(1-n)_{\nu}}{(n+2\ell+k+3)_{\nu}} \, \left[
    \frac{(\nu+2\ell+k+2)!}{(2\beta)^{k+3} \, \nu!} \right]^{1/2}
  \notag \\
  & \qquad \qquad \times \, 
  \prescript{}{k}{\Psi}_{\nu+\ell+1, \ell}^{m} (\beta, \bm{r}) \, .
\end{align}

With the help of (\ref{GusFun_Bfun}), (\ref{IntPowBfun}), and
(\ref{Bfun_GusFun}) it is possible to generalize the approach of Filter
and Steinborn \cite{Filter/Steinborn/1980}, which was based on the
convolution theorem (\ref{ConvInt_Bnlm_ESP}) of $B$ functions and which
produced expansions in terms of Lambda functions, to one-range addition
theorems for exponentially decaying functions that are expansions in
terms of Guseinov's functions $\prescript{}{k}{\Psi}_{n, \ell}^{m}
(\beta, \bm{r})$.

We cannot expect that these addition theorems are as compact as the
corresponding expansions in terms of Lambda functions, which can be
obtained by setting $k=0$ in Guseinov's functions. For example,
(\ref{GusFun_Bfun}) simplifies considerably for $k=-1$ and $k=0$
\cite[Eqs.\ (4.19) and (4.20)]{Weniger/1985}. Moreover, for $k=-1$ and
$k= 1, 2, \dots$, the weight function $r^k$ has to be absorbed via
(\ref{IntPowRBF_HI}), which produces an additional inner sum. So, unless
some additional simplifications can be found, one-range addition theorems
in terms of Guseinov's functions will have a more complicated structure
than the corresponding addition theorems in terms of Lambda functions.

This higher complexity is one reason why I am not interested in
explicitly constructing symmetrical one-range addition theorems that are
expansions in terms of Guseinov's functions with $k \ge 1$. I doubt that
these addition theorems would be useful enough to justify the effort. A
second reason is that I have -- as outlined in Section
\ref{Sec:LagTypeFun} -- severe doubts whether the weighted Hilbert spaces
$L_{r^k}^{2} (\mathbb{R}^3)$ with $k \neq 0$ are really suited for bound
state electronic structure calculations.

\typeout{==> Section: Guseinov's One-Range Addition Theorems for
  Slater-Type Functions}
\section{ Guseinov's One-Range Addition Theorems for Slater-Type
  Functions}
\label{Sec:AddTheor_STF}

As discussed in Section \ref{AddTheor_ETF}, symmetrical one-range
addition theorems of the type of (\ref{OneRangeAddTheor}) for
exponentially decaying functions can be derived comparatively easily via
the remarkably compact convolution theorem (\ref{ConvInt_Bnlm_ESP}) of
$B$ functions. However, Guseinov preferred to proceed differently.

In \cite{Guseinov/2001a}, Guseinov derived one-range addition theorems
for Slater-type functions with integral principal quantum numbers by
expanding them in terms of Sturmians and Lambda functions, and in
\cite{Guseinov/2002b} he did this for Slater-type functions with
nonintegral principal quantum numbers. In \cite{Guseinov/2002c}, Guseinov
introduced his functions $\prescript{}{k}{\Psi}_{n, \ell}^{m} (\beta,
\bm{r})$ defined by (\ref{Def_Psi_Guseinov}) and used them for the
construction of one-range addition theorems for Slater-type functions
with integral principal quantum numbers. Since Sturmians and Lambda
functions are special cases of Guseinov's functions with $k=-1$ and
$k=0$, respectively, these addition theorems generalize earlier
expansions in terms of Lambda functions and Sturmians derived in
\cite{Guseinov/2001a}. In \cite{Guseinov/2002d} Guseinov derived
one-range addition theorems in terms of his functions for so-called
``central and noncentral potentials'' which are nothing but special
Slater-type functions with integral and nonintegral principal quantum
numbers. In \cite{Guseinov/2003b}, Guseinov provided a ``unified
treatment'' of multicenter integrals of Slater-type functions with
integral and nonintegral principal quantum numbers by expanding
Slater-type functions in terms of his functions. In
\cite{Guseinov/2003c}, Guseinov constructed addition theorems for his
functions both in the coordinate as well as in the momentum
representation by expanding his functions in terms of Slater-type
functions with integral principal quantum numbers. In
\cite{Guseinov/2003d}, Guseinov used his addition theorems for
Slater-type functions to handle multicenter integrals of what he calls
``central and noncentral interaction potentials'' which are special
Slater-type functions with integral and nonintegral principal quantum
numbers. In \cite{Guseinov/2003e}, Guseinov provided again a ``unified
analytical treatment'' of multicenter integrals of ``central and
noncentral interaction potentials'' via one-range addition theorems of
Slater-type functions derived with the help of his functions. In
\cite{Guseinov/2004a}, Guseinov provided in this way a ``unified
analytical treatment'' of multicenter nuclear attraction, electric field
and electric field gradient integrals over Slater-type functions. In
\cite{Guseinov/2004b,Guseinov/2004e}, Guseinov provided another ``unified
treatment'' of essentially the same integrals as in
\cite{Guseinov/2004a}, but this type he emphasized the use of his
functions. In \cite{Guseinov/2004c}, we find another ``unified
treatment'' of multicenter integrals of Slater-type functions, but this
time the potentials are called ``integer and noninteger $u$ Yukawa-type
screened Coulomb type potentials'' (again, these potentials are nothing
but special Slater-type functions with integral and nonintegral principal
quantum numbers). In \cite{Guseinov/2004d}, Guseinov constructed
one-range addition theorems for derivatives of Slater-type functions, and
in \cite{Guseinov/2004f,Guseinov/2004i}, he considered again multicenter
integrals of ``central and noncentral interaction potentials''. In
\cite{Guseinov/2004k}, Guseinov provided another ``unified treatment'',
but this time of electronic attraction, electric field, and
electric-field gradient multicenter integrals of ``screened and
nonscreened Coulomb potentials''. In \cite{Guseinov/2005a}, Guseinov
derived one-range addition theorems for the Coulomb potential starting
from his addition theorems for the Yukawa potential which is essentially
a special Slater-type function $\chi_{N, L}^{M}$ with $N=L=M=0$. In
\cite{Guseinov/2005b,Guseinov/2006b}, Guseinov used the momentum space
addition theorems derived in \cite{Guseinov/2003c} for the construction
of momentum space addition theorems for Slater-type functions. In
\cite{Guseinov/2005c,Guseinov/2005d,Guseinov/2005e,Guseinov/2005g,%
  Guseinov/2006a}, Guseinov considered one-range addition theorems for
``Yukawa-like central and noncentral interaction potentials and their
derivatives'', for derivatives of his functions, for ``derivatives of
integer and noninteger $u$ Coulomb-Yukawa type central and noncentral
potentials'', and for ``combined Coulomb and Yukawa like central and
noncentral interaction potentials and their derivatives'', respectively.
In \cite{Guseinov/2007e}, Guseinov considered one-range expansions for
two-center charge densities of Slater-type functions with integral and
nonintegral principal quantum numbers. Finally, in \cite{Guseinov/2007f}
Guseinov provided another ``unified treatment'' of expansion theorems and
one-range addition theorems of complete orthonormal sets of functions in
coordinate, momentum and four-dimensional spaces.

These examples show that Guseinov used quite a few different names for
the functions he expanded, but with few exceptions he concentrated on
one-range addition theorems for Slater-type functions with integral and
nonintegral principal quantum numbers in the coordinate representation.
Consequently. I will also focus on these addition theorems and will
examine critically their derivation by Guseinov.

Guseinov exclusively derived unsymmetrical addition theorems of the type
of (\ref{OneRangeAddTheor}) or (\ref{OneRangeAddTheor_w}), in which the
two vectors $\bm{r}$ and $\bm{r}'$ are treated differently. He never
derived completely symmetrical addition theorems of the type of
(\ref{SymOneRangeAddTheor}), (\ref{SymOneRangeAddTheor_w}), or
(\ref{SymOneRangeAddTheor_GusFun}).

Moreover, Guseinov expanded Slater-type functions in terms of functions
with a \emph{different} scaling parameter. Normally, this is a bad idea
since overlap integrals with different scaling parameters are
significantly more complicated than overlap integrals with equal scaling
parameters (this is the reason why exclusively addition theorems with
equal scaling parameters were discussed in Section \ref{AddTheor_ETF}).
However, Guseinov needed this additional degree of freedom since he
wanted to construct one-range addition theorems for the Coulomb potential
by representing it as the limiting case of the Yukawa potential according
to $1/r = \lim_{\beta \to 0} \exp (- \beta r)/r$.

Guseinov's work on one-range addition theorems is based on expansions of
Slater-type functions with integral or nonintegral principal quantum
numbers in terms of his functions $\prescript{}{k}{\Psi}_{n, \ell}^{m}
(\gamma, \bm{r})$ with $k=-1, 0, 1, 2, \dots$ and $\gamma > 0$ (see for
example \cite[Eqs.\ (11) and (12)]{Guseinov/2002c}):
\begin{subequations}
  \label{Gus_OneRangeAddTheorSTF_k}
  \begin{align}
    \label{Gus_OneRangeAddTheorSTF_k_a}
    & \chi_{N, L}^{M} (\beta, \bm{r} \pm \bm{r}')
    \notag \\
    & \quad \; = \; \sum_{n \ell m} \, 
     \prescript{}{k}{\mathbf{X}}_{n, \ell, m}^{N, L, M} 
     (\gamma, \beta, \pm \bm{r}') \,
    \prescript{}{k}{\Psi}_{n, \ell}^{m} (\gamma, \bm{r}) \, ,
    \\
    \label{Gus_OneRangeAddTheorSTF_k_b}
    & \prescript{}{k}{\mathbf{X}}_{n, \ell, m}^{N, L, M} 
    (\gamma, \beta, \pm \bm{r}')
    \notag \\
    & \quad \; = \; \int \, \bigl[ \prescript{}{k}{\Psi}_{n, \ell}^{m}
    (\gamma, \bm{r}) \bigr]^{*} \, r^k \, \chi_{N, L}^{M} (\beta, \bm{r}
    \pm \bm{r}') \, \mathrm{d}^3 \bm{r} \, .
  \end{align}
\end{subequations}
As remarked above, Guseinov did not derive completely symmetrical
addition theorems of the type of (\ref{SymOneRangeAddTheor_GusFun}) by
expanding the overlap integrals (\ref{Gus_OneRangeAddTheorSTF_k_b}) in
terms of his functions. This is not completely satisfactory: In the
Slater-type function $\chi_{N, L}^{M} (\beta, \bm{r} \pm \bm{r}')$, the
vectors $\bm{r}$ and $\bm{r}'$ play an identical role, but not in the
addition theorems (\ref{Gus_OneRangeAddTheorSTF_k}).

Nevertheless, a definite assessment of the relative merits of symmetrical
and unsymmetrical one-range addition theorems is not so easy. Since we
cannot tacitly assume that one-range addition theorems necessarily lead
to rapidly convergent expansions for multicenter integrals, the crucial
question is whether it is easier to compute either overlap integrals like
(\ref{Gus_OneRangeAddTheorSTF_k_b}) or purely numerical coefficients as
they occur in (\ref{OverlapExpand_b}), (\ref{OverlapExpand_w_b}), and
(\ref{SymOneRangeAddTheor_GusFun_b}) both effectively and reliably for
large indices.

It also makes a big difference if one has to integrate over only one of
the two vectors $\bm{r}$ and $\bm{r}'$ or over both of them.  For
example, in \cite[Table I]{Guseinov/2005a} Guseinov considered
three-center one-electron nuclear attraction integrals
\begin{align}
  \label{ThreeCentNuclAttInt}
& \mathcal{I} (f, g; \bm{A}, \bm{B}, \bm{C})
\notag \\
& \qquad \; = \;
\int_{}^{} \, \bigl[ f (\bm{r} - \bm{A}) \bigr]^{*} \,
\frac{1}{\vert \bm{r} - \bm{B} \vert} \, g (\bm{r} - \bm{C}) \,
\mathrm{d}^3 \bm{r} \, ,
\end{align}
where $f$ and $g$ are Slater-type functions, and the vectors $\bm{A}$,
$\bm{B}$, and $\bm{C}$ are atomic centers. The integral
(\ref{ThreeCentNuclAttInt}) is the most complicated one-electron integral
occurring in molecular electronic structure calculations based on the
Hartree-Fock-Roothaan equations. By means of a shift of origin, one of
the three vectors $\bm{A}$, $\bm{B}$, and $\bm{C}$ can be made to vanish.
Thus, two addition theorems are needed to decouple the arguments in
(\ref{ThreeCentNuclAttInt}).

In the case of a one-electron integral like (\ref{ThreeCentNuclAttInt}),
the use of unsymmetrical one-range addition theorems of the type of
(\ref{Gus_OneRangeAddTheorSTF_k}) may well be feasible: It should not
matter too much whether the remaining two of the three vectors $\bm{A}$,
$\bm{B}$, and $\bm{C}$ occur in overlap integrals or in complete and
orthonormal functions. Nevertheless, a comparison of Guseinov's results
\cite[Table I]{Guseinov/2005a} with those reported by Bouferguene and
Jones \cite{Bouferguene/Jones/1998}, who had used two-range addition
theorems, would have been of considerable interest.

The most difficult multicenter integrals occurring in the LCAO-MO
approach are the notorious six-dimensional two-electron integrals
\begin{equation}
  \label{CouInt_f_g}
\mathcal{C} (f, g) \; = \; \int \! \int \, 
\frac{f^{*} (\bm{r}) \, g (\bm{r}')}{\vert \bm{r} - \bm{r}' \vert} \,
\mathrm{d}^{3} \bm{r} \, \mathrm{d}^{3} \bm{r}' \, ,
\end{equation}
which describe the Coulomb interaction of two in general nonclassical
charge distributions $[f (\bm{r})]^{*}$ and $g (\bm{r}')$ consisting of
products of effective one-particle wave functions located at different
atomic centers. We can only benefit in both integrations from the
simplifying power of orthogonality if we use in (\ref{CouInt_f_g})
symmetrical one-range addition theorems of the type of
(\ref{SymOneRangeAddTheor}), (\ref{SymOneRangeAddTheor_w}), or
(\ref{SymOneRangeAddTheor_GusFun}).

In principle, this applies also to multicenter integrals of those
functions which Guseinov and coworkers had called ``central and
noncentral interaction potentials''
\cite{Guseinov/2003b,Guseinov/2003d,Guseinov/2003e,Guseinov/2004c,%
  Guseinov/2004f,Guseinov/2004i,Guseinov/2005c,Guseinov/2005d,
  Guseinov/2005e,Guseinov/2005g,Guseinov/2006a,Guseinov/Mamedov/2004d,%
  Guseinov/Mamedov/2004e,Guseinov/Mamedov/2005d,Guseinov/Mamedov/2005g}
and which are nothing but special Slater-type functions with integral and
nonintegral principal quantum numbers. In the case of one-electron
integrals, unsymmetrical addition theorems of the type of
(\ref{Gus_OneRangeAddTheorSTF_k}) may well be sufficient, but if
two-electron integrals of these interaction potentials should indeed be
physically meaningful and have to be computed (see also
\cite{Novosadov/2001a}), I am skeptical about the feasibility of
Guseinov's approach based on the exclusive use of unsymmetrical addition
theorems.

The central computational problem of Guseinov's unsymmetrical one-range
addition theorems (\ref{Gus_OneRangeAddTheorSTF_k}) is the efficient and
reliable evaluation of the overlap integrals
(\ref{Gus_OneRangeAddTheorSTF_k_b}) between Slater-type functions and the
expansion functions $\prescript{}{k}{\Psi}_{n, \ell}^{m} (\gamma,
\bm{r})$ even for large indices.

Guseinov's solution is simple but not necessarily good. It follows at
once from the explicit expression (\ref{Def_GLagPol}) of the generalized
Laguerre polynomials that Guseinov's functions can be expressed as finite
sums of Slater-type functions with integral principal quantum numbers:
\begin{subequations}
  \label{GusFun2STF}
  \begin{align}
    \label{GusFun2STF_a}
    \prescript{}{k}{\Psi}_{n, \ell}^{m} (\gamma, \bm{r}) & \; = \;
    \sum_{\nu=0}^{n-\ell-1} \, \prescript{}{k}{\mathbf{G}}_{\nu}^{(n,
      \ell)} (\gamma) \, \chi_{\nu+\ell+1, \ell}^{m} (\gamma, \bm{r}) \, ,
    \\
    \label{GusFun2STF_b}
    \prescript{}{k}{\mathbf{G}}_{\nu}^{(n, \ell)} (\gamma) & \; = \;
    2^{\ell} \, \left[ \frac{(2\gamma)^{k+3} \,
        (n+\ell+k+1)!}{(n-\ell-1)!} \right]^{1/2}
    \notag \\
    & \qquad \times \, \frac{(-n+\ell+1)_{\nu} \,
      2^{\nu}}{(2\ell+k+\nu+2)! \, \nu!} \, .
  \end{align}
\end{subequations}
Accordingly, the overlap integrals (\ref{Gus_OneRangeAddTheorSTF_k_b})
can be expressed as finite sums of overlap integrals of Slater-type
functions: {\allowdisplaybreaks
\begin{subequations}
  \label{OverlapGusFun_STF}
  \begin{align}
    & \prescript{}{k}{\mathbf{X}}_{n, \ell, m}^{N, L, M} (\gamma, \beta,
    \pm \bm{r}') \; = \; \gamma^{-k}
    \notag \\
    & \quad \times \, \sum_{\nu=0}^{n-\ell-1} \,
    \prescript{}{k}{\mathbf{G}}_{\nu}^{(n, \ell)} (\gamma) \,
    \mathbf{S}_{\nu+\ell+k+1, \ell, m}^{N, L, M} (\gamma, \beta, \pm
    \bm{r}') \, ,
    \\
    & \mathbf{S}_{n, \ell, m}^{N, L, M} (\gamma, \beta, \pm \bm{r}')
    \notag \\
    & \quad \; = \; \int \, \bigl[ \chi_{n, \ell}^{m} (\gamma, \bm{r})
    \bigr]^{*} \, \chi_{N, L}^{M} (\beta, \bm{r} \pm \bm{r}') \,
    \mathrm{d}^3 \bm{r} \, .
  \end{align}
\end{subequations}}
Using this in the addition theorems (\ref{Gus_OneRangeAddTheorSTF_k})
yields:
\begin{align}
  \label{Gus_OneRangeAddTheorSTF_OvSTF_k}
  & \chi_{N, L}^{M} (\beta, \bm{r} \pm \bm{r}') \; = \; \gamma^{-k} \,
  \sum_{n \ell m} \, \prescript{}{k}{\Psi}_{n, \ell}^{m} (\gamma, \bm{r})
  \notag \\
  & \quad \, \times \, \sum_{\nu=0}^{n-\ell-1} \,
  \prescript{}{k}{\mathbf{G}}_{\nu}^{(n, \ell)} (\gamma) \,
  \mathbf{S}_{\nu+\ell+k+1, \ell, m}^{N, L, M} (\gamma, \beta, \pm
  \bm{r}') \, .
\end{align}
Superficially, this approach, which is mathematically completely
legitimate, seems to have the advantage that existing programs for
overlap integrals of Slater-type functions can be used for the evaluation
of the overlap integrals (\ref{Gus_OneRangeAddTheorSTF_k_b}) (this may be
the reason for Guseinov's approach). However, stability problems are
likely. The explicit expression (\ref{Def_GLagPol}) of a generalized
Laguerre polynomial $L_{n}^{(\alpha)} (x)$ in powers of $x$ becomes
numerical unstable for larger values of the index $n$. Because of these
stability problems, orthogonal polynomials are normally computed
recursively.

So, as long as the opposite is not explicitly proved, it certainly makes
sense to assume that an expression like (\ref{GusFun2STF}) inherits the
stability problems of (\ref{Def_GLagPol}). In addition, conventional
programs for overlap integrals of Slater-type functions, as they are for
instance used in semiempirical calculations, normally cannot be used in
the case of (very) large principal and angular momentum quantum numbers.
Thus, we need alternative expressions for the overlap integrals
(\ref{Gus_OneRangeAddTheorSTF_k_b}) not based on (\ref{GusFun2STF}),
which permit an efficient and reliable evaluation even for (very) large
indices.

In his desire to reduce his whole formalism of one-range addition
theorems to Slater-type functions, Guseinov even expressed the functions
$\prescript{}{k}{\Psi}_{n, \ell}^{m} (\gamma, \bm{r})$ on the right-hand
side of (\ref{Gus_OneRangeAddTheorSTF_OvSTF_k}) by Slater-type functions
according to (\ref{GusFun2STF}) (see for example \cite[Eq.\
(14)]{Guseinov/2002c}):
\begin{align}
  \label{Gus_OneRangeAddTheorSTF_STF_k}
  & \chi_{N, L}^{M} (\beta, \bm{r} \pm \bm{r}')
  \notag \\
  & \; = \; \gamma^{-k} \, \sum_{n \ell m} \, 
  \sum_{\nu'=0}^{n-\ell-1} \, 
  \prescript{}{k}{\mathbf{G}}_{\nu'}^{(n, \ell)} (\gamma) \, 
  \chi_{\nu'+\ell+1, \ell}^{m} (\gamma, \bm{r})
  \notag \\
  & \quad \; \times \,
    \sum_{\nu=0}^{n-\ell-1} \, 
    \prescript{}{k}{\mathbf{G}}_{\nu}^{(n, \ell)} (\gamma) \, 
    \mathbf{S}_{\nu+\ell+k+1, \ell, m}^{N, L, M} 
    (\gamma, \beta, \pm \bm{r}') \, .
\end{align}
This is still correct since the sum over $\nu'$ is nothing but the
function $\prescript{}{k}{\Psi}_{n, \ell}^{m} (\gamma, \bm{r})$ in
disguise. But Guseinov rearranged the order of summations in
(\ref{Gus_OneRangeAddTheorSTF_STF_k}), obtaining expansions for
Slater-type functions $\chi_{N, L}^{M} (\beta, \bm{r} \pm \bm{r}')$ with
integral or nonintegral principal quantum numbers in terms of Slater-type
functions $\chi_{n+\ell, \ell}^{m} (\beta, \bm{r})$ with integral
principal quantum numbers located at a different center (see for example
\cite[Eq.\ (15)]{Guseinov/2002c}): {\allowdisplaybreaks\begin{align}
  \label{Gus_RearrOneRangeAddTheorSTF_STF_k}
  & \chi_{N, L}^{M} (\beta, \bm{r} \pm \bm{r}') \; = \; 
  \gamma^{-k} \, \sum_{n \ell m} \, 
  \chi_{n+\ell, \ell}^{m} (\gamma, \bm{r})
  \notag \\
  & \qquad \times \,
  \sum_{n'=0}^{\infty} \, 
  \prescript{}{k}{\mathbf{G}}_{n-1}^{(n+n', \ell)} (\gamma) \,
    \sum_{\nu=0}^{n+n'-\ell-1} \, 
    \prescript{}{k}{\mathbf{G}}_{\nu}^{(n+n', \ell)} (\gamma)
  \notag \\
  & \qquad \quad \; \times \,
    \mathbf{S}_{\nu+\ell+k+1, \ell, m}^{N, L, M} 
    (\gamma, \beta, \pm \bm{r}') \, .
\end{align}}
This step is potentially disastrous. A rearrangement of the order of
summations of a double series is not always legitimate and can easily
lead to a divergent result (see for example \cite[Chapter
V]{Bromwich/1991}). So, when going from
(\ref{Gus_OneRangeAddTheorSTF_STF_k}) to
(\ref{Gus_RearrOneRangeAddTheorSTF_STF_k}), one cannot tacitly assume
that the rearranged expansion (\ref{Gus_RearrOneRangeAddTheorSTF_STF_k})
indeed converges.

As discussed in Section \ref{Sec:HilbertSpace}, a formal expansion of the
type of (\ref{Expand_f_CONS}) of a function $f$ in terms of a given
function set $\{ \varphi_n \}_{n=0}^{\infty}$ exists and converges in the
mean if $f$ belongs to the underlying Hilbert space and if the expansion
functions $\{ \varphi_n \}_{n=0}^{\infty}$ are complete and orthonormal
in this Hilbert space. If the expansion functions are only complete, but
not orthogonal, it is possible to construct finite approximations of the
type of (\ref{f_FinAppr}) by minimizing the mean square deviation $\Vert
f - f_N \Vert^2 = (f - f_N \vert f - f_N)$, but the existence of formal
expansions of the type of (\ref{Expand_f_formal}) in terms of
nonorthogonal functions is not guaranteed: These expansions may or may
not exist.
 
Slater-type functions are complete in all the Hilbert space considered in
this article (see for example \cite[Section 4]{Klahn/Bingel/1977b}), but
not orthogonal. Thus, it is not clear whether Guseinov's rearranged
addition theorems (\ref{Gus_RearrOneRangeAddTheorSTF_STF_k}) are
mathematically meaningful. This has to be checked.

Convergence and existence problems due to a rearrangement can be
illustrated via (\ref{GenPow2GLag}), which expresses a nonintegral power
$x^{\mu}$ with $\mu \in \mathbb{C} \setminus \mathbb{N}_0$ as an infinite
series of generalized Laguerre polynomials and which converges in the
mean in the Hilbert space (\ref{HilbertL^2_Lag}). If we insert the
explicit expression (\ref{Def_GLagPol}) of the generalized Laguerre
polynomials into (\ref{GenPow2GLag}) and interchange the order of
summations, we obtain after some algebra:
\begin{align}
  \label{Chk_x^m_GlagPol_6}
  x^{\mu} & \; = \; \frac{\Gamma (\mu+\alpha+1)}{\Gamma (\alpha+1)}
  \notag \\
  & \qquad \times \, \sum_{k=0}^{\infty} \, (-1)^k \,
  \frac{(-\mu)_k}{(\alpha+1)_k} \, \frac{x^k}{k!} \, {}_1 F_0 (k-\mu; 1)
  \, .
\end{align}
It looks as if we succeeded in constructing a power series expansion for
the nonintegral power $x^{\mu}$. However, the generalized hypergeometric
series ${}_1 F_0$ with unit argument is the limiting case $y \to -1$ of
the following binomial series satisfying \cite[p.\
38]{Magnus/Oberhettinger/Soni/1966}
\begin{equation}
  \label{BinomSer}
{}_1 F_0 (-a; -y) \; = \;
\sum_{m=0}^{\infty} \, \binom{a}{m} y^m \; = \; (1+y)^a \, ,
\qquad \vert y \vert < 1 \, .
\end{equation}
We then obtain for the hypergeometric series in
(\ref{Chk_x^m_GlagPol_6}):
\begin{align}
  \label{Lim_Ser_Chk_x^m_GlagPol_6}
&  {}_1 F_0 (k-\mu; 1) 
\notag \\
& \quad \; = \; \lim_{y \to -1} \, (1+y)^{\mu-k} \; = \;
{\begin{cases}
  \infty \, , \quad \mu < 0 \, , \\
  0 \, , \quad \; \, k < \mu \ge 0 \, , \\
  \infty \, , \quad k > \mu \ge 0 \, .
\end{cases}}
\end{align}
Thus, the power series (\ref{Chk_x^m_GlagPol_6}) does not exist and the
Laguerre series (\ref{GenPow2GLag}) for $x^{\mu}$ with $\mu \in
\mathbb{C} \setminus \mathbb{N}_0$ cannot be reformulated as a power
series in $x$. Of course, this is not really surprising: The general
power function $z^{\mu}$ with $z \in \mathbb{C}$ and $\mu \in \mathbb{C}
\setminus \mathbb{N}_0$ is not analytic at $z=0$ in the sense of complex
analysis. For $\mu=m$ with $m \in \mathbb{N}_0$, Taylor expansion of
$z^m$ around $z=0$ yields the trivial identity $z^m = z^m$.

One-range additions theorems for exponentially decaying functions are
fairly complicated mathematical objects. Consequently, \emph{explicit}
proofs of their convergence or divergence are very difficult.  It is
understandable that Guseinov was not interested in presenting such
proofs, although he should have done so in order to justify his
manipulations.  Fortunately, valuable insight can in some cases be gained
by considering not the addition theorems themselves, but their much
simpler one-center limits.

Let us therefore assume that we succeeded in constructing either 
symmetrical addition theorems of the type of
(\ref{SymOneRangeAddTheor_GusFun}) or also unsymmetrical addition
theorems involving overlap integrals for some function $f (\bm{r} \pm
\bm{r}')$ by expanding it in terms of Guseinov's functions. We now
consider the one-center limit by setting $\bm{r}' = \bm{0}$. Then, our
addition theorem must simplify to yield an identity for $f (\bm{r})$.
Under fortunate circumstances, the mathematical nature of this identity
can provide valuable insight.

First, we set $\beta=\gamma$ and $\bm{r}' = \bm{0}$ in the addition
theorems (\ref{Gus_OneRangeAddTheorSTF_k}) for $\chi_{N, L}^{M} (\beta,
\bm{r} \pm \bm{r}')$. After the cancellation of all common factors we
obtain expressions that are special cases of the expansion
(\ref{GenPow2GLag}) expressing an in general nonintegral power $x^{\mu}$
as an infinite series of generalized Laguerre polynomials. As shown in
(\ref{Chk_x^m_GlagPol_6}) - (\ref{Lim_Ser_Chk_x^m_GlagPol_6}), this
series can only be reformulated as a power series in $x$ if $\mu$ is a
nonnegative integer, $\mu=m$ with $m \in \mathbb{N}_0$, yielding the
trivial identity $x^m = x^m$. If we have instead $\mu \in \mathbb{R}
\setminus \mathbb{N}_0$, a rearranged power series in $x$ does not exist.

Thus, we can conclude that for $\beta=\gamma$, the one-center limit
$\bm{r}' = \bm{0}$ of the rearranged addition theorem
(\ref{Gus_RearrOneRangeAddTheorSTF_STF_k}) does not exist if the
principal quantum number $N$ of the Slater-type function $\chi_{N, L}^{M}
(\beta, \bm{r} \pm \bm{r}')$ is nonintegral, $N \in \mathbb{R} \setminus
\mathbb{N}$.

Next, we assume $\beta \neq \gamma$ in the addition theorem
(\ref{Gus_OneRangeAddTheorSTF_k}) and set $\bm{r}' = \bm{0}$. We then
obtain after the cancellation of all common factors expressions that are
special cases of the following expansion:
\begin{align}
  \label{ExpoPow2GLag}
  & x^{\mu} \, \mathrm{e}^{u x} \; = \; (1-u)^{-\alpha-\mu-1} \,
  \frac{\Gamma (\alpha+\mu+1)}{\Gamma (\alpha+1)}
  \notag \\
  & \quad \times \, \sum_{n=0}^{\infty} \, {}_2 F_1 \left(-n,
  \alpha+\mu+1; \alpha+1; \frac{1}{1-u} \right) \, L_{n}^{(\alpha)} (x) \, ,
  \notag \\
  & \qquad \mu \in \mathbb{R} \, , \quad \Re (\mu+\alpha) > - 1
  \, , \! \quad u \in (-\infty, 1/2) \, .
\end{align}  
This expansion, which can be derived with the help of (\ref{GR_7.414.7}),
converges in the mean with respect to the norm of the weighted Hilbert
space $L^{2}_{\mathrm{e}^{-x} x^{\alpha}} (\mathbb{R}_{+})$. For $u=0$,
(\ref{ExpoPow2GLag}) simplifies to give (\ref{GenPow2GLag}). This can be
shown easily with the help of Gauss' summation theorem \cite[p.\
40]{Magnus/Oberhettinger/Soni/1966}.

If we insert the explicit expression (\ref{Def_GLagPol}) of the
generalized Laguerre polynomials into (\ref{ExpoPow2GLag}) and
interchange the order of summations, we also obtain a formal power series
in $x$. Unfortunately, an analysis of the resulting power series becomes
very difficult because of the terminating Gaussian hypergeometric series
${}_2 F_1$ in (\ref{ExpoPow2GLag}) (probably, this would be a nontrivial
research project in its own right). However, we can argue that the
function $z^{\mu} \exp (u z)$ with $\mu, u, z \in \mathbb{C}$ is only
analytic at $z=0$ in the sense of complex analysis if $\mu$ is a
nonnegative integer, $\mu=m$ with $m \in \mathbb{N}_0$, yielding $z^m
\exp (u z) = \sum_{n=0}^{\infty} u^n z^{m+n}/n!$. If $\mu$ is
nonintegral, $\mu \in \mathbb{C} \setminus \mathbb{N}_0$, a power series
around $z=0$ does not exist.

Thus, we can conclude that also for $\beta \neq \gamma$, the one-center
limit $\bm{r}' = \bm{0}$ of the rearranged addition theorems
(\ref{Gus_RearrOneRangeAddTheorSTF_STF_k}) for $\chi_{N, L}^{M} (\beta,
\bm{r} \pm \bm{r}')$ does not exist if the principal quantum number $N$
is nonintegral, $N \in \mathbb{R} \setminus \mathbb{N}$.

From a mathematical point of view, a one-range addition theorem for a
function $f (\bm{r} \pm \bm{r}')$ is a mapping $\mathbb{R}^3 \times
\mathbb{R}^3 \to \mathbb{C}$. We are interested in \emph{one-range}
addition theorems because we would like to have a \emph{unique} (infinite
series) representation of $f (\bm{r} \pm \bm{r}')$ with \emph{separated}
variables $\bm{r}$ and $\bm{r}'$ that is valid for the \emph{whole}
argument set $\mathbb{R}^3 \times \mathbb{R}^3$. If we accept this
premise, then we have to conclude that Guseinov's manipulations, which
produced the rearranged addition theorems
(\ref{Gus_RearrOneRangeAddTheorSTF_STF_k}) for $\chi_{N, L}^{M} (\beta,
\bm{r} \pm \bm{r}')$, are at least in the case of nonintegral principal
quantum numbers $N \in \mathbb{R} \setminus \mathbb{N}$ a complete
failure.

This observation does not rule out the possibility that an appropriate
reinterpretation might make the rearranged addition theorems
(\ref{Gus_RearrOneRangeAddTheorSTF_STF_k}) with $N \in \mathbb{R}
\setminus \mathbb{N}$ mathematically meaningful in a restricted sense as
an approximation, although it does exist for the whole argument set
$\mathbb{R}^3 \times \mathbb{R}^3$. This has to be investigated.
Moreover, it is still open whether the rearranged addition theorems
(\ref{Gus_RearrOneRangeAddTheorSTF_STF_k}) exists for the whole argument
set $\mathbb{R}^3 \times \mathbb{R}^3$ if the principal quantum number
$N$ of the Slater-type function is a positive integer, $N \in
\mathbb{N}$. Also this remains has to be investigated. In all cases, the
burden of proof lies with Guseinov.

These observations should be particularly worrisome for those who
advocate the use of Slater-type functions with nonintegral principal
quantum numbers as basis functions in molecular electronic structure
calculations. It is generally accepted that a good basis set should
produce highly accurate approximations, but it should also lead to
multicenter integrals that can be computed efficiently and accurately at
tolerable computational costs. These two requirements have so far been
mutually exclusive.

Slater-type functions with integral principal quantum numbers or other
exponentially decaying function sets are well suited to produce highly
accurate approximations in electronic structure calculations, but their
multicenter integrals are notoriously difficult. In contrast, Gaussian
functions are nonphysical. Consequently, large basis sets are needed to
accomplish sufficiently accurate results. However, Gaussian functions
have one, albeit decisive advantage: Their multicenter integrals can be
computed comparatively easily.

If we nevertheless want to use physically better motivated exponentially
decaying basis functions, we should concentrate on those functions that
promise the simplest multicenter integrals, even if we should have to
sacrifice some accuracy. Superficially, Slater-type functions with
nonintegral principal quantum numbers look attractive since they promise
somewhat better results than Slater-type functions with integral
principal quantum numbers. But they achieve this at the cost of
significantly more complicated multicenter integrals. Therefore, we would
first need some fundamental mathematical breakthroughs to make
Slater-type functions with nonintegral principal quantum numbers
competitive with Slater-type functions with integral principal quantum
numbers or with other exponentially decaying functions. As far as I can
judge it, these breakthroughs are not in sight.

There is another, more general remark which I would like to make.
Addition theorems of exponentially decaying functions as well as their
multicenter integrals are fairly complicated mathematical objects.
Accordingly, it is usually very difficult and often even practically
impossible to justify nontrivial manipulations by rigorous mathematical
proofs. This is highly embarrassing. The only way out, which I see, is to
be very conservative and cautious in order to be on the safe side.

Let us assume that $f$ belongs to a suitable Hilbert space.  If we expand
$f (\bm{r} \pm \bm{r}')$ in terms of a function set, which is complete
and orthogonal in that Hilbert space, then the resulting one-range
addition theorem converges in the mean with respect to the norm of the
corresponding Hilbert space.  Moreover, in many cases the Schwarz
inequality (see for example \cite[Eq.\ (6) on p.\ 31]{Davis/1989})
suffices to guarantee that the use of this addition theorem in a
multicenter integral produces a convergent expansion. 

However, we are no longer on the safe side if we replace the complete and
orthogonal functions by other functions that are only complete but not
orthogonal. Formally, this may yield another one-range addition theorem
for $f (\bm{r} \pm \bm{r}')$, but we do not know whether this expansion
makes sense or not. This has to be proved. If we cannot prove this, then
we should rather avoid such a possibly dangerous manipulation. It is not
acceptable to ignore potential problems of that kind and hope for the
best, as it was done by Guseinov.

\typeout{==> Section: Weakly Convergent One-Range Addition Theorems for
  the Coulomb Potential}
\section{Weakly Convergent One-Range Addition Theorems for the Coulomb
  Potential}
\label{Sec:WeakConvAddTheor_CoulombPot}

Guseinov \cite{Guseinov/2005a} derived one-range addition theorems for
the Coulomb potential via addition theorems for the Yukawa potential
$\exp (- \beta r)/r$ \cite{Yukawa/1935}, which can be viewed to be an
exponentially screened Coulomb potential and which is also a special
Slater-type function according to
\begin{equation}
\frac{\mathrm{e}^{-\beta r}}{r} \; = \; (4\pi)^{1/2} \, \beta \, 
\chi_{0, 0}^{0} (\beta, \mathbf{r}) \, .
\end{equation}
Guseinov's idea was to set $N=L=M=0$ in his unsymmetrical one-range
addition theorems for Slater-type functions discussed in Section
\ref{Sec:AddTheor_STF} and to perform the limit $\beta \to 0$.
Unfortunately, things are more complicated and Guseinov's simplistic
approach is fundamentally flawed.

For $k = 0, 1, 2, \dots$, the Yukawa potential belongs to the weighted
Hilbert space $L_{r^k}^{2} (\mathbb{R}^3)$ defined by
(\ref{HilbertL_r^k^2}), but not for $k=-1$ (see for example \cite[p.\
410]{Homeier/Weniger/Steinborn/1992a}). This was apparently overlooked by
Guseinov who used his functions with unspecified $k$. Accordingly, the
addition theorems (\ref{Gus_OneRangeAddTheorSTF_k}) with $N=L=M=0$
converge for $k = 0, 1, 2, \dots$ in the mean with respect to the norm
(\ref{Norm_r^k_2}) of $L_{r^k}^{2} (\mathbb{R}^3)$, but not for $k=-1$.

Guseinov replaced in his one-range addition theorems
(\ref{Gus_OneRangeAddTheorSTF_k}) the overlap integrals
(\ref{Gus_OneRangeAddTheorSTF_k_b}) involving his complete and
orthonormal functions $\prescript{}{k}{\Psi}_{n, \ell}^{m} (\gamma,
\bm{r})$ by overlap integrals of Slater-type functions according to
(\ref{OverlapGusFun_STF}). This step, which yields the addition theorems
(\ref{Gus_OneRangeAddTheorSTF_OvSTF_k}), is mathematically legitimate,
but numerically dubious. Also the next step -- the replacement of the
expansion functions $\prescript{}{k}{\Psi}_{n, \ell}^{m} (\gamma,
\bm{r})$ on the right-hand side of
(\ref{Gus_OneRangeAddTheorSTF_OvSTF_k}) by Slater-type functions
according to (\ref{GusFun2STF}), which yields the addition theorems
(\ref{Gus_OneRangeAddTheorSTF_STF_k}) -- is in the case of the Yukawa
potential at least for $k = 0, 1, 2 \dots$ mathematically justified, but
only as long as the order of the infinite summations in
(\ref{Gus_OneRangeAddTheorSTF_STF_k}) is retained.

Guseinov's final step -- the rearrangement of the infinite summations in
(\ref{Gus_OneRangeAddTheorSTF_STF_k}) which formally produces the
addition theorems (\ref{Gus_RearrOneRangeAddTheorSTF_STF_k}) with
Slater-type functions as expansion functions -- is not legitimate: The
one-center limit of these addition theorems does not exist for $N=L=M=0$:
If we set $\bm{r}' = \bm{0}$ in the addition theorems
(\ref{Gus_OneRangeAddTheorSTF_k}) with $N=L=M=0$, we obtain expansions
that are special cases of the Laguerre expansion (\ref{ExpoPow2GLag}) for
$z^{\mu} \exp (u z)$ with $\mu=-1$. This Laguerre expansion cannot be
reformulated as a power series in $z$ since $\exp (u z)/z$ is not
analytic at $z=0$ in the sense of complex analysis. Accordingly, the
rearranged addition theorem (\ref{Gus_RearrOneRangeAddTheorSTF_STF_k})
with $N=L=M=0$ does not exist for the whole argument set $\mathbb{R}^3
\times \mathbb{R}^3$.

Thus, Guseinov's derivation \cite{Guseinov/2005a} of one-range addition
theorems for the Coulomb potential with Slater-type functions as
expansion functions fails since his starting point -- the unsymmetrical
one-range addition theorems (\ref{Gus_RearrOneRangeAddTheorSTF_STF_k})
with $N=L=M=0$ -- do not exist for all arguments $\bm{r}, \bm{r}' \in
\mathbb{R}^3$.

As a possible remedy, we could expand the Coulomb potential in terms of a
suitable complete and orthonormal set of functions. If we expand $1/
\vert \bm{r} - \bm{r}' \vert$ in terms of Guseinov's functions, we
formally obtain the following symmetrical one-range addition theorems
which are special cases of the general addition theorem
(\ref{SymOneRangeAddTheor_GusFun}):
{\allowdisplaybreaks\begin{subequations}
  \label{CP_OnerangeAddThm_GusFun_k}
  \begin{align}
    \label{CP_OnerangeAddThm_GusFun_k_a}
    & \frac{1}{\vert \bm{r} - \bm{r}' \vert} \; = \; \sum_{\substack{n
        \ell m \\ n' \ell' m'}} \, \prescript{}{k}{\mathbf{\Gamma}}_{n'
      \ell' m'}^{n \ell m} (\beta)
    \notag \\
    & \qquad \qquad \qquad \times \, 
    \prescript{}{k}{\Psi}_{n, \ell}^{m} (\beta, \bm{r}) \, 
    \prescript{}{k}{\Psi}_{n', \ell'}^{m'} (\beta, \bm{r}')
    \, ,
    \\
    \label{CP_OnerangeAddThm_GusFun_k_b}
    & \prescript{}{k}{\mathbf{\Gamma}}_{n' \ell' m'}^{n \ell m} (\beta)
    \notag \\
    & \qquad \; = \; \int \, \bigl[ \prescript{}{k}{\Psi}_{n',
      \ell'}^{m'} (\beta, \bm{r}') \bigr]^{*} \, r'^k \,
    \prescript{}{k}{\mathbf{C}}_{n, \ell}^{m} (\beta, \bm{r}') \,
    \mathrm{d}^3 \bm{r}' \, ,
    \\
    \label{CP_OnerangeAddThm_GusFun_k_c}
    & \prescript{}{k}{\bm{\mathcal{C}}}_{n, \ell}^{m} (\beta, \bm{r}')
    \; = \; \int \, \bigl[ \prescript{}{k}{\Psi}_{n, \ell}^{m} (\beta,
    \bm{r}) \bigr]^{*} \, \frac{r^k}{\vert \bm{r} - \bm{r}' \vert} \,
    \mathrm{d}^3 \bm{r} \, .
  \end{align}
\end{subequations}}%
Analogous symmetrical addition theorems can also be derived for the
Yukawa potential. 

Instead of Guseinov's functions, we could use in
(\ref{CP_OnerangeAddThm_GusFun_k}) or in analogous addition theorems for
the Yukawa potential any other function set that is complete and
orthonormal in an appropriate Hilbert space. An obvious, albeit
Gaussian-type choice would be the eigenfunctions $\Omega_{n, \ell}^{m}
(\beta, \bm{r})$ of the three-dimensional isotropic harmonic oscillator,
which are defined by (\ref{Def_OscillFun}) and which are complete and
orthonormal in $L^{2} (\mathbb{R}^3)$.

In the case of (\ref{CP_OnerangeAddThm_GusFun_k}) there is a principal
problem that was apparently overlooked by Guseinov: The Coulomb potential
does not belong to any of the Hilbert spaces considered in this article
or implicitly used by Guseinov, since they all involve an integration
over the whole $\mathbb{R}^3$. Thus, for $k = -1, 0, 1, 2, \dots$ the
addition theorems (\ref{CP_OnerangeAddThm_GusFun_k}) diverge in the mean
with respect to the norm (\ref{Norm_r^k_2}) of $L_{r^k}^{2}
(\mathbb{R}^3)$.

It is tempting to conclude that all attempts of constructing one-range
addition theorems for $1/ \vert \bm{r} - \bm{r}' \vert$ by expanding it
in terms of function sets, that are complete and orthonormal with respect
to an inner product involving an integration over the whole
$\mathbb{R}^3$, are doomed. However, this conclusion is premature, and
expansions of the type of (\ref{CP_OnerangeAddThm_GusFun_k}) may well be
our best chance of achieving our aim.

As discussed in Section \ref{Sec:HilbertSpace}, one-range addition
theorems are based on Hilbert space theory and thus rely heavily on
concepts from approximation theory. Loosely speaking, we may say that it
is the purpose of approximation theory to provide convenient expressions
-- for example series expansions -- that allow an efficient and reliable
evaluation of a given mathematical object.

There is, however, a very important difference between the use of
one-range addition theorems in multicenter integrals and more
conventional applications of approximation theory: We are not interested
in evaluating the functions that are represented by addition theorems. We
only use these addition theorems to simplify the integrations in
multicenter integral. Addition theorems are only intermediate results
which ultimately produce series expansions for multicenter integrals.

The use of convergent expansions in integrals has the undeniable
advantage that normally only comparatively mild assumptions are needed to
guarantee that integration and summation can be interchanged and that the
resulting expansions converge. Nevertheless, the use of convergent
expansions in integrals is to some extent a luxury and not strictly
necessary. We are free to use a divergent expansion in an integral and
interchange integration and summation if we can guarantee that the
resulting expansion converges to the correct result.

Obviously, such an approach gives us additional possibilities, but it
would be naive to expect a free lunch: It is grossly negligent to use
divergent series in integrals without knowing criteria of manageable
complexity that guarantee the convergence of the resulting expansions.

As discussed in Section \ref{Sec:Intro}, two-range addition theorems of
the type of the Laplace expansion (\ref{LapExp}) converge
\emph{pointwise} since they are essentially rearranged three-dimensional
Taylor series in the Cartesian components of the shift vector. By
relaxing the requirement of pointwise convergence, we arrive at one-range
addition theorems which in general converge \emph{in the mean} or
\emph{strongly} with respect to the norm of a suitable Hilbert space.

As a consequence, one-range addition theorems for functions, which do not
belong to a given Hilbert space, diverge with respect to its
corresponding norm. If we nevertheless insist on extending the formalism
of one-range addition theorems to functions that do not belong to this
Hilbert space, the requirement of \emph{convergence in the mean} or
\emph{strong convergence} is too demanding. We have to replace it by a
\emph{weaker} type of converge that is related to and inspired by the
theory of generalized functions in the sense of Schwartz
\cite{Schwartz/1966a}.

Quantum theory is largely based on functionals, which are mappings from a
function space to the complex numbers. Accordingly, the numerical values
and the local properties of wave functions are -- strictly speaking --
irrelevant, and convergence of functionals, also called \emph{weak
  convergence}, is all we really need. Stronger types of convergence such
as pointwise convergence or convergence in the mean are in principle pure
luxury.

Thus, the one-range addition theorems (\ref{CP_OnerangeAddThm_GusFun_k}),
which diverge for any $k=-1, 0, 1, 2, \dots$ with respect to the norm
(\ref{Norm_r^k_2}) of the Hilbert space $L_{r^k}^{2} (\mathbb{R}^3)$,
should be interpreted as \emph{weakly convergent} expansions that
produces correct results when used in suitable functionals.

In the context of multicenter integrals, weakly convergent expansions
have largely been a \emph{terra incognita}, but there is a detailed
mathematical literature both on weak convergence in general (see for
example \cite[Chapters 10 and 11]{Lax/2002}) as well as on the expansion
of generalized functions or distributions in terms of orthogonal
polynomials (see for example \cite{Walter/1965,Simon/1970b,Duran/1990}
and references therein).

Weakly convergent expansions can be quite useful. In \cite{Weniger/1985}
weakly convergent expansions for $\exp (\mathrm{i} \bm{p} \cdot \bm{r})$
in terms of complete orthonormal and biorthogonal function sets were
constructed. In some cases, these expansions simplify the evaluation of
Fourier transforms, and they can also be used for the construction of
one-range addition theorems (see \cite[Section VII]{Weniger/1985} or
\cite{Homeier/Weniger/Steinborn/1992a}). 

Expansions for the plane wave, that closely resemble those derived in
\cite{Weniger/1985}, were also constructed by Guseinov \cite[Eqs.\ (45) -
(46)]{Guseinov/2003c}. Guseinov, who did not mention the article
\cite{Weniger/1985} in \cite{Guseinov/2003c}, either overlooked or
ignored the obvious fact that that the plane wave does not belong to any
of the Hilbert spaces which he implicitly used. Accordingly, Guseinov's
expansion diverge in the mean and can only converge weakly. Guseinov's
oversight is hard to understand because he had cited \cite{Weniger/1985}
in several other articles
\cite{Guseinov/2002c,Guseinov/2004c,Guseinov/2004d,Guseinov/2005b,%
  Guseinov/2005c,Guseinov/2007g,Guseinov/Mamedov/2001c,%
  Guseinov/Mamedov/2002d,Guseinov/Mamedov/2004e}.

In my opinion, the essential features of weak convergence in contrast to
strong convergence can be explained most easily via the Euclidean vector
space $\mathbb{C}^{\infty}$ of infinite row or column vectors $\bm{u} =
(u_1, u_2, \dots)$ with complex coefficients $u_n$. By equipping the
vector space $\mathbb{C}^{\infty}$ with the inner product $(\bm{u} \vert
\bm{v}) = \sum_{n=1}^{\infty} [u_n]^{*} v_n$, we obtain the corresponding
Hilbert space $\ell^2 \subset \mathbb{C}^{\infty}$ with norm $\Vert
\bm{u} \Vert = \sqrt{(\bm{u} \vert \bm{u})}$. The condition $\Vert \bm{u}
\Vert < \infty$, which defines the Hilbert space $\ell^2$, can only be
satisfied if the coefficients $u_n \in \mathbb{C}$ of $\bm{u}$ decay
sufficiently fast as $n \to \infty$.

Let us now assume that some vector $\bm{w} \in \mathbb{C}^{\infty}$
cannot be normalized, i.e., $\Vert \bm{w} \Vert^2 = (\bm{w} \vert \bm{w})
= \sum_{n=1}^{\infty} \vert w_n \vert^2 < \infty$ does not hold. Thus,
$\bm{w} \notin \ell^2$, but this does not imply that all inner products
$(\bm{w}, \bm{u})$ with normalizable $\bm{u} \in \ell^2$ do not exist. If
the coefficients $u_n$ of $\bm{u}$ decay sufficiently rapidly as $n \to
\infty$, then the infinite series $(\bm{w} \vert \bm{u}) =
\sum_{n=1}^{\infty} [w_n]^{*} u_n$ may well converge to a finite result,
although the series $(\bm{w} \vert \bm{w}) = \sum_{n=1}^{\infty} \vert
w_n \vert^2$ diverges.

It is important to note that $\vert (\bm{w} \vert \bm{u}) \vert < \infty$
cannot hold for arbitrary $\bm{u} \in \ell^2$, but only for a suitably
restricted subspace $\mathcal{W} \subset \ell^2 \subset
\mathbb{C}^{\infty}$ of normalizable vectors such that $\bm{\omega} \in
\mathcal{W}$ implies $\vert (\bm{w} \vert \bm{\omega}) \vert < \infty$.
Accordingly, my interpretation of weak convergence resembles at least
conceptually the theory of rigged Hilbert spaces or Gelfand triplets (see
for example \cite{Boehm/1978,DeLaMadrid/2005} and references therein). A
very readable account of rigged Hilbert spaces from the perspective of
quantum mechanics and their relationship with Dirac's bra and ket
formalism can be found in the book by Ballentine \cite[Chapter
1.4]{Ballentine/1998}.

In principle, it is desirable to specify for a given $\bm{w}$ the whole
set $\mathcal{W}$, but this may be difficult.  In practice, it may well
be sufficient to specify only a sufficiently large subset of
$\mathcal{W}$. Let us for instance assume that the coefficients $w_n$ of
a vector $\bm{w} \notin \ell^2$ are all finite and satisfy $\vert w_n
\vert \sim n^{\beta}$ with $\beta \ge - 1/2$ as $n \to \infty$. Thus,
$\bm{w}$ cannot be normalized.  If, however, the coefficients $\omega_n$
of a vector $\bm{\omega} \in \ell^2$ are all finite and satisfy $\vert
\omega_n \vert \sim n^{\alpha}$ with $\alpha < - \beta - 1$ as $n \to
\infty$, then the infinite series $(\bm{w} \vert \bm{\omega}) =
\sum_{n=1}^{\infty} [w_n]^{*} \omega{}_n$ converges and the inner product
$(\bm{w} \vert \bm{\omega})$ makes sense.

However, the condition $\vert \omega_n \vert \sim n^{\alpha}$ with
$\alpha < - \beta - 1$ as $n \to \infty$ does not suffice to specify the
whole set $\mathcal{W}$ of vectors $\bm{\omega}$ with $\vert (\bm{w}
\vert \bm{\omega}) \vert < \infty$. It can happen that the series
$\sum_{n=1}^{\infty} [w_n]^{*} \omega{}_n$ diverges but is summable. If
summability techniques are included in our arsenal of mathematical
techniques, we could even discard the otherwise essential requirement
$\bm{\omega} \in \ell^2$ and try to make inner products $(\bm{w} \vert
\bm{\omega})$ with both $\bm{w}, \bm{\omega} \notin \ell^2$
mathematically meaningful. A (very) condensed review of the classic
summability methods associated with the names of Ces\`{a}ro, Abel, and
Riesz can be found in Zayed's book \cite[Chapter 1.11.1]{Zayed/1996}.
More detailed treatments of linear summability methods can be found in
specialized monographs such as the books by Boos \cite{Boos/2000}, Hardy
\cite{Hardy/1949}, Knopp \cite{Knopp/1964}, and Powell and Shah
\cite{Powell/Shah/1988}.

An interpretation of one-range addition theorems of the type of
(\ref{CP_OnerangeAddThm_GusFun_k}) as weakly convergent expansions
requires that we first specify the functional, in which addition theorems
are to be used. Then, regularity criteria have to be formulated whose
validity guarantee the convergence of the resulting expansions for this
functional.

The most difficult integrals, which occur in the LCAO-MO approach, are
the notorious two-electron integrals $\mathcal{C} (f, g)$ defined by
(\ref{CouInt_f_g}), whose evaluation can become extremely difficult if
$f$ and $g$ are nonclassical two-center charge distributions of
exponentially decaying functions. 

A theoretical analysis of the integrals $\mathcal{C} (f, g)$ is by no
means easy (see for example \cite[Chapter 9]{Lieb/Loss/1997}). In this
context, it is instructive to replace in (\ref{CouInt_f_g}) the Coulomb
potential by the Yukawa potential, yielding
\begin{equation}
  \label{YukawaInt_f_g}
\mathcal{Y} (f, g; \beta) \; = \; \int \! \int \, \frac
{f^{*} (\bm{r}) \, \mathrm{e}^{- \beta \vert \bm{r} - \bm{r}' \vert} \, 
g (\bm{r}')}{\vert \bm{r} - \bm{r}' \vert} \, 
\mathrm{d}^{3} \bm{r} \, \mathrm{d}^{3} \bm{r}'
\end{equation}
Obviously, we have $\mathcal{C} (f, g) = \lim_{\beta \to 0} \mathcal{Y}
(f, g; \beta)$. Unfortunately, it is not guaranteed that this limiting
process is continuous and produces a finite result. It can be shown with
the help of Fourier transformation (see for example \cite[Section
II]{Weniger/Grotendorst/Steinborn/1986b} and references therein) that
$\mathcal{Y} (f, g; \beta)$ exists for arbitrary densities $f, g \in
L^{2} (\mathbb{R}^3)$, but this does not suffice to guarantee the
existence of $\mathcal{C} (f, g)$. The densities $f$ and $g$ have to
satisfy more sophisticated conditions than square integrability to
guarantee this (see for example \cite[Section 4.3
(Hardy-Littlewood-Sobolev inequality)]{Lieb/Loss/1997} or \cite[Example 3
(Sobolev's inequality) on p.\ 31]{Reed/Simon/1975}).

If we want to apply weakly convergent addition theorems of the type of
(\ref{CP_OnerangeAddThm_GusFun_k}) for the evaluation of two-electron
integrals, we need additional criteria that guarantee the convergence of
the resulting expansions to the value of $\mathcal{C} (f, g)$ for
physically and mathematically reasonable charge densities $f$ and $g$.

If we insert the addition theorems (\ref{CP_OnerangeAddThm_GusFun_k})
into the integral (\ref{CouInt_f_g}), we obtain the following series
expansions: {\allowdisplaybreaks\begin{subequations}
  \label{Expand_C(f,g)}
  \begin{align}
    \label{Expand_C(f,g)_a}
    & \mathcal{C} (f, g) \; = \; \int \! \int \, 
     \frac{f^{*} (\bm{r}) \, g (\bm{r}')}{\vert \bm{r} - \bm{r}' \vert} 
     \, \mathrm{d}^{3} \bm{r} \, \mathrm{d}^{3} \bm{r}'
    \notag \\
    & \qquad \; = \; \sum_{\substack{n \ell m \\ n' \ell' m'}} \,
    \prescript{}{k}{\mathbf{\Gamma}}_{n' \ell' m'}^{n \ell m} (\beta) \,
    \prescript{}{k}{\mathbf{F}}_{n, \ell}^{m} (\beta) \,
    \prescript{}{k}{\mathbf{G}}_{n', \ell'}^{m'} (\beta) \, ,
    \\
    \label{Expand_C(f,g)_b}
    & \prescript{}{k}{\mathbf{F}}_{n, \ell}^{m} (\beta) \; = \; 
    \int \, f^{*} (\bm{r}) \, \prescript{}{k}{\Psi}_{n,
      \ell}^{m} (\beta, \bm{r}) \, \mathrm{d}^{3} \bm{r} \, ,
    \\
    \label{Expand_C(f,g)_c}
    & \prescript{}{k}{\mathbf{G}}_{n', \ell'}^{m'} (\beta)
    \; = \; \int \, g
    (\bm{r}') \, \prescript{}{k}{\Psi}_{n', \ell'}^{m'} (\beta, \bm{r}')
    \, \mathrm{d}^{3} \bm{r}' \, .
  \end{align}
\end{subequations}}
If only a finite number of the integrals $\prescript{}{k}{\mathbf{F}}_{n,
  \ell}^{m} (\beta)$ and $\prescript{}{k}{\mathbf{G}}_{n', \ell'}^{m'}
(\beta)$ are nonzero, then the use of (\ref{CP_OnerangeAddThm_GusFun_k})
in $\mathcal{C} (f, g)$ obviously produces the correct result. 

In general, there will be infinitely many nonzero integrals
$\prescript{}{k}{\mathbf{F}}_{n, \ell}^{m} (\beta)$ and
$\prescript{}{k}{\mathbf{G}}_{n', \ell'}^{m'} (\beta)$. In this case,
asymptotic estimates for the coefficients
$\prescript{}{k}{\mathbf{\Gamma}}_{n' \ell' m'}^{n \ell m} (\beta)$
defined by (\ref{CP_OnerangeAddThm_GusFun_k_b}) have to be constructed
that hold in the case of large indices. Subsequently, asymptotic
conditions on the decay of the integrals $\prescript{}{k}{\mathbf{F}}_{n,
  \ell}^{m} (\beta)$ and $\prescript{}{k}{\mathbf{G}}_{n', \ell'}^{m'}
(\beta)$, respectively, in the case of large indices can be formulated
which guarantee the convergence of the series expansion
(\ref{Expand_C(f,g)}).

It cannot be denied that such a proof of the convergence of the series
expansions (\ref{Expand_C(f,g)}) -- although most likely manageable --
would be highly pedestrian. Nontrivial technical difficulties are also
quite likely. Obviously, a more elegant convergence proof based on
concepts of functional analysis and suitable functions spaces would be
highly desirable. Unfortunately, this is currently not in sight.

There is also another problems. Kato \cite{Kato/1957} showed that the
singularities of the potential of atomic and molecular Hamiltonians
produce discontinuities of the wave functions commonly called Coulomb or
correlation cusps (see also \cite{Bingel/1967}). This implies that atomic
or molecular wave functions do not possess continuous derivatives of
arbitrary order at the locations of the nuclei. Consequently,
exponentially decaying functions, which cannot be differentiated
arbitrarily often at the origin, provide much better approximations than
infinitely differentiable Gaussian-type functions. According to
experience, a few exponentially decaying functions normally suffice to
accurately model the discontinuities of atomic and molecular wave
functions at the nuclei. But there is a price: Exponentially decaying
functions are fairly complicated mathematical objects whose multicenter
integrals are notoriously difficult.

It is quite likely that we also have to pay some price if we try to
analyze the weak convergence of the addition theorems
(\ref{CP_OnerangeAddThm_GusFun_k}) in appropriate functionals such as the
integral (\ref{CouInt_f_g}). As remarked above, weak convergence is
related to Schwartz's theory of generalized functions which applies to
integrals containing a distribution multiplied by a so-called test
function. To make these integrals well defined, the test functions must
possess a sufficient amount of mathematical ``niceness'' in order to
compensate the possibly highly irregular behavior of generalized
functions such as the Dirac delta ``function'' or their derivatives.

In mathematics, it is common to use test functions with highly idealized
properties because this greatly facilitates proofs. A frequently used
test function space is the Schwartz space $\mathcal{S} (\mathbb{R}^3)$ of
rapidly decreasing and infinitely differentiable functions $\psi \colon
\mathbb{R}^3 \to \mathbb{C}$, which satisfy
\begin{equation}
\sup_{\bm{r} \in \mathbb{R}^3} \, \left\vert x^k y^m z^n \, 
\left( \frac{\partial}{\partial x} \right)^u  
\left( \frac{\partial}{\partial y} \right)^v  
\left(\frac{\partial}{\partial z} \right)^w \, 
\psi (\bm{r}) \right\vert \; < \; \infty
\end{equation}
for all $k, m, n, u, v, w \in \mathbb{N}_0$ (see for example \cite[p.\
133]{Reed/Simon/1980}). The oscillator eigenfunctions $\Omega_{n,
  \ell}^{m} (\beta, \bm{r})$ defined by (\ref{Def_OscillFun}) or also
other Gaussian-type functions belong to $\mathcal{S} (\mathbb{R}^3)$, but
none of the exponentially decaying function sets considered in this
article. 

Even more idealized are test function $\psi \colon \mathbb{R}^3 \to
\mathbb{C}$ belonging to the space $\mathcal{D} (\mathbb{R}^3)$
consisting of infinitely differentiable functions with compact support
(see for example \cite[Chapter 6.2]{Lieb/Loss/1997}).

Because of their infinite differentiability, test functions belonging to
$\mathcal{S} (\mathbb{R}^3)$ or $\mathcal{D} (\mathbb{R}^3)$,
respectively, are extremely convenient analytical tools. Their use
greatly simply mathematical proofs. However, infinitely differentiable
functions are not ideally suited for the representation of effective
one-particle wave functions which are discontinuous and also decay
exponentially (see for example \cite{Agmon/1982,Agmon/1985} and
references therein).

The completeness of the oscillator eigenfunctions $\Omega_{n, \ell}^{m}
(\beta, \bm{r})$ in $L^2 (\mathbb{R}^3)$ implies that they can
approximate effective one-particle wave functions in the mean with an in
principle unlimited precision, but one should not expect rapid
convergence. If we instead use physically better motivated exponentially
decaying functions, which are not infinitely differentiable, we can
expect faster convergence, but all mathematical manipulations involving
these functions will be (much) more difficult. Most likely, this also
applies to the analysis of the properties of weakly convergent addition
theorems of the type of (\ref{CP_OnerangeAddThm_GusFun_k}).

Both the index $k$ as well as the scaling parameter $\beta$ of the
expansion functions $\prescript{}{k}{\Psi}_{n, \ell}^{m} (\beta, \bm{r})$
influence the rate of convergence of the infinite series in
(\ref{Expand_C(f,g)_a}). It is an obvious idea to try to maximize
convergence by optimizing both $k$ and $\beta$. Unfortunately, our
current level of understanding does not allow detailed predictions. Most
likely, the ``optimal'' values of $k$ and $\beta$ depend quite strongly
on the properties of the two charge densities $f$ and $g$. It is also
possible that for given $f$ and $g$ only some values of $k$ will lead to
a convergent series expansion (\ref{Expand_C(f,g)_a}).

The use of a nonzero $k$ can at least potentially lead to stability
problems. For $k=0$, the integrals $\prescript{}{k}{\mathbf{F}}_{n,
  \ell}^{m} (\beta)$ and $\prescript{}{k}{\mathbf{G}}_{n', \ell'}^{m'}
(\beta)$ are the coefficients of expansions of $f$ and $g^{*}$ in terms
of Lambda functions, which are according to (\ref{Lambda_OrthoNor})
orthonormal with respect to an integration over the whole $\mathbb{R}^3$.
Since the coefficients of orthogonal expansions satisfy Parseval's
equality (\ref{ParsevalEquality}), the integrals
$\prescript{}{k}{\mathbf{F}}_{n, \ell}^{m} (\beta)$ and
$\prescript{}{k}{\mathbf{G}}_{n', \ell'}^{m'} (\beta)$ with $k=0$ are
bounded in magnitude and vanish for large indices. For $k \neq 0$,
Guseinov's functions $\prescript{}{k}{\Psi}_{n, \ell}^{m} (\beta,
\bm{r})$ are not orthogonal with respect to an integration over the whole
$\mathbb{R}^3$, but satisfy (\ref{Psi_Guseinov_OrthoNor}). Thus,
$\prescript{}{k}{\mathbf{F}}_{n, \ell}^{m} (\beta)$ and
$\prescript{}{k}{\mathbf{G}}_{n', \ell'}^{m'} (\beta)$ with $k \neq 0$
are not the coefficients of orthogonal expansions and do not satisfy
Parseval's equality. Accordingly, it is not guaranteed that they are
bounded in magnitude and that they vanish for large indices.

Let me emphasize that the convergence or divergence of a one-range
addition theorem and the convergence or divergence of the resulting
expansion for a multicenter integral are not directly related. In
\cite[pp.\ 212 - 213]{Guseinov/2005a} Guseinov claimed that the
convergence of his addition theorems for the Coulomb potential can be
demonstrated via the convergence of the resulting series expansions for
three-center integrals of the type of (\ref{ThreeCentNuclAttInt}). This
is of course wrong: Weak convergence does not imply convergence in the
mean, let alone pointwise convergence (see for instance \cite[\S
29]{RieszSzNagy/1990}). Moreover, Guseinov should know that an observed
agreement of different floating point computations up to a certain number
of digits does not necessarily prove anything (see for example
\cite[Table 2]{Barnett/2002}).

But even if the one-range addition theorems
(\ref{CP_OnerangeAddThm_GusFun_k}) or related addition theorems should
lead to divergent series expansions for multicenter integrals, the
resulting expansions can nevertheless be computationally useful. In the
book by Bornemann, Laurie, Wagon, and Waldvogel \cite[p.\
225]{Bornemann/Laurie/Wagon/Waldvogel/2004}, there is the following
instructive remark:
\begin{quote}
  \sl The question whether a series converges is largely irrelevant when
  the reason for using a series is to approximate its sum numerically.
\end{quote}
Thus, even wildly divergent series can be used for computational purposes
if suitable summation methods can be found.

I am aware of several predominantly theoretical articles dealing with the
summation of divergent Laguerre expansions (see for example
\cite{Stempak/2000,Burmistrova/2006} and references therein). However,
from a practical point of view, it is probably more effective to use
purely numerical summation methods. I am also skeptical about the
classical summability methods mentioned above and discussed in books by
Boos \cite{Boos/2000}, Hardy \cite{Hardy/1949}, Knopp \cite{Knopp/1964},
and Powell and Shah \cite{Powell/Shah/1988}. Instead, I suggest to use
nonlinear sequence transformations, which can also be applied in the case
of slowly convergent sequences and series and which often achieve
spectacular improvements of convergence. In that context, it may be of
interest to note that the most recent (third) edition of the book
\emph{Numerical Recipes} \cite{Press/Teukolsky/Vetterling/Flannery/2007}
now also discusses nonlinear sequence transformations (for further
details, see also \cite{Weniger/2007d}).

The best known and most often used sequence transformations are the
so-called Pad\'{e} approximants \cite{Pade/1892} which accomplish an
acceleration of convergence or a summation by converting the partial sums
of a power series to a doubly indexed sequence of rational functions. As
documented by the long list of successful applications in the book by
Baker and Graves-Morris \cite{Baker/Graves-Morris/1996}, Pad\'{e}
approximants have become the standard tool in theoretical physics and in
applied mathematics to overcome problems with slowly convergent or
divergent power series. 

It is, however, not so well known among non-specialists that alternative
sequence transformations as for example Wynn's epsilon and rho algorithm
\cite{Wynn/1956a,Wynn/1956b}, Brezinski's theta algorithm
\cite{Brezinski/1971}, or Levin's transformation \cite{Levin/1973} and
later generalizations (see for example
\cite{Homeier/2000a,Weniger/1989,Weniger/2004} and references therein)
can at least for certain computational problems be much more effective
than Pad\'{e} approximants.

As emphasized already several times, we have no \emph{a priori} reason to
assume that one-range addition theorems necessarily lead to rapidly
convergent expansions for multicenter integrals. Accordingly, our ability
of evaluating infinite series \emph{effectively} and \emph{reliably} is
crucial for the practical usefulness of addition theorems.

The conventional process of adding up the terms of a series successively
is at least in principle able to produce approximations with unlimited
accuracy for a convergent series, but it completely fails in the case of
a divergent series. Moreover, adding up the terms successively is in far
too many cases prohibitively inefficient. Therefore, it makes sense to
try to use sequence transformations for the evaluation of series
expansions for multicenter integrals whenever possible. This is not a new
idea. The oldest article using sequence transformations for the
evaluation of multicenter integrals, which I am aware of, was published
in 1967 by Petersson and McKoy \cite{Petersson/McKoy/1967}.

Personally, I became interested in sequence transformations during my PhD
thesis \cite{Weniger/1982}, in which series expansions for multicenter
integrals played a major role. Since it (too) often happened that my
series expansions converged slowly (see for example \cite[Table
II]{Weniger/Steinborn/1983b}), it was a natural idea to speed up
convergence with the help of sequence transformations. During my PhD
thesis, I only knew linear series transformations as described in the
classic, but now outdated book by Knopp \cite{Knopp/1964}, which turned
out to be ineffective. I was completely ignorant of the more powerful
nonlinear transformations, which often accomplish spectacular
improvements of convergence. Only later, I used nonlinear sequence
transformations with considerable success for the evaluation of
multicenter integrals of exponentially decaying functions
\cite{Grotendorst/Weniger/Steinborn/1986,Homeier/Weniger/1995,%
  Steinborn/Weniger/1990,Weniger/Grotendorst/Steinborn/1986a,%
  Weniger/Steinborn/1988}.

The usefulness of sequence transformations in the context of multicenter
integrals is not limited to the evaluation of infinite series
representations. Particularly noteworthy seems to be Safouhi's approach.
Starting from the Fourier transform (\ref{FT_B_Fun}) of $B$ functions,
Safouhi converts complicated multicenter integrals of $B$ or Slater-type
functions to multidimensional integral representations that have to be
evaluated by numerical quadrature.

At first sight, this does not look like a good idea because the
oscillatory nature of the multidimensional integral representations makes
the straightforward application of conventional quadrature methods
difficult. However, these problems can be overcome by combining
quadrature schemes with suitable nonlinear sequence transformations.
Based on previous work of Sidi \cite{Sidi/1980c} and of Levin and Sidi
\cite{Levin/Sidi/1981}, Safouhi succeeded in developing some
extrapolation techniques specially suited to his needs. This permits a
remarkably efficient and reliable evaluation of complicated multicenter
integrals via oscillatory integral representations (see for example
\cite{Berlu/Safouhi/2004,Duret/Safouhi/2007,Safouhi/2000,Safouhi/2004a,%
  Safouhi/2004b,Safouhi/Bouferguene/2006b,Safouhi/Bouferguene/2007a} and
references therein).

In my opinion, Safouhi's work is a convincing demonstration of the
practical usefulness of extrapolation and convergence acceleration
techniques in electronic structure calculations. Safouhi's work also
shows that there is no reason for despair if existing sequence
transformations turn out to be not powerful enough to solve certain
problems. It may well be possible to construct new transformations that
can do the job.

Sequence transformations can be useful also in completely different
contexts. For example, I have applied sequence transformations
successfully in such diverse fields as the evaluation of special
functions and related objects
\cite{Weniger/1989,Jentschura/Gies/Valluri/Lamm/Weniger/2002,%
  Jentschura/Mohr/Soff/Weniger/1999,Weniger/1990,Weniger/1994a,%
  Weniger/1994b,Weniger/1996d,Weniger/2001,Weniger/2003,%
  Weniger/Cizek/1990,Weniger/Steinborn/1989a}, the summation of strongly
divergent quantum mechanical perturbation expansions
\cite{Weniger/2001,Bender/Weniger/2001,Cizek/Vinette/Weniger/1991,%
  Cizek/Vinette/Weniger/1993a,Cizek/Weniger/Bracken/Spirko/1996,%
  Jentschura/Becher/Weniger/Soff/2000,Jentschura/Weniger/Soff/2000,%
  Weniger/1990,Weniger/1992,Weniger/1994b,Weniger/1996a,Weniger/1996b,%
  Weniger/1996c,Weniger/1996e,Weniger/1997,%
  Weniger/Cizek/Vinette/1991,Weniger/Cizek/Vinette/1993}, the prediction
of unknown series coefficients
\cite{Bender/Weniger/2001,Jentschura/Becher/Weniger/Soff/2000,%
  Jentschura/Weniger/Soff/2000,Weniger/1997,Weniger/2000b}, and the
extrapolation of quantum chemical crystal orbital and cluster electronic
structure calculations for oligomers to their infinite chain limits of
stereoregular \emph{quasi}-onedimensional organic polymers
\cite{Weniger/Liegener/1990,Cioslowski/Weniger/1993,Weniger/Kirtman/2003}.
Many other applications of sequence transformations are listed in my
Habilitation thesis \cite{Weniger/1994b} and in \cite{Weniger/2004}.

These examples should convince even a skeptical reader that sequence
transformations are extremely useful computational tools, and that it is
worth while to invest time and effort to understand their power as well
as their shortcomings. Of course, sequence transformations are no
panacea, and not all my attempts were successful. However, even my
failures often turned out to be fruitful in the long run, since they
frequently provided new insight which ultimately paved the way for the
construction of new convergence acceleration and summation techniques
\cite{Weniger/1989,Weniger/2004,Homeier/Weniger/1995,Weniger/1994a,%
  Weniger/2001,Weniger/2003,Weniger/1992,Weniger/1991,Weniger/2007a}.

In my opinion, it is hard to understand that there are still researchers
like Guseinov who try to evaluate series expansions for multicenter
integrals without the help of sequence transformations. Multicenter
integrals of exponentially decaying functions are notoriously
complicated, and a nice explicit expression for a multicenter integral
does not necessarily permit its efficient and reliable evaluation. It is
also necessary to employ powerful and sophisticated numerical techniques.
The conventional process of adding up the terms of a series successively
until convergence is achieved does not fall into this category.

The currently most complete reference on Pad\'{e} approximants is the
impressive book by Baker and Graves-Morris
\cite{Baker/Graves-Morris/1996}. The most recent monograph on sequence
transformations is the book by Sidi \cite{Sidi/2003}. It contains a
wealth of information and is undeniably useful for specialists, but
presentation and choice of topics makes it basically unsuited for novices
(compare also my book reviews
\cite{Brezinski/Weniger/2004a,Brezinski/Weniger/2004b}). So, the best
choice may well be the book by Brezinski and Redivo Zaglia
\cite{Brezinski/RedivoZaglia/1991a}. Some people also like my long review
\cite{Weniger/1989}. As a first introduction for novices, I can recommend
Appendix A of the book by Bornemann, Laurie, Wagon, and Waldvogel
\cite{Bornemann/Laurie/Wagon/Waldvogel/2004}.

\typeout{==> Section: Differentiation Techniques and the Spherical Tensor
  Gradient Operator}
\section{Differentiation Techniques and the Spherical Tensor Gradient
  Operator}
\label{Sec:DiffTechYlmNabla}

Often, it comparatively easy to obtain explicit analytical expressions
for multicenter integrals over scalar functions, which are irreducible
spherical tensors of rank zero with respect to their local (atomic)
coordinate systems. If, however, the functions occurring in the
multicenter integral are irreducible spherical tensors of higher ranks,
the derivation of explicit expressions can easily become extremely
difficult or even practically impossible.

It is also an empirical fact that it is usually much easier to
differentiate than to integrate (integration is an art, but every fool
and in particular also computer algebra systems like Maple or Mathematica
can differentiate). Accordingly, it is an obvious idea to try to generate
an explicit expression for a difficult multicenter integral over
nonscalar functions by differentiating the simpler multicenter integral
over scalar functions -- preferably the simplest scalar functions -- with
respect to scaling parameters and/or nuclear coordinates (see for example
\cite[Section IV]{Weniger/Steinborn/1983a}). This approach is not
restricted to the notoriously difficult multicenter integrals of
exponentially decaying functions. It is also used in the case of
multicenter integrals of the so-called spherical Gaussian functions (see
for example
\cite{Dunlap/1990,Dunlap/2001,Dunlap/2002,Dunlap/2003,Dunlap/2005} and
references therein).

It is relatively easy to generate multicenter integrals of higher scalar
functions by differentiating the simplest scalar functions with respect
to their scaling parameters. In the case of a $1s$ Slater-type or
Gaussian function, we can construct higher scalar functions easily by
repeatedly using the relationships $\partial \exp (-\alpha r)/\partial
\alpha = -r \exp (-\alpha r)$ or $\partial \exp (-\alpha r^2)/\partial
\alpha = -r^2 \exp (-\alpha r^2)$. 

The generation of anisotropic functions, which are irreducible spherical
tensors of rank $\ell$, from scalar functions is less straightforward,
but also here an highly developed mathematical technology based on the
differential operator $\mathcal{Y}_{\ell}^{m} (\nabla)$ is available (see
for example \cite{Weniger/2005} and \cite[Section 3]{Martin/2006} and
references therein).

In \cite[Section 3]{Guseinov/2005a}, Guseinov used differentiation
techniques for the derivation of more complicated addition theorems by
differentiating his one-range addition theorems for $1/\vert \bm{r} -
\bm{r}' \vert$ with respect to the Cartesian components of either $\bm{r}
= (x, y, z)$ or $\bm{r}' = (x', y', z')$. Guseinov pursued essentially
analogous approaches in his articles
\cite{Guseinov/2004a,Guseinov/2004b,Guseinov/2004c,Guseinov/2004d,%
  Guseinov/2004e,Guseinov/2004k,Guseinov/2005c,Guseinov/2005d,%
  Guseinov/2005e,Guseinov/2005g,Guseinov/2006a,Guseinov/Mamedov/2004d,%
  Guseinov/Mamedov/2004e,Guseinov/Mamedov/2005c,Guseinov/Mamedov/2005d}.

It is not a new idea to generate more complicated addition theorems of
anisotropic functions by differentiating simpler addition theorems of
scalar functions. In \cite[Sections IV and VI]{Weniger/Steinborn/1985},
it was shown that the two-range addition theorems of the irregular solid
harmonic and the modified Helmholtz harmonic, respectively, can be
derived by differentiating the two-range addition theorems for the
Coulomb potential and the Yukawa potential, respectively. Similar ideas
were pursued in \cite{Weniger/Steinborn/1989b}.

Unfortunately, this does not imply that Guseinov's approach is
computationally efficient or at least mathematically sound. This Section
first describes how irreducible spherical tensors can be differentiated
comparatively easily with respect to the Cartesian components of their
argument vectors. The second topic is less technical and refers to more
fundamental mathematical problems: In
\cite{Guseinov/2004a,Guseinov/2004b,Guseinov/2004c,Guseinov/2004d,%
  Guseinov/2004e,Guseinov/2004k,Guseinov/2005a,Guseinov/2005c,%
  Guseinov/2005d,Guseinov/2005e,Guseinov/2005g,Guseinov/2006a,%
  Guseinov/Mamedov/2004d,Guseinov/Mamedov/2004e,Guseinov/Mamedov/2005c,%
  Guseinov/Mamedov/2005d}, Guseinov consistently ignored all questions of
convergence and existence of his addition theorems obtained by
differentiating simpler addition theorems. 

This article considers exclusively addition theorems of irreducible
spherical tensors of the type of (\ref{Def_IrrSphericalTensor}).
Moreover, all functions occurring in these addition theorems are also
irreducible spherical tensors of a given rank $\ell \ge 0$.  Irreducible
spherical tensors $F_{\ell}^{m} (\bm{r} \pm \bm{r}')$, that are of
interest in the context of electronic structure calculations, can all be
differentiated at least a finite number of times with respect to the
Cartesian components of either $\bm{r}$ or $\bm{r}'$. Because of the
convenient orthonormality properties of the spherical harmonics, it is
highly desirable to express the angular part of Cartesian
differentiations in terms of spherical harmonics. The necessary algebra
can be done, but in particular for large angular momentum quantum numbers
of the irreducible spherical tensors and for large orders of the
differential operators, we would be confronted with messy expressions and
nontrivial technical problems. There is also the danger that delta
function contributions, which can occur in multicenter integrals and
addition theorems (see for example the articles by Pitzer Kern, and
Liscomb \cite{Pitzer/Kern/Liscomb/1962} and by Kay, Todd, and Silverstone
\cite{Kay/Todd/Silverstone/1969a} or the book by Judd \cite[Chapter
5.3]{Judd/1975}), are easily overlooked if we differentiate with respect
to Cartesian components (see also
\cite{Blinder/2003,Hu/2004,Gsponer/2007a,Gsponer/2007b}).

A much more convenient approach is possible that completely avoids all
differentiations of irreducible spherical tensors $F_{\ell}^{m} (\bm{r}
\pm \bm{r}')$ with respect to Cartesian components of the argument
vectors and only requires differentiations with respect to the radial
variables $r = \vert \bm{r} \vert$ and $r' = \vert \bm{r}' \vert$,
respectively. This is accomplished by reformulating the original
differential operators with unspecified transformation properties as
finite sums of differential operators that are irreducible spherical
tensors of a given integral rank.

Let us assume that $\mathcal{P}_n (\bm{r})$ is a polynomial of degree $n$
in the Cartesian components of $\bm{r} = (x, y, z)$:
\begin{equation}
  \label{Def:P_n(x,y,z)}
\mathcal{P}_n (\bm{r}) \; = \; \sum_{u, v, w \ge 0}^{u+v+w \leq n} \,
\mathcal{C}_{u v w}^{(n)} \, x^u \, y^v \, z^w \, .
\end{equation}
The regular solid harmonic ${\mathcal{Y}}_{\ell}^{m} (\bm{r})$ is
according to (\ref{YlmHomPol}) a homogeneous polynomial of degree $\ell$
in the Cartesian components of $\bm{r} = (x, y, z)$. Thus, the
completeness and the orthonormality of the surface spherical harmonic
$Y_{\ell}^{m} (\theta, \phi)$ with respect to an integration over the
unit sphere in $\mathbb{R}^3$ implies that the polynomial $\mathcal{P}_n
(\bm{r})$ can be expressed as a finite sum of solid harmonics in $\bm{r}$
multiplied by even powers of $r = \vert \bm{r} \vert$ (see also \cite[\S
96 on pp.\ 147 - 148]{Hobson/1965}):
\begin{equation}
  \label{ExpandYlm_P_n(x,y,z)}
\mathcal{P}_n (\bm{r}) \; = \; 
\sum_{\nu, \lambda \ge 0}^{2\nu+\lambda \leq n} \, 
\, \sum_{\mu=-\lambda}^{\lambda} \, 
\mathbf{C}_{\nu \lambda \mu}^{(n)} \, r^{2\nu} \, 
{\mathcal{Y}}_{\lambda}^{\mu} (\bm{r}) \, .
\end{equation}
This is a relationship among polynomials in the Cartesian components of
an essentially arbitrary three-dimensional vector $\bm{r}$. Thus,
(\ref{ExpandYlm_P_n(x,y,z)}) also holds if we replace $\bm{r}$ by $\nabla
= (\partial/\partial x, \partial/\partial y, \partial/\partial z)$, and
any differential operator, which is polynomial of degree $n$ in the
Cartesian components of $\nabla$, can be expressed as a finite sum of
products of integral powers of the Laplacian $\nabla^2 =
\partial^2/\partial x^2 + \partial^2/\partial y^2 + \partial^2/\partial
z^2$ multiplied by a solid harmonic in $\nabla$:
\begin{align}
  \label{ExpandYlm_P_n(Nabla)}
  \mathcal{P}_n (\nabla) & \; = \;
  \sum_{u, v, w \ge 0}^{u+v+w \leq n} \, 
   \mathcal{C}_{u v w}^{(n)} \, 
   \left( \frac{\partial}{\partial x} \right)^u \, 
   \left( \frac{\partial}{\partial y} \right)^v \, 
   \left(\frac{\partial}{\partial z} \right)^w 
  \notag \\
  & \; = \; \sum_{\nu, \lambda \ge 0}^{2\nu+\lambda \leq n} \, 
  \sum_{\mu=-\lambda}^{\lambda} \, \mathbf{C}_{\nu \lambda \mu}^{(n)} \, 
  \nabla^{2\nu} \, {\mathcal{Y}}_{\lambda}^{\mu} (\nabla) \, .
\end{align}

We obtain an explicit expression for the spherical tensor gradient
operator $\mathcal{Y}_{\ell}^{m} (\nabla)$ by replacing in
(\ref{YlmHomPol}) the Cartesian components of $\bm{r}$ by those of
$\nabla$:
\begin{align}
  \label{Def:YlmNabla}
  & \mathcal{Y}_{\ell}^{m} (\nabla) \; = \; \left[
    \frac{2\ell+1}{4\pi} (\ell+m)!(\ell-m)! \right]^{1/2}
  \notag \\[1.5\jot]
  & \quad \times \, \sum_{k \ge 0} \, \frac
  {\left(-\frac{\partial}{\partial x} -
      \mathrm{i}\frac{\partial}{\partial y}\right)^{m+k} \,
    \left(\frac{\partial}{\partial x} -
      \mathrm{i}\frac{\partial}{\partial y}\right)^{k} \, \left(
      \frac{\partial}{\partial z} \right)^{\ell-m-2k}}
      {2^{m+2k} (m+k)! k! (\ell-m-2k)!} \, .
\end{align}
In Martin's book \cite[p.\ 62]{Martin/2006}, $\mathcal{Y}_{\ell}^{m}
(\nabla)$ is called the \emph{Erd\'{e}lyi operator}. I prefer instead the
name \emph{spherical tensor gradient operator} because the expressions
for products $\mathcal{Y}_{\ell_1}^{m_1} (\nabla) F_{\ell_2}^{m_2}
(\bm{r})$, that will be mentioned later, generalize -- as emphasized by
Bayman \cite[p.\ 2558]{Bayman/1978} -- the well known gradient formula in
angular momentum theory (see for example \cite[Chapters 5.7 and
5.9]{Edmonds/1974} or \cite[Chapter II.11]{Rose/1955}).

In Martin's book \cite[Section 3]{Martin/2006}, the mathematical theory
of $\mathcal{Y}_{\ell}^{m} (\nabla)$ as well as numerous applications
predominantly in classical physics are discussed. In \cite{Weniger/2005},
one also finds a detailed discussion of the mathematical properties of
$\mathcal{Y}_{\ell}^{m} (\nabla)$ as well as numerous applications, but
this time emphasis is on electronic structure theory.

A new differential operator is not necessarily a useful thing, let alone
a major achievement. As remarked above, differentiating an irreducible
spherical tensor $F_{\ell}^{m} (\bm{r})$ of the type of
(\ref{Def_IrrSphericalTensor}) with respect to the Cartesian components
of $\bm{r}$ produces messy expressions. So, if we look at
(\ref{ExpandYlm_P_n(Nabla)}) and take into account (\ref{Def:YlmNabla}),
we may get the impression that we replaced something complicated -- the
left-hand side of (\ref{ExpandYlm_P_n(Nabla)}) -- by something even more
complicated.

We thus arrive at the paradoxical conclusion that the differential
operators $\mathcal{P}_n (\nabla)$ and ${\mathcal{Y}}_{\ell}^{m}
(\nabla)$ are practically useful only if it is not necessary to
differentiate an irreducible spherical tensor $F_{\ell}^{m} (\bm{r})$
with respect to the Cartesian components of $\bm{r}$ via the defining
explicit expression (\ref{ExpandYlm_P_n(Nabla)}) and
(\ref{Def:YlmNabla}). Fortunately, this is the case. Differentiations
with respect to Cartesian components can be avoided completely. The
Laplacian $\nabla^2$ is an irreducible spherical tensor of rank zero, and
${\mathcal{Y}}_{\ell}^{m} (\nabla)$ is just like the corresponding
regular solid harmonic ${\mathcal{Y}}_{\ell}^{m} (\bm{r})$ an irreducible
spherical tensor of rank $\ell$ (a formal proof can be found in \cite[p.\
312]{Biedenharn/Louck/1981a}). Consequently, we can hope for substantial
computational and technical benefits if products involving
${\mathcal{Y}}_{\ell}^{m} (\nabla)$ and other irreducible spherical
tensors are handled via the powerful machinery of angular momentum
coupling.

The first article on the differential operator ${\mathcal{Y}}_{\ell}^{m}
(\nabla)$, which I am aware of, is due to Hobson who derived in 1892 a
very consequential theorem on the differentiation of functions $f \colon
\mathbb{R}^n \to \mathbb{C}$ \cite[p.\ 67]{Hobson/1892}. This theorem is
also discussed in Hobson's book \cite[pp.\ 124 - 129]{Hobson/1965}, in
Martin's book \cite[Section 3.5]{Martin/2006}, and in \cite[Section
3]{Weniger/2005}.

With the help of Hobson's theorem, it is comparatively easy to obtain
explicit expressions for the product of $\mathcal{Y}_{\ell}^{m} (\nabla)$
and a radially symmetric function $\varphi \colon \mathbb{R}^{3} \to
\mathbb{}C$ that only depends on $r = \vert \bm{r} \vert$ (see for
example \cite[Eq.\ (3.4)]{Weniger/2005}):
\begin{equation}
  \label{YlmNabla_phi}
\mathcal{Y}_{\ell}^{m} (\nabla) \, \varphi (r) \; = \; \biggl[
\left( \frac{1}{r} \, \frac{\mathrm{d}}{\mathrm{d} r} \right)^{\ell}
\, \varphi (r) \biggr] \, {\mathcal{Y}}_{\ell}^{m} (\bm{r}) \, .
\end{equation}
Thus, ${\mathcal{Y}}_{\ell}^{m} (\nabla)$ is a generating differential
operator that transforms an irreducible spherical tensor of rank zero to
an irreducible spherical tensor of rank $\ell$.

With the help of Hobson's theorem it is also possible to construct
explicit expressions for products ${\mathcal{Y}}_{\ell_1}^{m_1} (\nabla)
F_{\ell_2}^{m_2} (\bm{r})$. For that purpose, let us assume that the
irreducible spherical tensor $F_{\ell_2}^{m_2} (\bm{r})$ satisfies a
relation of the kind of (\ref{YlmNabla_phi}), i.e., it can be generated
by applying ${\mathcal{Y}}_{\ell_2}^{m_2} (\nabla)$ to a suitable scalar
function $\Phi_{\ell_2} (r)$:
\begin{equation}
  \label{Def_Phi_l2m2}
F_{\ell_2}^{m_2} (\bm{r}) \; = \;
{\mathcal{Y}}_{\ell_2}^{m_2} (\nabla) \, \Phi_{\ell_2} (r) \, .
\end{equation}
In view of ${\mathcal{Y}}_{\ell_1}^{m_1} (\nabla) F_{\ell_2}^{m_2}
(\bm{r}) = {\mathcal{Y}}_{\ell_1}^{m_1} (\nabla)
{\mathcal{Y}}_{\ell_2}^{m_2} (\nabla) \, \Phi_{\ell_2} (r)$, we need an
explicit expression of manageable complexity for the product
${\mathcal{Y}}_{\ell_1}^{m_1} (\nabla) {\mathcal{Y}}_{\ell_2}^{m_2}
(\nabla)$. This can be accomplished easily. Since Gaunt coefficients
defined by (\ref{Def_Gaunt}) linearize the product of two spherical
harmonics, multiplication of (\ref{Ylm_lin}) by $r^{\ell_1+\ell_2}$
yields the linearization formula for the product of the regular solid
harmonics $\mathcal{Y}_{\ell_1}^{m_1} (\bm{r})$ and ${Y}_{\ell_2}^{m_2}
(\bm{r})$ (see for example \cite[Eq.\ (3.6)]{Weniger/2005}). In this
linearization formula, we only have to replace the Cartesian components
of $\bm{r}$ by those of $\nabla$ and obtain (see for example \cite[Eq.\
(3.7)]{Weniger/2005}):
\begin{align}
  \label{YlmNabla_lin}
  \mathcal{Y}_{\ell_1}^{m_1} (\nabla) \,
  \mathcal{Y}_{\ell_2}^{m_2} (\nabla) & \; = \;
  \sum_{\ell=\ell_{\mathrm{min}}}^{\ell_{\mathrm{max}}} \! {}^{(2)}
  \, \langle \ell m_1+m_2 \vert \ell_1 m_1 \vert \ell_2 m_2 \rangle)
  \notag \\
  & \qquad \quad \times \,
  \nabla^{2 \Delta \ell} \, \mathcal{Y}_{\ell}^{m_1+m_2}
  (\nabla) \, .
\end{align}
The abbreviation $\Delta \ell$, which is either a positive integer or
zero, is defined by (\ref{Def_Del_l}).

By combining (\ref{YlmNabla_phi}), (\ref{Def_Phi_l2m2}), and
(\ref{YlmNabla_lin}) we obtain (see for example \cite[Eq.\
(3.8)]{Weniger/2005}):
\begin{align}
  \label{YlmNabla_PHIlm}
  & {\mathcal{Y}}_{\ell_1}^{m_1} (\nabla) \, F_{\ell_2}^{m_2}
  (\bm{r}) \; = \;
  \sum_{\ell=\ell_{\mathrm{min}}}^{\ell_{\mathrm{max}}} \! {}^{(2)}
  \, \langle \ell m_1+m_2 \vert \ell_1 m_1 \vert \ell_2 m_2 \rangle
  \notag \\
  & \qquad \times \, \nabla^{2 \Delta \ell} \, \biggl[ \left(
    \frac{1}{r} \, \frac{\mathrm{d}}{\mathrm{d} r} \right)^{\ell} \,
  \Phi_{\ell_2} (r) \biggr] \, {\mathcal{Y}}_{\ell}^{m_1+m_2} (\bm{r})
  \, .
\end{align}

In principle, (\ref{YlmNabla_PHIlm}) should suffice for our purposes
since the scalar function $\Phi_{\ell_2} (r)$ in (\ref{Def_Phi_l2m2}) can
according to (\ref{YlmNabla_phi}) be obtained from the scalar function
$f_{\ell_2} (r)$ in (\ref{Def_IrrSphericalTensor}) by repeated
integration with respect to $r$. However, repeated integrations can at
least potentially lead to nontrivial technical problems. Alternative
expressions for the product ${\mathcal{Y}}_{\ell_1}^{m_1} (\nabla)
F_{\ell_2}^{m_2} (\bm{r})$ are thus desirable.

By systematically exploiting the tensorial nature of the spherical tensor
gradient operator, the product ${\mathcal{Y}}_{\ell_1}^{m_1} (\nabla)
F_{\ell_2}^{m_2} (\bm{r})$ can be expressed as a finite linear
combination of Gaunt coefficients, radial functions $\gamma _{\ell_1
  \ell_2}^{\ell} (r)$, and spherical harmonics \cite[Eq.\
(4.7)]{Weniger/Steinborn/1983c}:
\begin{align}
  \label{YlmNab2Flm_GenStruc}
  & {\mathcal{Y}}_{\ell_1}^{m_1} (\nabla) \, F_{\ell_2}^{m_2} (\bm{r})
  \notag \\
  & \qquad \; = \; 
  \sum_{\ell=\ell_{\mathrm{min}}}^{\ell_{\mathrm{max}}} \! {}^{(2)} \, 
  \langle \ell m_1+m_2 \vert \ell_1 m_1 \vert \ell_2 m_2 \rangle
  \notag \\
  & \qquad \qquad \times \, \gamma _{\ell_1 \ell_2}^{\ell} (r) \,
  Y_{\ell}^{m_1+m_2} (\bm{r}/r) \, .
\end{align}
The functions $\gamma _{\ell_1 \ell_2}^{\ell} (r)$ in
(\ref{YlmNab2Flm_GenStruc}) can be obtained by differentiating the radial
part $f_{\ell_2} (r)$ of the spherical tensor $F_{\ell_2}^{m_2} (\bm{r})$
with respect to $r = \vert \bm{r} \vert$ (see \cite[Eqs.\ (3.29), (4.15)
- (4.18), and (4.24)]{Weniger/Steinborn/1983c} or \cite[Eqs.\ (4.11) -
(4.16)]{Weniger/2005}): 
{\allowdisplaybreaks\begin{align}
  \label{f2gamma_1}
  & \gamma _{\ell_1 \ell_2}^{\ell} (r) \; = \; \sum_{q=0}^{\Delta \ell}
  \, \frac{(-\Delta \ell)_q (-\sigma (\ell)-1/2)_q} {q!} \, 2^q \,
  r^{\ell_1 + \ell_2 - 2 q}
  \notag \\
  & \qquad \quad \times \, \left( \frac{1}{r}
    \frac{\mathrm{d}}{\mathrm{d} r} \right)^{\ell_1 - q} \,
  \frac{f_{\ell_2} (r)}{r^{\ell_2}}
  \\[1.5\jot]
  \label{f2gamma_2}
  & \quad \; = \; r^{-\ell-1} \, \left( \frac{1}{r}
    \frac{\mathrm{d}}{\mathrm{d} r} \right)^{\Delta \ell} \,
  r^{\ell_1+\ell_2+\ell+1} \, \left( \frac{1}{r}
    \frac{\mathrm{d}}{\mathrm{d} r} \right)^{\Delta \ell_2} \,
  \frac{f_{\ell_2} (r)}{r^{\ell_2}}
  \\[1.8\jot]
  \label{f2gamma_3}
  & \quad \; = \; r^{\ell} \, \left( \frac{1}{r}
    \frac{\mathrm{d}}{\mathrm{d} r} \right)^{\Delta \ell_2} \,
  r^{\ell_1-\ell_2-\ell-1} \, \left( \frac{1}{r}
    \frac{\mathrm{d}}{\mathrm{d} r} \right)^{\Delta \ell} \,
  r^{\ell_2+1} \, f_{\ell_2} (r)
  \\[1.9\jot]
  \label{f2gamma_4}
  & \quad \; = \; r^{-\ell-1} \, \left( \frac{1}{r}
    \frac{\mathrm{d}}{\mathrm{d} r} \right)^{\Delta \ell_2} \,
  r^{\ell_1-\ell_2+3\ell+1} \, \left( \frac{1}{r}
    \frac{\mathrm{d}}{\mathrm{d} r} \right)^{\Delta
    \ell_2} \notag \\[1.5\jot]
  & \qquad \quad \times \, r^{-2\ell-1} \, \left( \frac{1}{r}
    \frac{\mathrm{d}}{\mathrm{d} r} \right)^{\ell_2-\ell} r^{\ell_2+1}
  \, f_{\ell_2} (r)
  \\[1.5\jot]
  \label{f2gamma_5}
  & \quad \; = \; r^{\ell} \, \left( \frac{1}{r}
    \frac{\mathrm{d}}{\mathrm{d} r} \right)^{\Delta \ell} \,
  r^{\ell_1+\ell_2-3\ell-1} \, \left( \frac{1}{r}
    \frac{\mathrm{d}}{\mathrm{d} r} \right)^{\Delta
    \ell} \notag \\[1.5\jot]
  & \qquad \quad \times \, r^{2\ell+1} \, \left( \frac{1}{r}
    \frac{\mathrm{d}}{\mathrm{d} r} \right)^{\ell-\ell_2}
  \frac{f_{\ell_2} (r)}{r^{\ell_2}}
  \\[1.5\jot]
  \label{f2gamma_6}
  & \quad \; = \; \sum_{s=0}^{\Delta \ell_2} \, \frac{(-\Delta
    \ell_2)_s (\Delta \ell_1 + 1/2)_s}{s!} \,
  2^s \, r^{\ell_1 - \ell_2 - 2 s - 1}\notag \\[1.5\jot]
  & \qquad \quad \times \, \left( \frac{1}{r}
    \frac{\mathrm{d}}{\mathrm{d} r} \right)^{\ell_1 - s} \, r^{\ell_2 +
    1} \, f_{\ell_2} (r) \, .
\end{align}}%
The abbreviations $\Delta l$, $\Delta l_1$, $\Delta l_2$, and $\sigma
(\ell)$ are defined by (\ref{Def_Del_l}) - (\ref{Def_sigma_l}).

Other expressions for the product ${\mathcal{Y}}_{\ell_1}^{m_1} (\nabla)
F_{\ell_2}^{m_2} (\bm{r})$ can be found in articles by Bayman
\cite{Bayman/1978}, Santos \cite{Santos/1973}, Stuart \cite{Stuart/1981},
Niukkanen \cite{Niukkanen/1983a}, and Rashid \cite{Rashid/1986}.

There are some radially symmetric functions of considerable relevance in
electronic structure theory that lead to remarkably simple expressions if
${\mathcal{Y}}_{\ell}^{m} (\nabla)$ is applied to them via
(\ref{YlmNabla_phi}). The classic example is the Coulomb potential $1/r$.
Hobson \cite{Hobson/1892} showed that the irregular solid harmonic
$\mathcal{Z}_{\ell}^{m} (\bm{r}) = r^{-\ell-1} Y_{\ell}^{m} (\theta,
\phi)$ is generated by applying ${\mathcal{Y}}_{\ell}^{m} (\nabla)$ to
$1/r$ (further details can be found in Hobson's book \cite[pp.\ 124 -
129]{Hobson/1965}). In modern notation, Hobson's result can be expressed
as follows (see for example \cite[Eq.\ (4.16)]{Weniger/Steinborn/1983a}):
\begin{equation}
   \label{YllNabla_Coul}
\mathcal{Z}_{\ell}^{m} (\bm{r}) \; = \;
\frac{(-1)^{\ell}}{(2 \ell -  1)!!}
\, \mathcal{Y}_{\ell}^{m} (\nabla) \, \frac{1}{r} \, .
\end{equation} 
Thus, (\ref{ExpandYlm_P_n(Nabla)}) and (\ref{YllNabla_Coul}) imply that
Cartesian derivatives of the Coulomb potential as considered by Guseinov
in \cite[Section 3]{Guseinov/2005a} or by Guseinov and Mamedov
\cite{Guseinov/Mamedov/2005c} can be expressed by linear combinations of
irregular solid harmonics $\mathcal{Z}_{\ell}^{m}$ multiplied by integral
powers of the Laplacian $\nabla^2$. If a nonzero power of the Laplacian
acts on an irregular solid harmonic, we obtain (see for example
\cite[Eq.\ (29)]{Rowe/1978}):
\begin{equation}
  \label{PoissonEq_Zlm}
\nabla^2 \, \mathcal{Z}_{\ell}^{m} (\bm{r}) \; = \;
- 4 \pi \, \delta_{\ell}^{m} (\bm{r}) \, .
\end{equation}
Here, $\delta_{\ell}^{m}$ is the so-called spherical delta function (see
for example \cite[Eq.\ (30)]{Rowe/1978}):
\begin{equation}
  \label{Def_delta_lm}
\delta_{\ell}^{m} (\bm{r}) \; = \;
\frac {(-1)^{\ell}} {(2\ell - 1)!!} \,
\mathcal{Y}_{\ell}^{m} (\nabla) \, \delta (\bm{r}) \, .
\end{equation}
If we set in (\ref{PoissonEq_Zlm}) $\ell=0$, we obtain the well known
Poisson equation of a unit point charge:
\begin{equation}
  \label{PoissonEq_CP}
\nabla^2 \, \frac{1}{r} \; = \; - 4\pi \, \delta (\bm{r}) \, .
\end{equation}
Thus, the spherical delta function $\delta_{\ell}^{m}$ can be viewed as a
generalized solution of the Poisson equation of a unit multipole charge.

In \cite{Guseinov/2004b,Guseinov/2004c,Guseinov/2004d,%
  Guseinov/2004e,Guseinov/2004k,Guseinov/2005c,Guseinov/2005d,%
  Guseinov/2005e,Guseinov/2005g,Guseinov/2006a,Guseinov/Mamedov/2004d,%
  Guseinov/Mamedov/2004e,Guseinov/Mamedov/2005d}, Guseinov considered
Cartesian derivatives of the Yukawa potential and more generally of
Slater-type functions. Also in these cases, is is advantageous to rewrite
differentiations with respect to Cartesian components in tensorial form
according to (\ref{ExpandYlm_P_n(Nabla)}).

In this context, it is highly advantageous that Slater-type functions
with integral principal quantum numbers can according to
(\ref{STF->Bfun}) be expressed as finite linear combinations of $B$
functions. The reason is that it is remarkably easy to apply the
spherical tensor gradient operator to $B$ functions. For example,
application of ${\mathcal{Y}}_{\ell}^{m} (\nabla)$ to a scalar $B$
function simply produces a nonscalar $B$ function \cite[Eq.\
(4.12)]{Weniger/Steinborn/1983a}:
\begin{equation}
  \label{STGO_Bn00}
\mathcal{Y}_{\ell}^{m} (\nabla) \, B_{n+\ell,0}^{0} (\alpha, \bm{r}) 
\; = \; \frac{(-\alpha)^{\ell}}{(4\pi)^{1/2}}  \,
B_{n,\ell}^{m} (\alpha, \bm{r}) \, .
\end{equation}
It is also fairly easy to apply the spherical tensor gradient operator to
a nonscalar $B$ function  \cite[Eq.\
(6.25)]{Weniger/Steinborn/1983c}:
\begin{align}
  \label{STGO_Bnlm}
  & \mathcal{Y}_{\ell_1}^{m_1} (\nabla) \, B_{n_2,\ell_2}^{m_2} (\alpha,
  \bm{r})
  \notag \\
  & \qquad \; = \; (-\alpha)^{\ell_1} \,
  \sum_{\ell=\ell_{\mathrm{min}}}^{\ell_{\mathrm{max}}} \! {}^{(2)} \,
  \langle \ell m_1+m_2 \vert \ell_1 m_1 \vert \ell_2 m_2 \rangle
  \notag \\
  & \qquad \qquad \times \, \sum_{t=0}^{\Delta \ell} \, (-1)^t \,
  {\binom{\Delta \ell} {t}} \, B_{n_2+\ell_2-\ell-t,\ell}^{m_1+m_2}
  (\alpha, \bm{r}) \, .
\end{align}

If we set $n=\ell=m=0$ in (\ref{STF->Bfun}), we see that the Yukawa
potential is also a special $B$ function:
\begin{equation}
  \label{YukawaPot_Bfun}
\frac{\mathrm{e}^{-\beta r}}{r} \; = \;
\beta \, \hat{k}_{-1/2} (\beta r) \; = \;
(4\pi)^{1/2} \, \beta \,  B_{0, 0}^{0} (\beta, \bm{r}) \, .
\end{equation}
Thus, (\ref{STGO_Bn00}) implies that the application of the spherical
tensor gradient operator to the Yukawa potential yields the so-called
modified Helmholtz harmonic \cite[Eq.\ (6.9)]{Weniger/Steinborn/1985}:
\begin{equation}
B_{-\ell, \ell}^{m} (\beta, \bm{r}) \; = \;
(4\pi)^{1/2} \, (-\beta)^{-\ell} \, \mathcal{Y}_{\ell}^{m} (\nabla) \,
B_{0, 0}^{0} (\beta, \bm{r}) \, .
\end{equation}
In the limit of vanishing screening ($\beta \to 0$), the modified
Helmholtz harmonic approaches an irregular solid harmonic according to
\cite[Eq.\ (3.10)]{Weniger/Grotendorst/Steinborn/1986b}
\begin{equation}
\mathcal{Z}_{\ell}^{m} (\bm{r}) \; = \; [(2\ell-1)!!]^{-1} \,
\lim_{\beta \to 0} \,
\bigl[ \beta^{\ell+1} B_{-\ell, \ell}^{m} (\beta, \bm{}r) \bigr] \, .
\end{equation}
Moreover, the modified Helmholtz harmonic satisfies the following
exponentially screened variant of the anisotropic Poisson equation
(\ref{PoissonEq_Zlm}) \cite[Eq.\ (6.18)]{Weniger/2005}:
\begin{align}
  & \bigl[ 1 - \beta^{-2} \nabla^2 \bigr] \, B_{-\ell, \ell}^{m}
  (\beta, \bm{r})
  \notag \\
  & \qquad \; = \; (-1)^{\ell} \, \frac{4\pi}{\beta^{\ell+3}} \,
  \mathcal{Y}_{\ell}^{m} (\nabla) \, \delta (\bm{r})
  \notag \\
  & \qquad \; = \;
  \frac{4\pi}{\beta^{\ell+3}} \, (2\ell-1)!! \,
  \delta_{\ell}^{m} (\bm{r}) \, .
\end{align}
In view of this relationship and also because of \cite[Eq.\
(5.6)]{Weniger/Steinborn/1983c}
\begin{equation}
  \label{Shift_n_Blm}
[1 - \beta^{-2} \nabla^2] \, B_{n,\ell}^{m} (\beta, \bm{r}) \; = \;
B_{n-1,\ell}^{m} (\beta, \bm{r}) \, ,
\end{equation}
which shows that the differential operator $1 - \beta^{-2} \nabla^2$ of
the modified Helmholtz equation acts as a ladder operator in the case of
$B$ functions, it makes sense to define a distributional $B$ function as
the following derivative of the three-dimensional Dirac delta function
\cite[Eq.\ (6.20)]{Weniger/Steinborn/1983c}:
\begin{align}
  \label{Def_Distrib_B_Fun}
B_{-k-\ell,\ell}^{m} (\beta, \bm{r}) & \; = \;
\frac{(2\ell-1)!! \, 4\pi}{\beta^{\ell+3}} \,
\left[1 - \beta^{-2} \nabla^2 \right]^{k-1} \,
\delta_{\ell}^{m} (\bm{r}) \, ,
\notag \\
& \qquad k \in \mathbb{N} \, .
\end{align}
It is also easy to apply an integral power of the Laplacian to a $B$
function.  The binomial expansion of $\beta^{-2\nu} \, \nabla^{2\nu}$ in
powers of $1 - \beta^{-2} \nabla^2$ in combination with
(\ref{Shift_n_Blm}) yields \cite[Eq.\ (5.7)]{Weniger/Steinborn/1983c}:
\begin{equation}
  \label{Delta2nB_Fun}
\frac{\nabla^{2\nu}}{\beta^{2\nu}} \, B_{n,\ell}^{m} (\beta, \bm{r})
\; = \; \sum_{t=0}^{\nu} \, (-1)^t \, {\binom{\nu} {t}} \,
B_{n-t,\ell}^{m} (\beta, \bm{r}) \, .
\end{equation}

The $B$ function relationships given above show that the application of
the spherical tensor gradient operator multiplied by integral powers of
the Laplacian to the Yukawa potential leads to remarkably compact
expressions. Since (\ref{Delta2nB_Fun}) also holds for distributional $B$
functions of the type of (\ref{Def_Distrib_B_Fun}), it is almost
trivially simple to keep track of delta function contributions.

If only first and second order derivatives with respect to the Cartesian
components of the Coulomb or the Yukawa potential are needed, it is not
really necessary to do the differentiations via
(\ref{ExpandYlm_P_n(Nabla)}), and it is also not too difficult to keep
track of delta function contributions. If, however, differentiations of
(very) high orders have to be done, the tensorial version
(\ref{ExpandYlm_P_n(Nabla)}) offers substantial advantages.

For example, the two-range addition theorems discussed in
\cite{Weniger/2000a,Weniger/2002} and in \cite[Section 7]{Weniger/2005}
were derived via the following decomposition of the translation operator
in (\ref{ExpDifOp}) in terms of tensorial invariants:
\begin{align}
  \label{ST_TransOp}
\mathrm{e}^{\bm{r}_{<} \cdot \nabla_{>}} & \; = \;
2 \pi \, \sum_{\ell=0}^{\infty} \, \sum_{m=-\ell}^{\ell} \,
\left[ \mathcal{Y}_{\ell}^{m} (\bm{r}_{<}) \right]^{*} \,
\mathcal{Y}_{\ell}^{m} (\nabla_{>})
\notag \\
& \qquad \times \, \sum_{k=0}^{\infty} \,
\frac{\bm{r}_{<}^{2k} \, \nabla_{>}^{2k}}
{2^{\ell+2k} k! (1/2)_{\ell+k+1}} \, .
\end{align}
Apparently, this expansion was first published by Santos \cite[Eq.\
(A.6)]{Santos/1973}, who emphasized that it should be useful for the
derivation of addition theorems, although he never used it for that
purpose. However, the addition theorems derived in
\cite{Weniger/2000a,Weniger/2002} and in \cite[Section 7]{Weniger/2005}
show that the tensorial decomposition (\ref{ST_TransOp}) is indeed a
practically useful mathematical tool, although it involves irreducible
spherical tensors of arbitrary order.

After this essentially technical digression on Guseinov's way of
differentiating functions $f \colon \mathbb{R}^{3} \to \mathbb{C}$ with
respect to the Cartesian components of the argument vector, more
fundamental questions such as the existence and convergence of
derivatives of one-range addition theorems will be discussed.

As emphasized already several times, two-range addition theorems are
essentially rearranged three-dimensional Taylor expansions in the
Cartesian components of the shift vector, which converge pointwise and in
suitable open sets even uniformly. Accordingly, two-range addition
theorems represent ordinary functions, which are locally smooth and can
be differentiated. Thus, it is possible to generate more complicated
two-range addition theorems by differentiating simpler ones (see for
example \cite{Weniger/Steinborn/1989b,Weniger/Steinborn/1985}).

In contrast, Guseinov's one-range addition theorems discussed in Sections
\ref{Sec:AddTheor_STF} and \ref{Sec:WeakConvAddTheor_CoulombPot} -- if
they exist at all -- either converge in the mean with respect to the norm
of an appropriate Hilbert space or they converge weakly, i.e., they only
make sense in suitable functionals restricted to subsets of certain
Hilbert spaces. As is well known, smoothness is neither necessary for
strong or weak convergence nor implied by it. Thus, it is not clear
whether Guseinov's one-range addition theorems represent sufficiently
smooth functions that can be differentiated, and if yes, in which sense
they can be differentiated.

These problems are not uncommon in mathematics. Often, it is
comparatively easy to prove convergence in the mean, but very difficult
or even impossible to prove pointwise convergence or even uniform
convergence directly. Scenarios like this one obviously raise the
question, under which conditions uniform convergence is implied by
convergence in the mean. A condensed review of these issues and a
discussion of some elementary results can be found in a recent article by
Ford and Pennline \cite{Ford/Pennline/2007}. Related topics, albeit with
an emphasis on distribution theory, are discussed in the book by
Strichartz \cite[Chapter 8]{Strichartz/2003}.

Guseinov's derivatives of strongly convergent one-range addition theorems
for Slater-type functions can be made mathematically meaningful if it can
be shown that these addition theorems do not only converge in the mean,
but also pointwise and uniformly. 

Probably, there is some hope in the case of the one-range addition
theorems (\ref{Gus_OneRangeAddTheorSTF_k}), which are expansions in terms
of Guseinov's functions $\prescript{}{k}{\Psi}_{n, \ell}^{m} (\gamma,
\bm{r})$ and which converge with respect to the norm (\ref{Norm_r^k_2})
of the weighted Hilbert space $L_{r^k}^{2} (\mathbb{R}^3)$, if the
principal quantum number $N$ of the Slater-type function $\chi_{N, L}^{M}
(\beta, \bm{r} \pm \bm{r}')$ is a positive integer $N \in \mathbb{N}$.
We would have to check whether the additional conditions discussed in
Szeg\"{o}'s book \cite[Theorem 9.1.5 on p.\ 246]{Szegoe/1967} are
satisfied in this case.  Such an approach may also be successful in the
case of the derivatives of the one-range addition theorems for Guseinov's
functions $\prescript{}{k}{\Psi}_{n, \ell}^{m} (\beta, \bm{r})$
considered in \cite{Guseinov/2005d,Guseinov/2005e}.

I am, however, very skeptical about all addition theorems for Slater-type
functions $\chi_{N, L}^{M} (\beta, \bm{r} \pm \bm{r}')$ with nonintegral
principal quantum number $N \in \mathbb{R} \setminus \mathbb{N}$.

The situation is more complicated if weakly convergent one-range addition
theorems of the type of (\ref{CP_OnerangeAddThm_GusFun_k}) are to be
differentiated. Weakly convergent addition theorems are essentially
expansions of generalized functions in the sense of Schwartz
\cite{Schwartz/1966a}, whose derivatives also have to be interpreted in
this sense. If $f$ is a generalized function and $\psi \colon
\mathbb{R}^3 \to \mathbb{C}$ is a suitable test function, derivatives of
$f$ are defined by the following functional \cite[Chapter
II]{Schwartz/1966a}:
\begin{align}
  \label{Def:DiffGenFun}
  & \int \left[ \left( \frac{\partial}{\partial x} \right)^u \left(
      \frac{\partial}{\partial y} \right)^v
    \left(\frac{\partial}{\partial z} \right)^w f (\bm{r}) \right] \psi
  (\bm{r}) \, \mathrm{d}^3 \bm{r}
  \notag \\
  & \quad = \; (-1)^{u+v+w}
  \notag \\
  & \quad \times \, \int f (\bm{r}) \left[ \left(
      \frac{\partial}{\partial x} \right)^u \left(
      \frac{\partial}{\partial y} \right)^v
    \left(\frac{\partial}{\partial z} \right)^w \, \psi (\bm{r}) \right]
  \, \mathrm{d}^3 \bm{r} \, .
\end{align}
If $\psi \in \mathcal{S} (\mathbb{R}^3)$ or $\psi \in \mathcal{D}
(\mathbb{R}^3)$, $\psi$ is infinitely differentiable and it is possible
to define \emph{weak} derivatives of arbitrary order of a generalized
function $f$ via (\ref{Def:DiffGenFun}).

There is a detailed mathematical literature on derivatives of generalized
functions and their interpretation and application. For example, there is
a monograph by Ziemer on weakly differentiable functions
\cite{Ziemer/1989}. Unfortunately, the mathematical literature is not
particularly helpful if we want to apply derivatives of one-range
addition theorems in multicenter integrals and if we insist on using
exponentially decaying functions. Mathematicians almost exclusively use
infinitely differentiable test functions belonging to either $\mathcal{S}
(\mathbb{R}^3)$ or $\mathcal{D} (\mathbb{R}^3)$. However, Kato's work on
cusps \cite{Kato/1957} shows that atomic and molecular wave functions are
not infinitely differentiable. This is also true if we approximate atomic
and molecular wave functions in variational calculations by exponentially
decaying functions.

If we want to use one-range addition theorems, which are derivatives of
generalized functions, in multicenter integrals, we are confronted with
the annoying problem that the remainder of the integrand, which assumes
the role of the test function, not only has to decay sufficiently rapidly
for large arguments, but also has to possess continuous derivatives of a
sufficiently high order. Obviously, this complicates considerably our
attempts to prove that the resulting expansions converge to the correct
results.  Mathematicians know why they prefer infinitely differentiable
test functions belonging to either $\mathcal{S} (\mathbb{R}^3)$ or
$\mathcal{D} (\mathbb{R}^3)$.

To illuminate these problems, let us apply the spherical tensor gradient
operator $\mathcal{Y}_{L}^{M} (\nabla) = \mathcal{Y}_{L}^{M}
(\nabla_{\bm{r}})$ to the one-range addition theorems
(\ref{CP_OnerangeAddThm_GusFun_k}) for the Coulomb potential. Then, we
formally obtain with the help of (\ref{YllNabla_Coul}) one-range addition
theorems for the irregular solid harmonic:
\begin{align}
  \label{YllNabla_CoulAddThm}
  & \mathcal{Z}_{L}^{M} (\bm{r}-\bm{r}') \; = \; 
   \frac{(-1)^{L}}{(2L-1)!!} \, \mathcal{Y}_{L}^{M} (\nabla) \, 
   \frac{1}{\vert \bm{r}-\bm{r}' \vert}
  \notag \\
  & \quad \; = \; 
   \frac{(-1)^{L}}{(2 L - 1)!!} \, \sum_{\substack{n \ell m \\ n'
    \ell' m'}} \, \prescript{}{k}{\mathbf{\Gamma}}_{n' \ell' m'}^{n
    \ell m} (\beta)
  \notag \\
  & \quad \quad \times \, \left[ \mathcal{Y}_{L}^{M} (\nabla) \,
    \prescript{}{k}{\Psi}_{n, \ell}^{m} (\beta, \bm{r}) \right] \, 
    \prescript{}{k}{\Psi}_{n', \ell'}^{m'} (\beta, \bm{r}') \, .
\end{align}
An essentially identical expression -- the factor $(-1)^{L}$ would be
missing -- can be derived by differentiating $1/\vert \bm{r}-\bm{r}'
\vert$ instead with respect to $\bm{r}'$.

If we use the addition theorems (\ref{YllNabla_CoulAddThm}) in a
multicenter integral, we essentially have two options: Either, we can try
to construct an explicit expression for the product $\mathcal{Y}_{L}^{M}
(\nabla) \, \prescript{}{k}{\Psi}_{n, \ell}^{m} (\beta, \bm{r})$, or we
could use (\ref{Def:DiffGenFun}) and apply $\mathcal{Y}_{L}^{M} (\nabla)$
to the $\bm{r}$-dependent part of the remainder of the integrand, which
assumes the role of the test function.

If we apply $\mathcal{Y}_{L}^{M} (\nabla)$ to $\prescript{}{k}{\Psi}_{n,
  \ell}^{m} (\beta, \bm{r})$ and use one of the numerous expressions for
$\gamma _{\ell_1 \ell_2}^{\ell} (r)$ defined by
(\ref{YlmNab2Flm_GenStruc}), it should be possible to construct a closed
form expression for this product. However, the resulting functions of
$\bm{r}$ will no longer be orthogonal and they may also have a (much)
more complicated structure. 

Alternatively, we can use (\ref{Def:DiffGenFun}) and shift
$\mathcal{Y}_{L}^{M} (\nabla)$ to the remainder to the $\bm{r}$-dependent
part of the integrand. If this happens to be a nonclassical two-center
density, we have to apply the Leibniz theorem of the spherical tensor
gradient operator (see \cite[Eq.\ (4.22)]{Weniger/2005} and references
therein). Messy expressions are then likely, possibly involving singular
or distributional contributions. In either case, nontrivial technical
difficulties are to be expected. One must not forget that we still have
to prove that the resulting series expansion for the multicenter integral
converges to the correct result.

If we should indeed need weakly convergent one-range addition theorems of
the type of (\ref{CP_OnerangeAddThm_GusFun_k}) for the irregular solid
harmonic $\mathcal{Z}_{L}^{M} (\bm{r}-\bm{r}')$, then it is probably
easier to construct it from the scratch by expanding
$\mathcal{Z}_{L}^{M}$ in terms of Guseinov's functions
$\prescript{}{k}{\Psi}_{n, \ell}^{m} (\beta, \bm{r})$. I strongly suspect
that in this case it would be much easier to prove convergence of the
resulting expansion to the correct result.

One the basis of our current level of knowledge (or rather the lack of
it), the use of derivatives of weakly convergent addition theorems in
multicenter integrals would be purely experimental.
 
\typeout{==> Section: Summary and Conclusions}
\section{Summary and Conclusions}
\label{Sec:SummConclu}

The efficient and reliable evaluation of multicenter integrals is among
the oldest mathematical and computational problems of molecular
electronic structure theory (a review of the older literature can be
found in an article by Dalgarno \cite{Dalgarno/1954}). In spite of the
efforts of numerous researchers including Guseinov and coworkers, no
completely satisfactory solution has been found yet. The situation is
particularly unsatisfactory in the case of the notoriously difficult
integrals of exponentially decaying functions, but even in the case of
the much simpler integrals of Gaussians, there is still active research
going on (see for example \cite{ChowChiu/Moharerrzadeh/2001,Dunlap/1990,%
  Dunlap/2001,Dunlap/2002,Dunlap/2003,Dunlap/2005,%
  Reine/Tellgren/Helgaker/2007} and references therein).

Different centers in the integrand make the evaluation of an integral
difficult. They prevent the straightforward separation of a multicenter
into products of simpler integrals. As discussed in Section
\ref{Sec:Intro}, addition theorems, which are expansions of a given
function $f (\bm{r} \pm \bm{r}')$ with $\bm{r}, \bm{r}' \in \mathbb{R}^3$
in products of other functions depending on either $\bm{r}$ or $\bm{r}'$,
are principal tools that can accomplish a separation of variables, albeit
at the cost of infinite series expansions.

In electronic structure calculations, predominantly those addition
theorems have played a major role that depend on $\bm{r}$ and $\bm{r}'$
only indirectly via $\bm{r}_{<}$ and $\bm{r}_{>}$ and thus possess a
two-range form. The prototype of such an addition theorem is the Laplace
expansion (\ref{LapExp}) of the Coulomb potential. Unfortunately, the
use of two-range addition theorems in multicenter integrals can easily
lead to nontrivial technical problems. This explains why many authors
have tried to construct one-range addition theorems that can be applied
more easily.

Two-range addition theorems can be constructed by applying the
translation operator $\mathrm{e}^{\bm{r}_{<} \cdot \nabla_{>}}$ in its
tensorially invariant form (\ref{ST_TransOp}) to $f (\bm{r}_{>})$, which
is by assumption analytic at $\bm{r}_{>}$. Accordingly, two-range
addition theorems are essentially rearranged three-dimensional Taylor
expansions in the Cartesian components of the shift vector $\bm{r}_{<}$
that converge pointwise and in suitable open sets even uniformly. If $f$
is not analytic at the origin, its addition theorem converges only if it
has a two-range form.

Classical and complex analysis is mainly concerned with power series that
converge pointwise. However, pointwise convergence is a very demanding
requirement, and its scope is too limited to cover all cases of interest.
If we want to avoid the troublesome two-range form of addition theorems,
we have to replace pointwise convergence by a weaker form of convergence.

One-range addition theorems can be constructed by exploiting Hilbert
space theory, whose basic facts are reviewed in Section
\ref{Sec:HilbertSpace}. In this approach, a function $f (\bm{r} \pm
\bm{r}')$ belonging to a suitable Hilbert space is according to
(\ref{OneRangeAddTheor}) expanded in terms of a function set $\{
\varphi_{n, \ell}^{m} (\bm{r}) \}_{n, \ell, m}$ that is complete and
orthonormal in this Hilbert space. In general, such an expansion
converges only in the mean, i.e., with respect to the norm of the
underlying Hilbert space, but not pointwise.

The natural Hilbert space for electronic structure calculations based on
effective one-particle wave functions is the Hilbert space $L^{2}
(\mathbb{R}^3)$ of square integrable functions defined by
(\ref{HilbertL^2}). As discussed in Section \ref{Sec:OneRanAddTheor}, it
is also possible to construct one-range addition theorems that converge
with respect to the norm of a suitable weighted Hilbert space $L_{w}^{2}
(\mathbb{R}^3)$ defined by (\ref{HilbertL_w^2}). Here, $w (\bm{r}) \ge 0$
is a suitable weight function. There is, however, a principle problem: In
general, we neither have $L^{2} (\mathbb{R}^3) \subset L_{w}^{2}
(\mathbb{R}^3)$ nor $L_{w}^{2} (\mathbb{R}^3) \subset L^{2}
(\mathbb{R}^3)$. Accordingly, it is by no means easy to find a nontrivial
weight function $w (\bm{r}) \ne 1$ that leads to substantial
computational benefits and simultaneously avoids major drawbacks.

If we demand that the complete and orthonormal functions $\{ \varphi_{n,
  \ell}^{m} (\bm{r}) \}_{n, \ell, m}$ are also irreducible spherical
tensors of the type of (\ref{Def_IrrSphericalTensor}), we more or less
automatically arrive at functions based on the generalized Laguerre
polynomials. Section \ref{Sec:LagTypeFun} discusses Lambda functions
$\Lambda_{n, \ell}^{m} (\beta, \bm{r})$ defined by (\ref{Def_LambdaFun}),
Sturmians $\Psi_{n, \ell}^{m} (\beta, \bm{r})$ defined by
(\ref{Def_SturmFun}), and Guseinov's functions $\prescript{}{k}{\Psi}_{n,
  \ell}^{m} (\beta, \bm{r})$ with $k = -1, 0, 1, 2, \dots$ defined by
(\ref{Def_Psi_Guseinov}). 

Guseinov's functions satisfy the orthonormality relationship
(\ref{Psi_Guseinov_OrthoNor}) involving the weight function $w (\bm{r}) =
r^k$. Accordingly, they are complete and orthonormal in the weighted
Hilbert space $L_{r^k}^{2} (\mathbb{R}^3)$ with $k = -1, 0, 1, 2, \dots$
defined by (\ref{HilbertL_r^k^2}). Guseinov's functions provide some
unification since they contain for $k=0$ and $k=-1$, respectively, Lambda
functions and Sturmians as special cases. However, this unification does
not apply to approximation processes. For $k \ne 0$, we neither have
$L_{r^k}^{2} (\mathbb{R}^3) \subset L^{2} (\mathbb{R}^3)$ nor $L^{2}
(\mathbb{R}^3) \subset L_{r^k}^{2} (\mathbb{R}^3)$. Thus, $f \in L^{2}
(\mathbb{R}^3)$ does not imply that an expansion of $f$ in terms of
Guseinov's functions with $k \ne 0$ necessarily converges in $L_{r^k}^{2}
(\mathbb{R}^3)$.
 
One-range addition theorems for exponentially decaying functions had been
considered by several researchers before Guseinov. The so far most
compact results were obtained by Filter and Steinborn
\cite{Filter/Steinborn/1980} who derived the symmetrical one-range
addition theorems (\ref{OneRangeAddTheor_Lambda}) and
(\ref{OneRangeAddTheor_B}) for Lambda and $B$ functions, respectively, by
expanding them in terms of Lambda functions, which are complete and
orthonormal in $L^{2} (\mathbb{R}^3)$. In Section \ref{AddTheor_ETF}, it
is shown that the approach of Filter and Steinborn, which was based on
the convolution theorem (\ref{ConvInt_Bnlm_ESP}) of $B$ functions, works
also in case of Slater-type functions. The approach of Filter and
Steinborn can be generalized to expansions in terms of Guseinov's
functions. However, these addition theorems have a more complicated
structure than the corresponding expansions in terms of Lambda functions.

Section \ref{Sec:AddTheor_STF} describes Guseinov's mathematically
dubious attempts of deriving one-range addition theorems for Slater-type
functions. Guseinov first derived the expansions
(\ref{Gus_OneRangeAddTheorSTF_k}) of a Slater-type function $\chi_{N,
  L}^{M} (\beta, \bm{r} \pm \bm{r}')$ with integral or nonintegral
principal quantum number $N$ in terms of his functions
$\prescript{}{k}{\Psi}_{n, \ell}^{m} (\gamma, \bm{r})$ with $\gamma > 0$.
Guseinov's results are mathematically correct, although not necessarily
optimal. In a later step, Guseinov replaced his orthonormal functions by
nonorthogonal Slater-type functions with integral principal quantum
numbers according to (\ref{GusFun2STF}) and rearranged the order of
summations in his addition theorems. In this way, Guseinov obtained the
expansions (\ref{Gus_RearrOneRangeAddTheorSTF_STF_k}) of $\chi_{N, L}^{M}
(\beta, \bm{r} \pm \bm{r}')$ in terms of Slater-type functions $\chi_{n,
  \ell}^{m} (\gamma, \bm{r})$ with integral principal quantum numbers $n
\in \mathbb{N}$.

As reviewed in Section \ref{Sec:HilbertSpace}, a function $f$ belonging
to a given Hilbert space can be expanded in terms of complete and
orthogonal functions. Such an expansion converges with respect to the
norm of the Hilbert space, and the coefficients of this expansion satisfy
Parseval's equality (\ref{ParsevalEquality}). Thus, the expansion
coefficients are bounded in magnitude and they also vanish for large
indices.

If the orthogonal functions are replaced by nonorthogonal functions, it
is possible to construct finite approximations of the type of
(\ref{f_FinAppr}). Unfortunately, it is not guaranteed that an infinite
expansion of the type of (\ref{Expand_f_formal}) exists, and if it does,
it can happen that the coefficients of this expansion are unbounded and
do not vanish for large indices. 

Accordingly, Guseinov's manipulations, which produced the rearranged
addition theorem (\ref{Gus_RearrOneRangeAddTheorSTF_STF_k}) from
(\ref{Gus_OneRangeAddTheorSTF_k}), are dangerous. Thus, he should have
proved that his manipulations are legitimate and lead to a meaningful
result.

Addition theorems of exponentially decaying functions are fairly
complicated mathematical objects. Therefore, it is very difficult or
practically even impossible to prove explicitly that a given addition
theorem converges or diverges.  This applies also to Guseinov's
rearranged addition theorem (\ref{Gus_RearrOneRangeAddTheorSTF_STF_k})
for Slater-type functions $\chi_{N, L}^{M} (\beta, \bm{r} \pm \bm{r}')$.
However, as shown in Section \ref{Sec:AddTheor_STF}, the one-center limit
$\bm{r}' = \bm{0}$ of Guseinov's addition theorem
(\ref{Gus_RearrOneRangeAddTheorSTF_STF_k}) does not exist if the
principal quantum $N$ is not a positive integer.  Thus, for $N \in
\mathbb{R} \setminus \mathbb{N}$ the rearranged addition theorem
(\ref{Gus_RearrOneRangeAddTheorSTF_STF_k}) does not exist for the whole
argument set $\mathbb{R}^3 \times \mathbb{R}^3$. The remaining question,
whether the rearranged addition theorem
(\ref{Gus_RearrOneRangeAddTheorSTF_STF_k}) exists for $N \in \mathbb{N}$,
cannot be decided in this way and is still open.

The nonexisting rearranged addition theorems
(\ref{Gus_RearrOneRangeAddTheorSTF_STF_k}) with $N=L=M=0$ were the
starting point for Guseinov's construction \cite{Guseinov/2005a} of
one-range addition theorems for the Coulomb potential, but this is not
the only problem. The Coulomb potential does not belong to any of the
Hilbert spaces implicitly used by Guseinov, since they all involve an
integration over the whole $\mathbb{R}^3$.  Accordingly, expansions of
$1/ \vert \bm{r} - \bm{r}' \vert$ in terms of Guseinov's functions
$\prescript{}{k}{\Psi}_{n, \ell}^{m} (\beta, \bm{r})$, which formally
yield the symmetrical one-range addition theorems
(\ref{CP_OnerangeAddThm_GusFun_k}), diverge for any $k = -1, 0, 1, 2,
\dots$ in the mean with respect to the norm of the weighted Hilbert space
$L_{r^k}^{2} (\mathbb{R}^3)$.

This observation seems to imply that all attempts of constructing a
symmetrical one-range addition theorem for $1/ \vert \bm{r} - \bm{r}'
\vert$ by expanding it in terms of functions, that are complete and
orthonormal with respect to an inner product involving an integration
over the whole $\mathbb{R}^3$, are futile. Although seemingly obvious,
this conclusion is premature and the situation is better than it may
look.

It is possible to use divergent one-range addition theorems like
(\ref{CP_OnerangeAddThm_GusFun_k}) in multicenter integrals in a
mathematically rigorous way. However, convergence in the mean, which in
the mathematical literature is frequently called strong convergence, is
too demanding for that. We have to replace it by an even weaker form of
convergence. Instead, we require that a possibly divergent one-range
addition theorem produces meaningful results when used in a multicenter
integral. Obviously, this approach is inspired by the theory of
generalized functions in the sense of Schwartz \cite{Schwartz/1966a}.

Weak convergence of addition theorems is all we really need in
variational electronic structure calculations. Pointwise convergence or
convergence in the mean -- although undeniably convenient -- are not
really necessary. Thus, the use of weakly convergent addition theorems
offers new perspectives and simplifies or even solves certain technical
problems. However, it would be overly optimistic to expect that we could
get all these benefits without having to pay a price.

The key problem, which makes the application of weakly convergent
addition theorems difficult, is the formulation of sufficiently simple
criteria which guarantee the convergence of the resulting expansions for
multicenter integrals. In that respect, a lot of work remains to be done.

Regularity conditions -- or rather their absence -- also play a major
role in Section \ref{Sec:DiffTechYlmNabla}. In several articles, Guseinov
had constructed more complicated one-range addition theorems by
differentiating simpler one-range addition theorems with respect to the
Cartesian components of their argument vectors. Guseinov's one-range
addition theorems -- if they exist at all -- either converge in the mean
with respect to the norm of an appropriate weighted Hilbert space or they
converge weakly, i.e., they only make sense as generalized functions in
suitable functionals. It is thus not at all clear whether Guseinov's
addition represent functions in the ordinary sense that are locally
smooth and can be differentiated. We need to know much more before we can
safely apply derivatives of one-range addition theorems in multicenter
integrals.

Section \ref{Sec:DiffTechYlmNabla} also discusses a more convenient
alternative to the troublesome differentiation of irreducible spherical
tensors of the type of (\ref{Def_IrrSphericalTensor}) with respect to the
Cartesian components of their argument vectors. The alternative proposed
here is based on the reformulation (\ref{ExpandYlm_P_n(Nabla)}) of a
differential operator as a finite sum of integral powers of the Laplacian
multiplied by irreducible spherical tensor gradient operators
$\mathcal{Y}_{\ell}^{m} (\nabla)$ defined by (\ref{Def:YlmNabla}). This
differential operator plays a decisive role in the derivation of
two-range addition theorems as rearranged Taylor expansions
\cite{Weniger/2000a,Weniger/2002,Weniger/2005}.

At first sight, the use of (\ref{ExpandYlm_P_n(Nabla)}) looks like a bad
idea. However, $\mathcal{Y}_{\ell}^{m} (\nabla)$ is an irreducible
spherical tensor of rank $\ell$. Consequently. products
${\mathcal{Y}}_{\ell_1}^{m_1} (\nabla) F_{\ell_2}^{m_2} (\bm{r})$ can be
simplified considerably with the help of the powerful machinery of
angular momentum coupling, yielding the finite linear combination
(\ref{YlmNab2Flm_GenStruc}) of Gaunt coefficients, radial functions, and
spherical harmonics. Thus, no differentiations with respect to the
Cartesian components of $\bm{r}$ are needed. We only have to
differentiate with respect to the radial variable $r = \vert \bm{r}
\vert$.

It is the intention of this article to demonstrate that the analytical
tools of classical and complex analysis do not suffice for a successful
treatment of one-range addition theorems. More modern and also more
sophisticated concepts such as Hilbert spaces and generalized functions
are indispensable. 

In particular, we need different concepts of convergence. If a function
$f$ belongs to a suitable Hilbert space, one-range addition theorems can
be constructed that converge in the mean with respect to the norm of that
Hilbert space, even if $f$ is not analytic everywhere in $\mathbb{R}^3$.
If, however, $f$ does mot belong to that Hilbert space, one-range
addition theorems can only converge weakly in the sense of generalized
functions.

Analogous considerations apply also with respect to numerical techniques.
In the context of multicenter integrals, addition theorems are
essentially mathematical recipes that generate infinite series
expansions. We have no \emph{a priori} reason to assume that these series
expansions converge rapidly.  Thus, our assessment of the usefulness of
addition theorems depends crucially on our ability of evaluating slowly
convergent or even divergent series effectively and reliably.

The conventional process of adding up the terms of an infinite series
successively until convergence is finally achieved is in far too many
cases hopelessly inefficient. It is thus an obvious idea to use
convergence acceleration and summation techniques. Of course, we do not
know whether the currently known convergence acceleration and summation
techniques, which were mentioned shortly in Section
\ref{Sec:WeakConvAddTheor_CoulombPot}, are powerful enough to produce
satisfactory results or whether it will be necessary to construct new and
hopefully more powerful transformations. This has to be investigated.

Let me emphasize: If we want to evaluate even difficult multicenter
integrals effectively and reliably via series expansion, then there is no
alternative to the systematic and intelligent use of convergence
acceleration and summation techniques.

It is dangerous to rely too much on completeness. The completeness of a
function set in a Hilbert space only implies that any $f$ belonging to
this Hilbert space can be expanded in terms of these functions and that
this expansion converges in the mean. Unfortunately, convergence is a
very weak statement. In particular, the convergence of an infinite series
does not imply that this series is numerically useful. A simple and yet
striking example is provided by the Dirichlet series $\zeta (s) =
\sum_{n=0}^{\infty} (n+1)^{-s}$ for the Riemann zeta function. This
series converges if $\Re (s) > 1$, but is notorious for extremely slow
convergence if $\Re (s)$ is only slightly larger than 1.  For example, in
\cite[p.\ 194]{Weniger/Kirtman/2003} it was estimated that in the order
of $10^{600}$ terms of the Dirichlet series would be needed to compute
$\zeta (1.01)$ with an accuracy of 6 decimal digits.

\begin{appendix}
\typeout{==> Appendix: Spherical Harmonics and Gaunt Coefficients}
\section{Spherical Harmonics and Gaunt Coefficients}
\label{App:YlmGaunt}

If we choose the phase convention of Condon and Shortley \cite[Chapter
III.4]{Condon/Shortley/1970}, the spherical harmonic can be expressed as
follows \cite[p.\ 69]{Biedenharn/Louck/1981a}:
\begin{align}
  \label{Def_Ylm}
Y_{\ell}^{m} (\theta, \phi) & \; = \; \mathrm{i}^{m + \vert m \vert} \,
\left[ \frac{(2 \ell + 1)(\ell - \vert m \vert)!}
{4 \pi (\ell + \vert m \vert)!} \right]^{1/2}
\notag \\
& \qquad \times \,
P_{\ell}^{\vert m \vert} (\cos \theta) \,
{\mathrm{e}}^{\mathrm{i} m \phi} \, .
\end{align}
Here, $P_{\ell}^{\vert m \vert} (\cos \theta)$ is an associated Legendre
polynomial \cite[p.\ 155]{Condon/Odabasi/1980}. Alternative phase
conventions for the spherical harmonics are discussed in \cite[pp. 17 -
22]{Steinborn/Ruedenberg/1973}.

The spherical harmonics $Y_{\ell}^{m} (\theta, \phi)$ are often called
surface harmonics because the angles $\theta$ and $\phi$ characterize a
point $\bm{r}/r$ on the surface of the three-dimensional unit sphere. In
the literature, it is common to introduce the so-called regular and
irregular solid harmonics
\begin{align}
  \label{Def_RegSolHar}
  {\mathcal{Y}}_{\ell}^{m} (\bm{r}) & \; = \;
  r^{\ell} Y_{\ell}^{m} (\theta, \phi) \, ,
  \\
  \label{Def_IrregSolHar}
  {\mathcal{Z}}_{\ell}^{m} (\bm{r}) & \; = \;
  r^{-\ell-1} Y_{\ell}^{m} (\theta, \phi) \, .
\end{align}
The regular solid harmonic ${\mathcal{Y}}_{\ell}^{m} (\bm{r})$ is a
homogeneous harmonic polynomial of degree $\ell$ in the Cartesian
components of $\bm{r} = (x, y, z)$ \cite[p.\ 71]{Biedenharn/Louck/1981a}:
\begin{align}
  \label{YlmHomPol}
  & \mathcal{Y}_{\ell}^{m} (\bm{r}) \; = \; \left[ \frac{2\ell+1}{4\pi}
    (\ell+m)!(\ell-m)! \right]^{1/2}
  \notag \\
  & \qquad \times \, \sum_{k \ge 0} \, \frac {(-x-\mathrm{i}y)^{m+k}
    (x-\mathrm{i}y)^{k} z^{\ell-m-2k}} {2^{m+2k} (m+k)! k! (\ell-m-2k)!}
  \, .
\end{align}
Moreover, $\mathcal{Y}_{\ell}^{m} (\bm{r})$ is for all $\bm{r} \in
\mathbb{R}^3$ a solution of the homogeneous three-dimensional Laplace
equation
\begin{equation}
  \label{HomogLaplaceEqn}
\nabla^2 \, f (\bm{r}) \; = \;
\left[ \frac{\partial^2}{\partial x^2} +
\frac{\partial^2}{\partial y^2} +
\frac{\partial^2}{\partial z^2} \right] \, f (\bm{r}) \; = \; 0 \, ,
\end{equation}
whereas the irregular solid harmonics are generalized solutions of the
Poisson equation (\ref{PoissonEq_Zlm}) of a unit multipole charge,
yielding the spherical delta function (\ref{Def_delta_lm}).

The so-called Gaunt coefficient \cite{Gaunt/1929} is the integral of the
product of three spherical harmonics over the surface of the unit sphere
in ${\mathbb{R}}^3$:
\begin{align}
  \label{Def_Gaunt}
  & \langle \ell_3 m_3 \vert \ell_2 m_2 \vert \ell_1 m_1 \rangle
  \notag \\
  & \qquad \; = \; \int \, \bigl[ Y_{\ell_3}^{m_3} (\Omega) \bigr]^{*} \,
  Y_{\ell_2}^{m_2} (\Omega) \, Y_{\ell_1}^{m_1} (\Omega) \, {\mathrm{d}}
  \Omega \, .
\end{align}
It follows from the orthonormality of the spherical harmonics that Gaunt
coefficients linearize the product of two spherical harmonics:
\begin{align}
  \label{Ylm_lin}
& Y_{\ell_1}^{m_1} (\Omega) \, Y_{\ell_2}^{m_2} (\Omega)
\notag \\
& \quad = \,
\sum_{\ell=\ell_{\mathrm{min}}}^{\ell_{\mathrm{max}}}
\! {}^{(2)} \,
\langle \ell m_1+m_2 \vert \ell_1 m_1 \vert \ell_2 m_2 \rangle
Y_{\ell}^{m_1+m_2} (\Omega) \, .
\end{align}
The symbol $\sum \! {}^{(2)}$ indicates that the summation proceeds in
steps of two. The summation limits in (\ref{Ylm_lin}) are given by
\cite[Eq.\ (3.1)]{Weniger/Steinborn/1982}
\begin{subequations}
  \label{SumLims}
\begin{align}
  \label{SumLims_a}
  \ell_{\mathrm{max}} & \; = \; \ell_1 + \ell_2 \, ,
  \\
  \label{SumLims_b}
  \ell_{\mathrm{min}} & \; = \;
\begin{cases}
  \lambda_{\mathrm{min}} \, , \qquad \; \, \, \text{if} \;
  \ell_{\mathrm{max}} + \lambda_{\mathrm{min}} \; \text{is even} \, ,
  \\[1.5\jot]
  \lambda_{\mathrm{min}} + 1 \, , \quad \text{if} \; \ell_{\mathrm{max}}
  + \lambda_{\mathrm{min}} \; \text{is odd} \, ,
\end{cases} \\
   \label{SumLims_c}
   \lambda_{\mathrm{min}} & \; = \; \max (\vert \ell_1 - \ell_2 \vert,
   \vert m_1 + m_2 \vert) \, .
\end{align}
\end{subequations}
A compact review of the properties of Gaunt coefficients and additional
references can be found in \cite[Appendix C]{Weniger/2005}.

In this article, the following abbreviations are used:
\begin{align}
  \label{Def_Del_l}
\Delta \ell & \; = \; (\ell_1 + \ell_2 - \ell)/2 \, ,
\\
  \label{Def_Del_l_1}
\Delta \ell_1 & \; = \; (\ell - \ell_1 + \ell_2)/2 \, ,
\\
  \label{Def_Del_l_2}
\Delta \ell_2 & \; = \; (\ell + \ell_1 - \ell_2)/2 \, ,
\\
  \label{Def_sigma_l}
\sigma (\ell) & \; = \; (\ell_1 + \ell_2 + \ell)/2 \, .
\end{align}
If the three orbital angular momentum quantum numbers $\ell_1$, $\ell_2$,
and $\ell$ satisfy the summation limits (\ref{SumLims}), then these
quantities are either positive integers or zero.


\end{appendix}

%
%
%
{\small

\providecommand{\SortNoop}[1]{} \providecommand{\OneLetter}[1]{#1}
  \providecommand{\SwapArgs}[2]{#2#1}

}
%

\end{multicols}

\end{document}